\documentclass[11pt,letterpaper]{article}
\pdfoutput=1
\usepackage{jheppub}
\usepackage{color}
\usepackage{graphicx}
\usepackage{tabularx}
\usepackage{xspace}
\usepackage{empheq}
\usepackage{verbatim}
\usepackage{amsmath}
\usepackage{amssymb}
\usepackage[caption=false]{subfig}
\usepackage{url}
\usepackage{bbold}
\usepackage{slashed}
\usepackage{array}
\usepackage{aas_macros}
\usepackage{subfig}
\usepackage{graphicx}
\usepackage{subcaption}
\usepackage{gensymb}
\usepackage{multirow}
\usepackage{threeparttable}
\usepackage{paralist}
\usepackage{xspace}
\usepackage{upgreek}
\usepackage{lipsum}
\usepackage[T1]{fontenc}
\usepackage{scalerel}

\usepackage[framemethod=TikZ]{mdframed}
\usepackage{setspace}
\usepackage{mathtools}

\allowdisplaybreaks

\newcommand{\be}{\begin{equation}}
\newcommand{\bea}{\begin{equation}\begin{aligned}}
\newcommand{\ee}{\end{equation}}
\newcommand{\eea}{\end{aligned}\end{equation}}

\def\nn{{\nonumber}}

\DeclareMathAlphabet\mathbfcal{OMS}{cmsy}{b}{n}
\mathchardef\mhyphen="2D 

\newcommand{\aW}{\alpha_{\scriptscriptstyle W}}

\newcommand{\mW}{m_{\scriptscriptstyle W}}
\newcommand{\mZ}{m_{\scriptscriptstyle Z}}
\newcommand{\smallZ}{{\scriptscriptstyle Z}}
\newcommand{\smallW}{{\scriptscriptstyle W}}
\newcommand{\smallX}{{\scriptscriptstyle X}}

\newcommand{\thetaW}{\theta_{\scriptscriptstyle W}}
\newcommand{\sW}{s_{\scriptscriptstyle W}}
\newcommand{\cW}{c_{\scriptscriptstyle W}}

\newcommand{\mdm}{M_{\chi}}

\begin{document}

\preprint{\hbox{MIT-CTP/5893}}

\title{Testing Real WIMPs with CTAO}

\author[1]{\small Matthew Baumgart,}
\author[2]{\small Salvatore Bottaro,}
\author[3]{\small Diego Redigolo,}
\author[4,5]{\small Nicholas L. Rodd,}
\author[6]{\small and Tracy R. Slatyer}

\affiliation[1]{\footnotesize Department of Physics, Arizona State University, Tempe, AZ 85287, USA}
\affiliation[2]{\footnotesize Raymond and Beverly Sackler School of Physics and Astronomy, Tel-Aviv 69978, Israel}
\affiliation[3]{\footnotesize INFN, Sezione di Firenze, Via Sansone 1, 50019 Sesto Fiorentino, Italy}
\affiliation[4]{\footnotesize Theory Group, Lawrence Berkeley National Laboratory, Berkeley, CA 94720, USA}
\affiliation[5]{\footnotesize Berkeley Center for Theoretical Physics, University of California, Berkeley, CA 94720, USA}
\affiliation[6]{\footnotesize Center for Theoretical Physics -- a Leinweber Institute, Massachusetts Institute of Technology, Cambridge, MA 02139, USA}

\emailAdd{Matt.Baumgart@asu.edu}
\emailAdd{salvatoreb@tauex.tau.ac.il}
\emailAdd{diego.redigolo@fi.infn.it}
\emailAdd{nrodd@lbl.gov}
\emailAdd{tslatyer@mit.edu}

\abstract{We forecast the reach of the upcoming Cherenkov Telescope Array Observatory (CTAO) to the full set of real representations within the paradigm of minimal dark matter.
We employ effective field theory techniques to compute the annihilation cross section and photon spectrum that results when fermionic dark matter is the neutral component of an arbitrary odd and real representation of SU(2), including the Sommerfeld enhancement, next-to-leading log resummation of the relevant electroweak effects, and the contribution from bound states.
We also compute the corresponding signals for scalar dark matter, with the exception of the bound state contribution. 
Results are presented for all real representations from the $\sim$3\,TeV triplet (or wino), a $\mathbf{3}$ of SU(2), to the $\sim$300\,TeV tredecuplet, a $\mathbf{13}$ of SU(2) that is at the threshold of the unitarity bound.
Using these results, we forecast that with 500\,hrs of Galactic Center observations and assuming background systematics are controlled at the level of ${\cal O}(1\%)$, then should no signal emerge, CTAO could exclude all representations up to the $\mathbf{11}$ of SU(2) in even the most conservative models for the dark-matter density in the inner galaxy, in both the fermionic and scalar dark matter cases.
Assuming the default CTAO configuration, the tredecuplet will marginally escape exclusion, although we outline steps that CTAO could take to test even this scenario.
In summary, CTAO appears poised to make a definitive statement on whether real WIMPs constitute the dark matter of our universe. 
}

\maketitle
\setcounter{page}{2}

\section{Introduction}

What is the simplest extension one can make to the standard model (SM) that explains dark matter (DM)?
Whilst there is an enormous degree of subjectivity in any approach to this question, one compelling answer is provided by the framework of minimal dark matter (MDM)~\cite{Cirelli:2005uq}.
Within this paradigm, DM is the neutral component of a single new state appended to the SM that has an arbitrary charge under the electroweak gauge group.
Further, although the motivation for MDM is primarily bottom up, many well motivated UV scenarios fall into its classification, including the wino and higgsino DM of split supersymmetry~\cite{Arkani-Hamed:2004ymt,Giudice:2004tc,Arkani-Hamed:2004zhs,Arvanitaki:2012ps} (assuming the remaining supersymmetric states are effectively decoupled).
In many senses, MDM distills the essence of the WIMP hypothesis.

MDM is extremely predictive: once the representation is chosen, all interaction strengths are fixed by the known SM couplings which further determine the DM mass by invoking a minimal thermal cosmology.
While the possibility that the DM sits in a electroweak multiplet has been extensively explored~\cite{Cirelli:2005uq,Cirelli:2007xd,Cirelli:2009uv,Hambye:2009pw}, there is an increasing recognition that the next generation of experimental facilities—spanning colliders, direct detection, and indirect detection—could provide a definitive answer on this possibility see e.g. Refs.~\cite{Bottaro:2021snn,Bottaro:2022one,Bloch:2024suj}.
Given its simplicity, fully testing the predictions of MDM will represent a key milestone in the search for the nature of DM.

In the present work we demonstrate that the upcoming Cherenkov Telescope Array Observatory (CTAO) is well placed to test core predictions of MDM.\footnote{Although we do not consider it here, another instrument that is likely to have powerful sensitivity to MDM is the Large Array of imaging atmospheric Cherenkov Telescopes (LACT) currently being constructed at the site of the LHAASO detector~\cite{Zhang:2024ztw}.}
CTAO is the next generation imaging air-Cherenkov telescope (IACT)~\cite{CTAConsortium:2017dvg}.
The observatory will be made of two arrays: a northern site in La Palma that is already partially operational~\cite{CTA-LSTProject:2023haa} and a southern site in Chile that will become operational in the coming years~\cite{eric}.\footnote{For the searches proposed in the current work, the improved reach CTAO achieves over existing experiments is the result of combining multiple ${\cal O}(1)$ improvements.
The observatory is forecast to improve the background rejection efficiency, have a larger effective area and observational field of view, and further reconstruct photon energies more accurately.
See Ref.~\cite{Rodd:2024qsi} for a quantitative analysis of this point.}
Just as the annihilations of heavy WIMPs lead to a thermal abundance emerging from the early universe, those same annihilations in the present universe can lead to an observable signature of high energy photons emerging from regions of large DM density such as the Galactic Center.
Consequently, the southern CTAO site is particularly well placed to search for indirect signatures of heavy DM, with the center of the Milky Way readily visible overhead to its instruments.\footnote{Although we focus on the southern observatory in this work, we note that searches with the northern array should also be possible as recently emphasized in Ref.~\cite{Abe:2025lci}.}
We focus our attention on real fermionic WIMPs: the purely SU(2) or real representations of MDM.
These include a $\mathbf{3}$ (triplet/wino), $\mathbf{5}$ (quintuplet), up to the $\mathbf{13}$ (tredecuplet) and no higher as dictated by unitarity~\cite{Bottaro:2021snn}.

Our conclusions are as follows.
We find that with 500\,hrs of observations of the Galactic Center, CTAO-South -- assuming the Alpha Configuration of the instrument -- will provide a definitive test of the fermionic triplet, quintuplet, and septuplet ($\mathbf{7}$) scenarios based solely on the expected line plus endpoint signal of hard photons with $E_\gamma \simeq M_\chi$, computed at next-to-leading logarithmic (NLL) accuracy.
When adding the expected contribution from lower energy continuum photons, as arises from final state $W$ and $Z$ bosons and the decay of intermediate bound states, CTAO could probe all the fermionic electroweak multiplets up to the undecuplet ($\mathbf{11}$), even under the most pessimistic assumptions about the DM density in the inner Galaxy.
The sole exception to this statement is the tredecuplet, for which CTAO is expected to exclude more than 90\% of the thermal mass range, with the exception of two narrow intervals—445–450 TeV and 464–475 TeV.
Even in these regions, the predicted limits lie very close to the exclusion threshold, and a 30\% increase in CTAO’s sensitivity would be sufficient to achieve full coverage and deliver a conclusive test.
We further test the robustness of our conclusions with respect to background systematics.
In particular, we show that a percent-level control of the background is required to probe all electroweak multiplets up to the tredecuplet; although challenging, this is a level that has been achieved by HESS~\cite{HESS:2022ygk}.
By contrast, a 10\% systematic uncertainty on the background would only allow a complete test of the triplet and quintuplet.
Lastly, in App.~\ref{app:scalar} we show that a similar conclusion holds to the real scalar MDM representations.

Underpinning these forecasts is a detailed prediction of the gamma-ray indirect detection signal associated with annihilation of DM for an arbitrary real MDM representation.
Obtaining this prediction requires an accounting of the full set of effects that impact the DM annihilation rate and resultant photon spectrum: Sommerfeld enhancement~\cite{Hisano:2003ec,Hisano:2004ds,Cirelli:2007xd,Arkani-Hamed:2008hhe,Blum:2016nrz}, resummation of large electroweak and endpoint logarithms using effective field theory (EFT) techniques,\footnote{Specifically, our approach is built upon soft-collinear effective theory~\cite{Bauer:2000yr,Bauer:2001ct,Bauer:2001yt} and heavy quark EFT~\cite{Manohar:2000dt}.} and bound state contributions~\cite{Asadi:2016ybp,Mitridate:2017izz,Harz:2018csl}.
To account for these effects we extend the heavy DM EFT formalism of Refs.~\cite{Baumgart:2014vma,Bauer:2014ula,Ovanesyan:2014fwa,Baumgart:2014saa,Baumgart:2015bpa,Ovanesyan:2016vkk,Baumgart:2017nsr,Baumgart:2018yed} that was specifically developed for the wino, an SU(2) triplet.\footnote{An alternative approach, which has also been extended to the complex representation of the higgsino, was developed in Refs.~\cite{Beneke:2018ssm,Beneke:2019vhz,Beneke:2022eci,Beneke:2022pij}.}
More recently, in Ref.~\cite{Baumgart:2023pwn} that approach was extended to the quintuplet, a $\mathbf{5}$ of SU(2).
That work exemplified the power of the EFT approach, demonstrating that all of the core results for the wino could be factorized off from the DM representation, which could then be accounted for by minor modifications to the appropriate group theory factors and Sommerfeld contributions.
In the present work we continue in that spirit by extending the calculation to arbitrary real representations of SU(2).
This EFT formalism applies almost unchanged to the cases of fermionic and scalar MDM, but we will focus on the fermionic case in the main text and discuss the scalar case in App.~\ref{app:scalar}.

We divide the bulk of our discussion into two sections.
In the first of these, Sec.~\ref{sec:spectra}, we detail the computation of the photon spectra that emerges from the annihilation of two real MDM states, including all of the aforementioned effects.
Secondly, in Sec.~\ref{sec:forecasts}, we explain our procedure for forecasting the sensitivity of CTAO to these scenarios and justify our conclusion that the experiment is well placed to test the real WIMP hypothesis.

\section{The Annihilation Spectrum of Real WIMPs}
\label{sec:spectra}

\begin{figure}[htp!]
\centering
\includegraphics[width=0.47\textwidth]{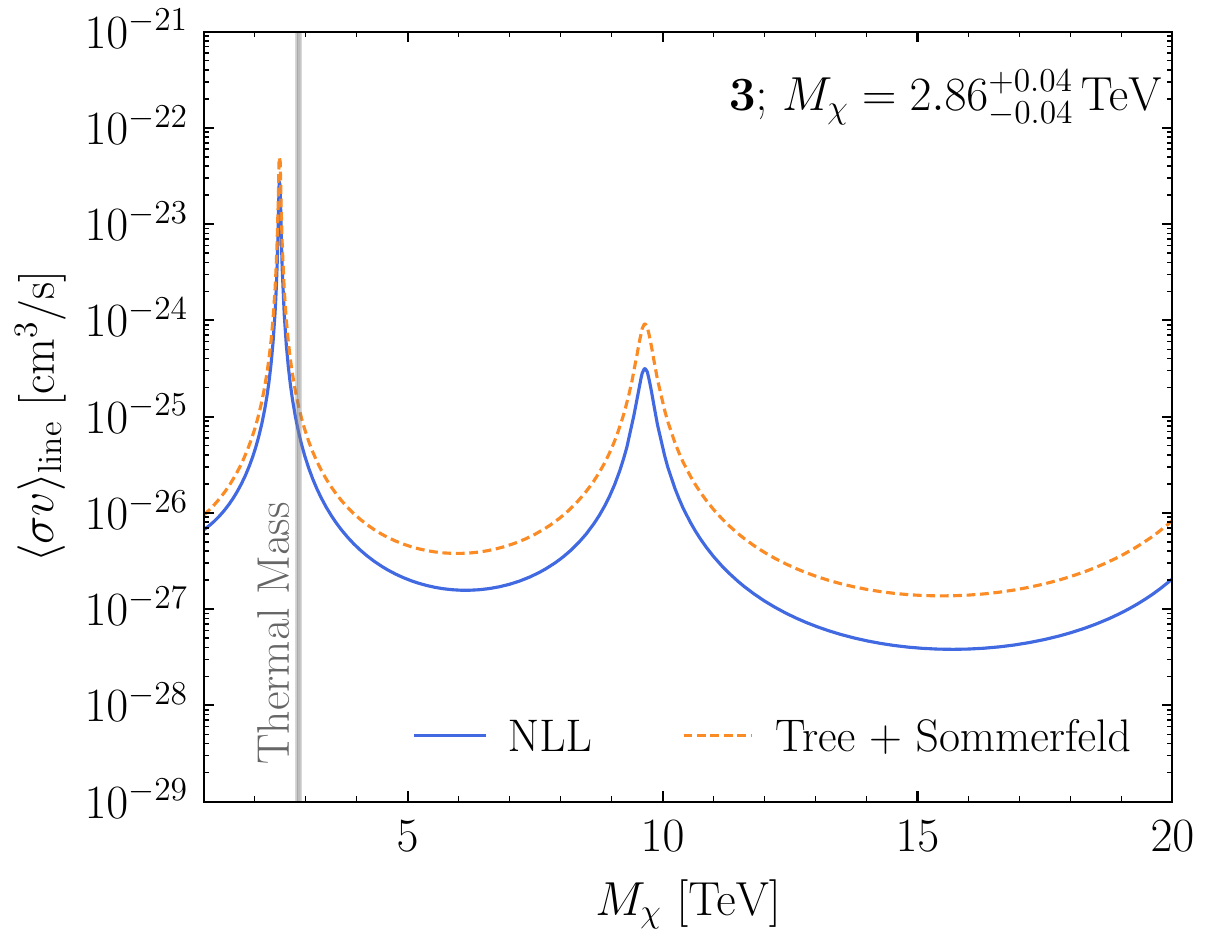}
\quad\includegraphics[width=0.47\textwidth]{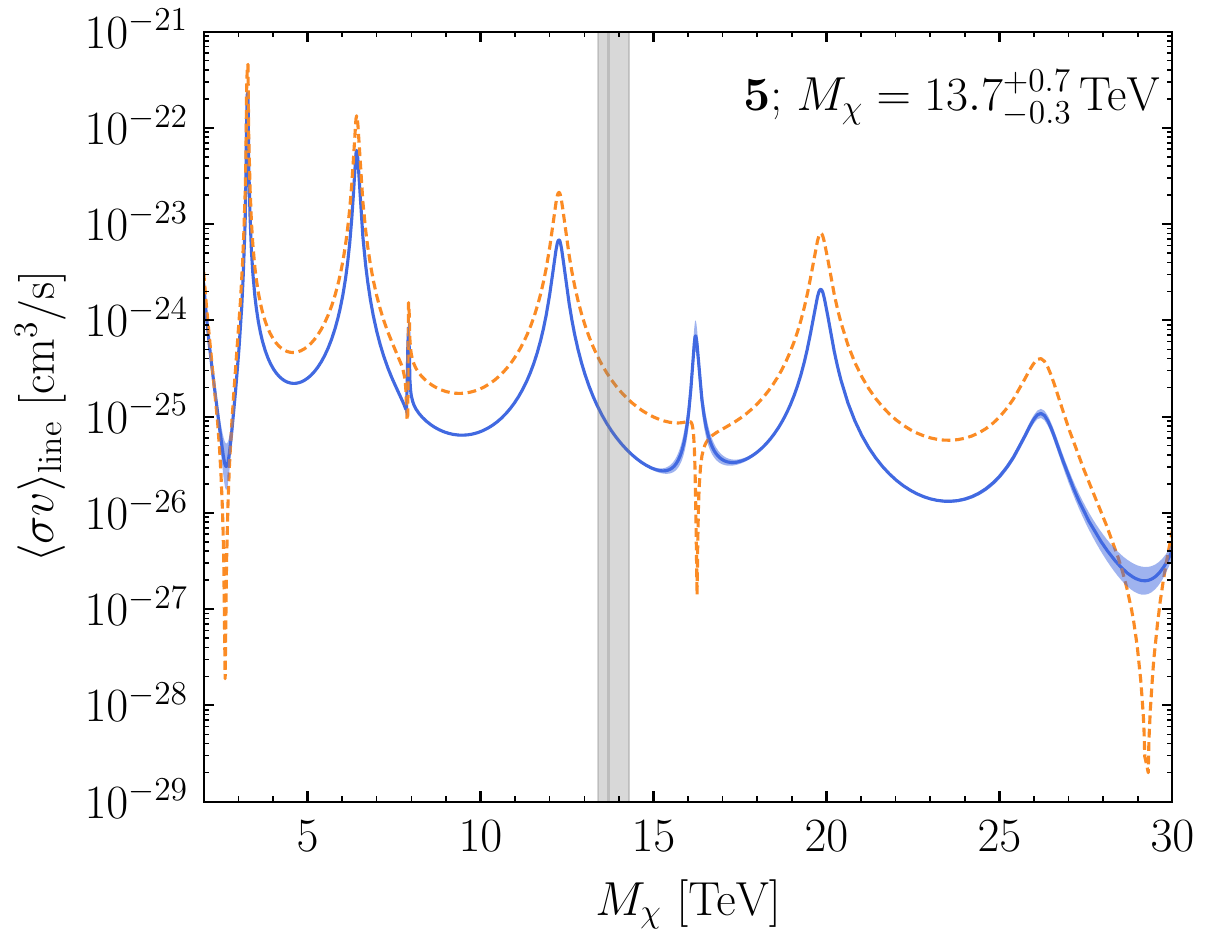}\\
\includegraphics[width=0.47\textwidth]{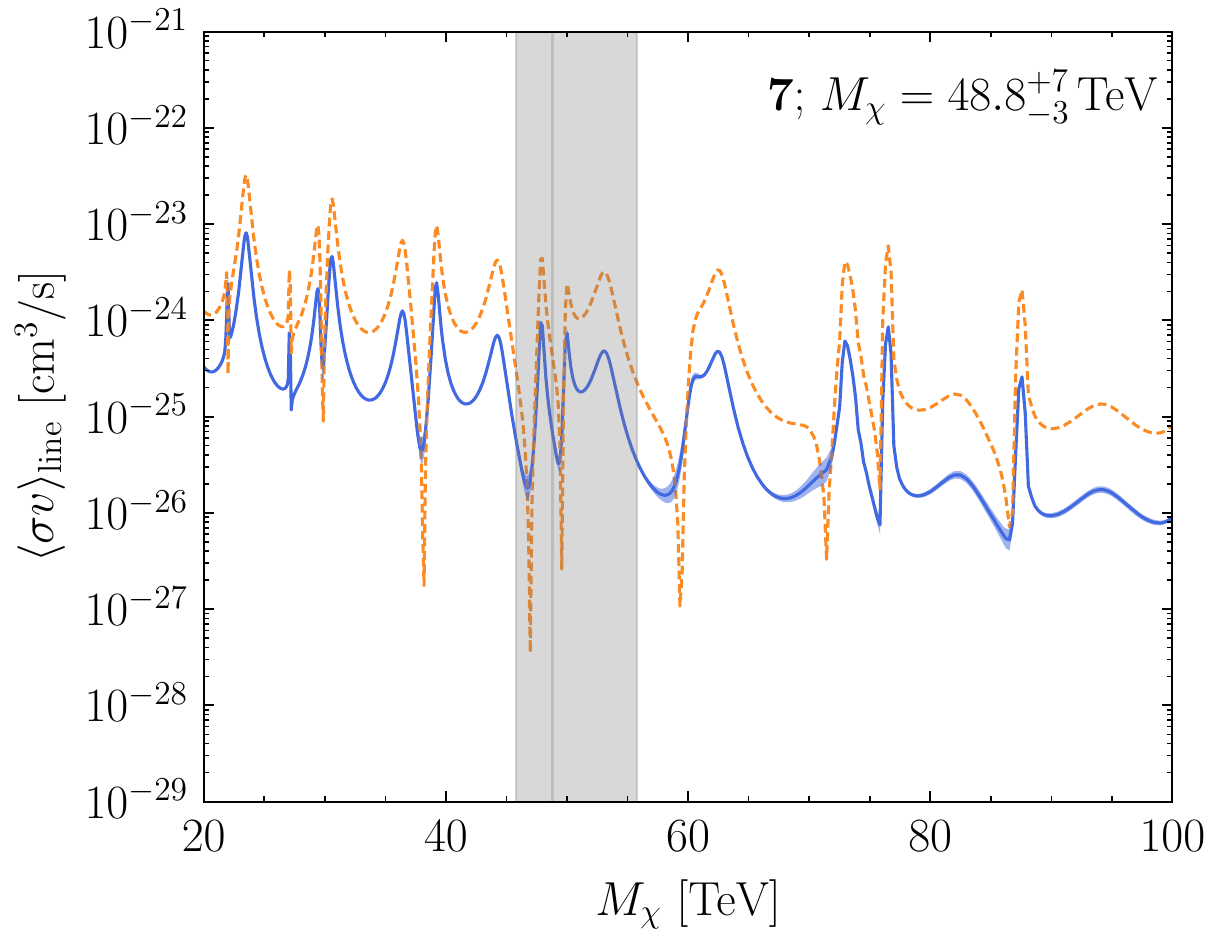}
\quad\includegraphics[width=0.47\textwidth]{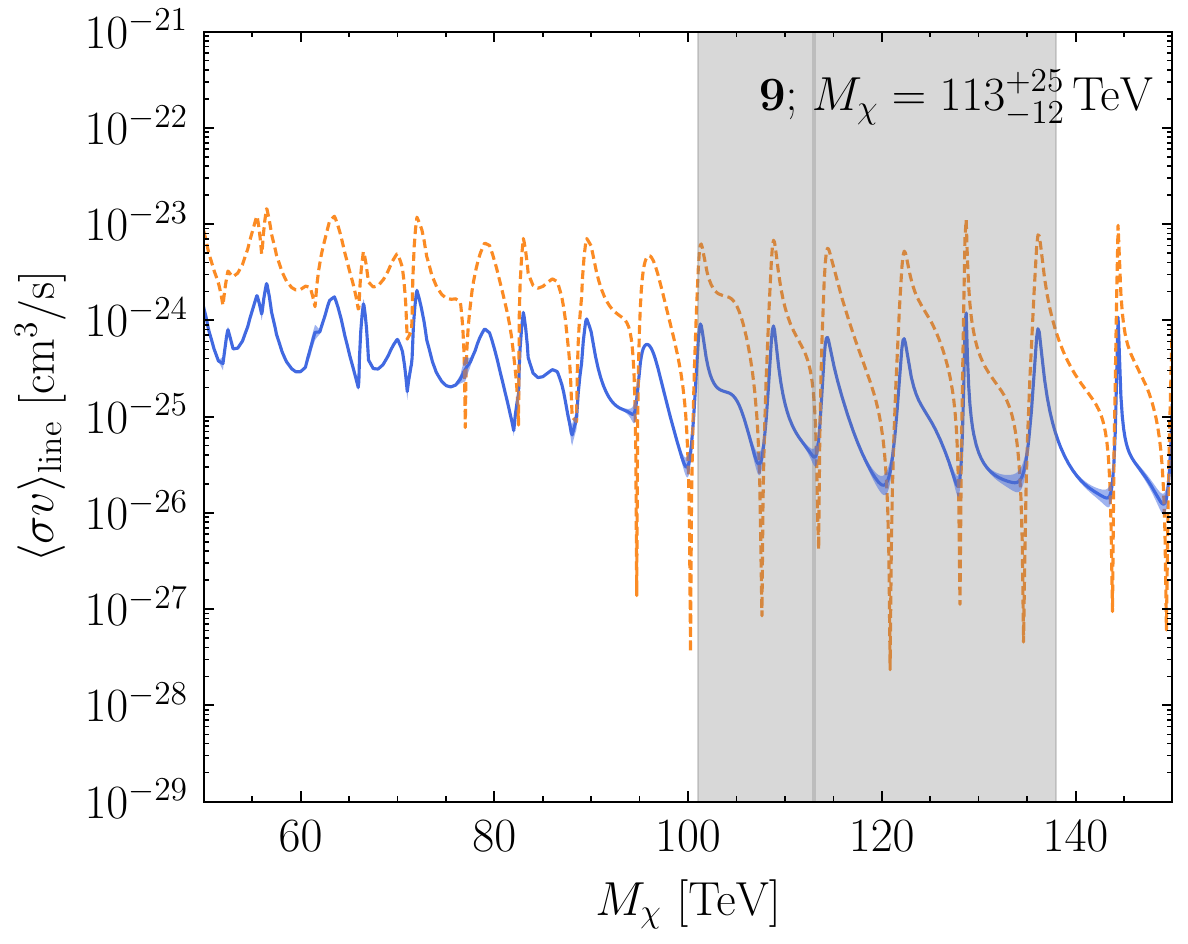}\\
\includegraphics[width=0.47\textwidth]{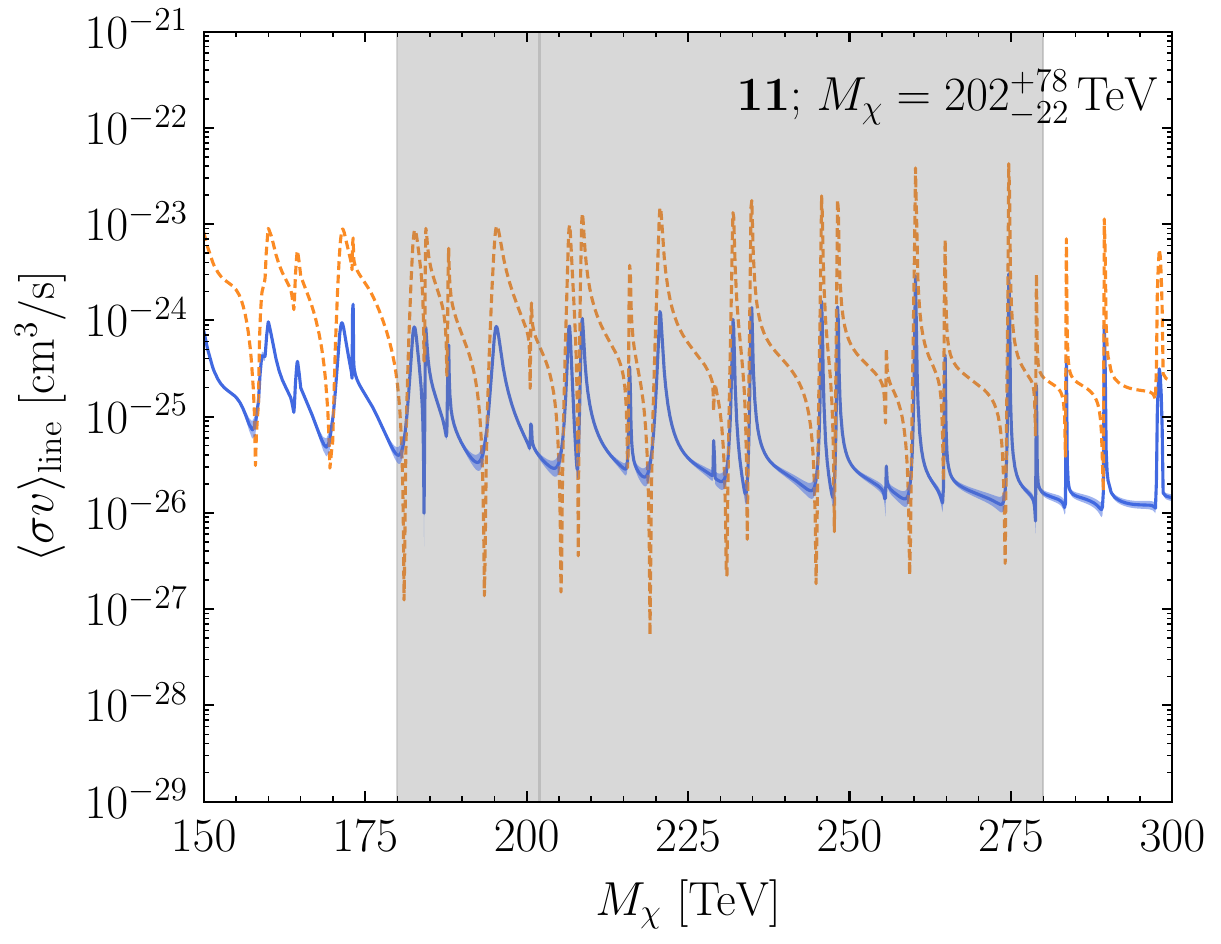}
\quad\includegraphics[width=0.47\textwidth]{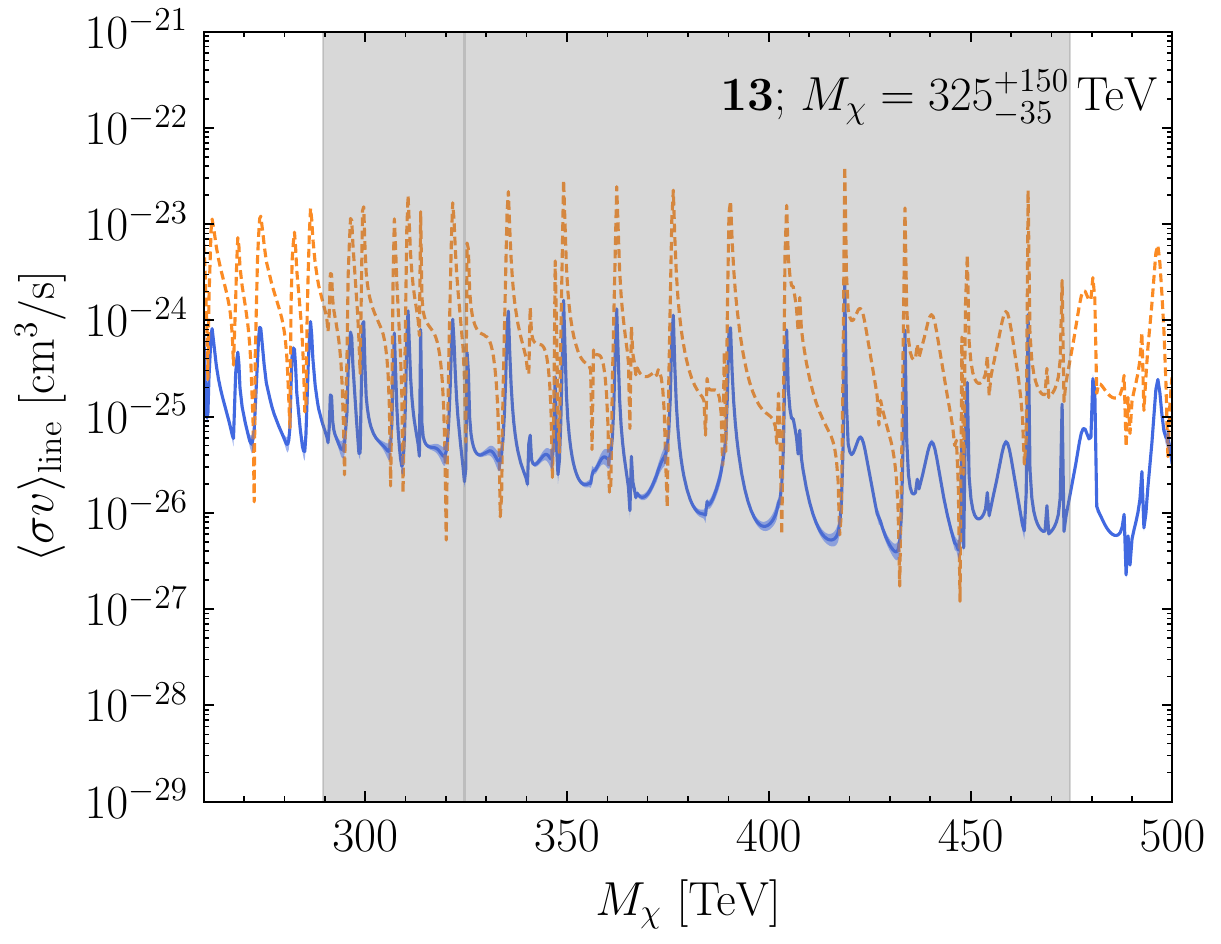}
\vspace{-0.2cm}
\caption{Line cross-sections for all real multiplets as a function of the DM mass.
The blue band shows the NLL prediction with associated uncertainties, whereas the dashed orange curve corresponds to the Sommerfeld enhanced tree-level prediction.
The gray shaded area shows the predicted range for the thermal mass~\cite{Bottaro:2021snn,Bottaro:2023wjv}.}
\vspace{-0.2cm}
\label{fig:linexsec}
\end{figure}

\begin{figure*}[!t]
\centering
\includegraphics[width=0.45\textwidth]{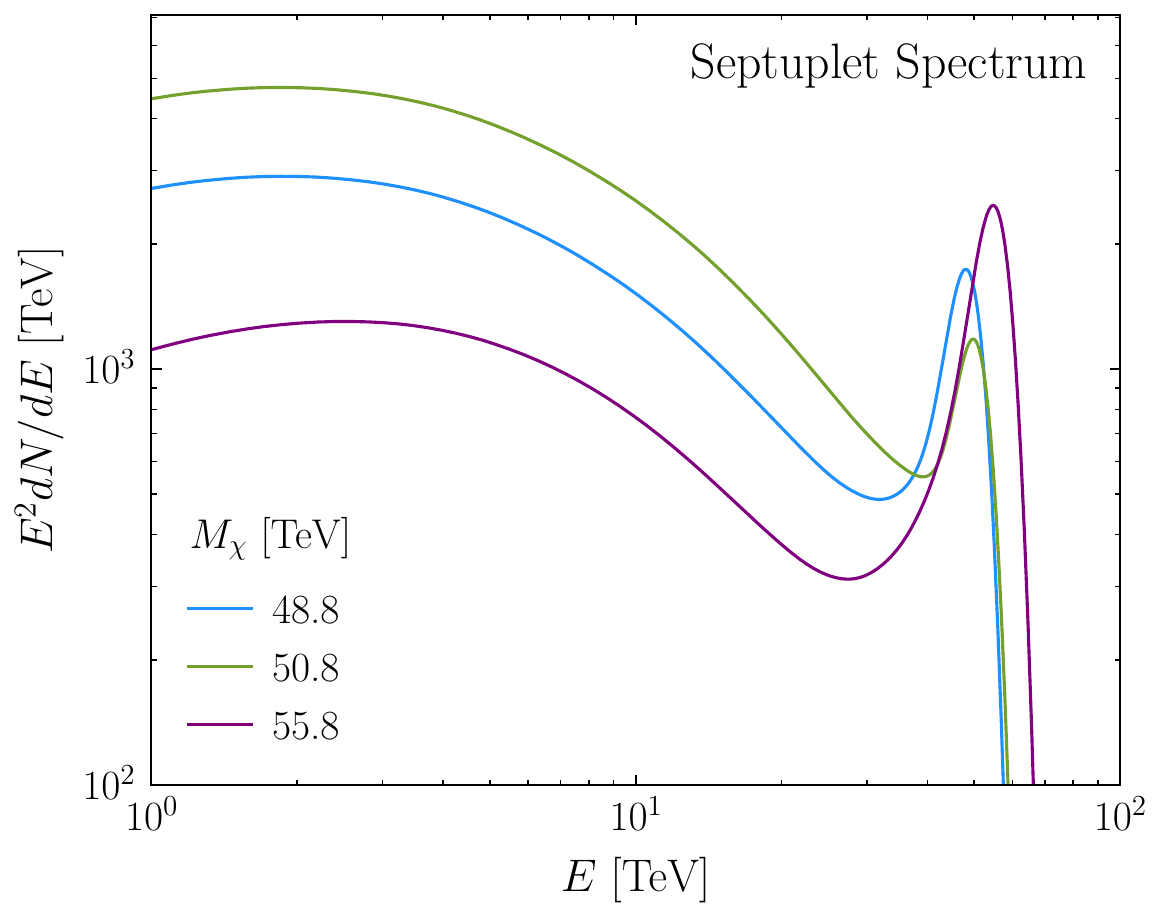}
\hspace{0.5cm}
\includegraphics[width=0.45\textwidth]{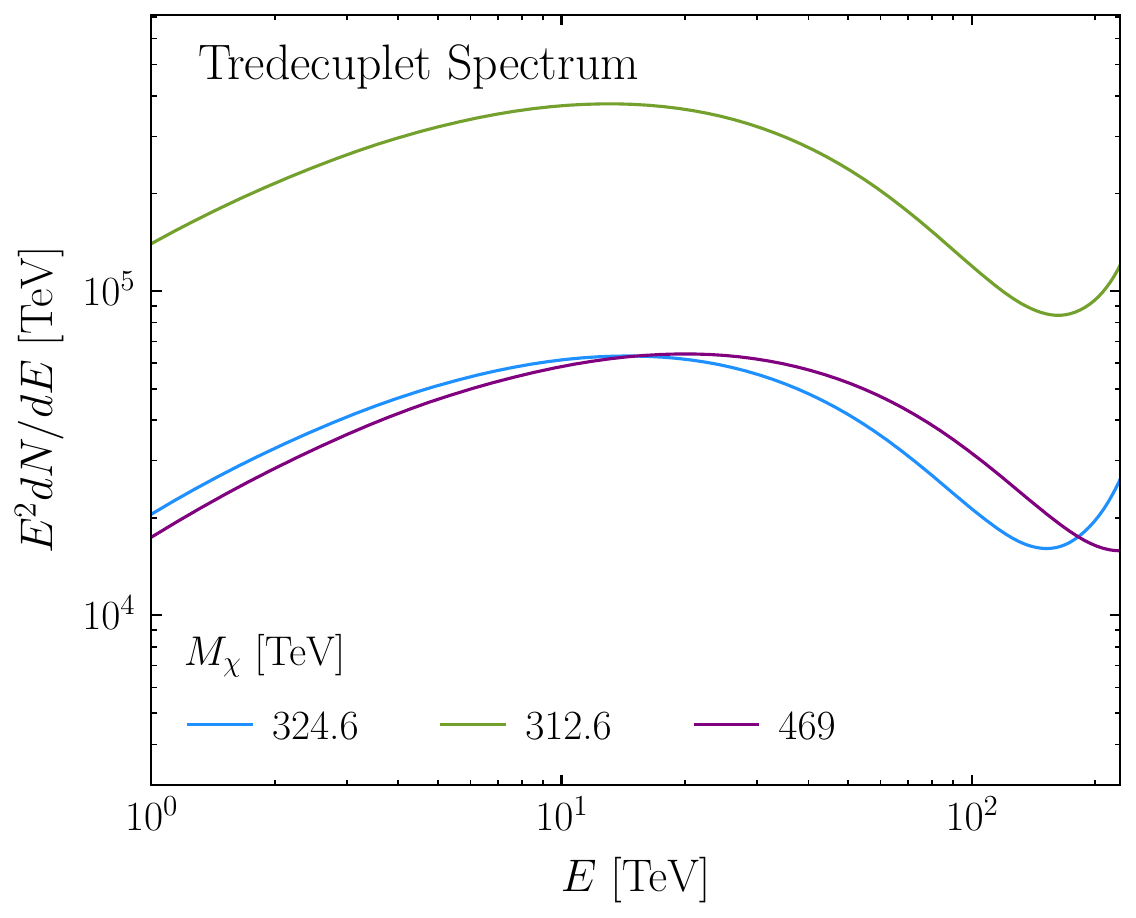}
\vspace{-0.2cm}
\caption{The spectrum of the septuplet ($\mathbf{7}$) and tredecuplet ($\mathbf{13}$) for three masses around their central thermal value.
The CTAO forecasts in this work will only assume sensitivity up to 200\,TeV photons and therefore we do not show the tredecuplet spectrum up to the energies where one would resolve its line-like feature.
The definition of the spectrum is provided in Eq.~\eqref{eq:spectra} and the results here have been smeared by the CTAO energy resolution.}
\vspace{-0.2cm}
\label{fig:Spectra}
\end{figure*}

In this section we compute the differential photon cross section for DM, $d\langle \sigma v \rangle/dE$, under the hypothesis that DM is the neutral fermionic component of a single real MDM representation.
This quantity encodes both the cross section for DM annihilation and further the differential distribution of photons that emerge per annihilation.
These results are deployed in Sec.~\ref{sec:forecasts} to forecast the sensitivity of CTAO to these models.

It is convenient to divide calculation of the differential cross section into four steps.
Firstly, in Sec.~\ref{ssec:DA} we determine the production rate of hard photons with $E_{\gamma} \simeq M_{\chi}$.
This calculation is performed to NLL accuracy by extending the existing EFT framework developed for the wino and quintuplet to an arbitrary real representation of SU(2).
That result depends on the Sommerfeld factors for each representation, which we compute (as the second step) in Sec.~\ref{ssec:SE}.
Thirdly, the formation of electroweak bound states is an independent and physically important contribution to the annihilation spectrum and is discussed in Sec.~\ref{ssec:BS}.
Finally, lower energy continuum photons, those with $E_{\gamma} \ll M_{\chi}$ that emerge from both the direct annihilation and decay of bound states, can also be observed at CTAO and therefore we compute their contribution in Sec.~\ref{ssec:Cnt}.

Let us preview several of the results of these calculations.
A common metric to quantify the overall strength of the annihilation is the rate for the direct production of two photons, denoted $\langle \sigma v \rangle_{\rm line}$.
For all six real representations we show $\langle \sigma v \rangle_{\rm line}$ in Fig.~\ref{fig:linexsec}.
The results contrast with what would occur if one performed a tree-level calculation augmented with the Sommerfeld enhancement as opposed to the full NLL computation; we show that the differences can be significant, well outside  the band denoting the remaining theoretical uncertainties estimated from scale variation.
A striking feature of the results is that as we move to higher representations, the structure of resonances arising from Sommerfeld enhancement becomes increasingly rich.
This behavior also suggests a correspondingly richer spectrum of bound states, as we confirm in Sec.~\ref{ssec:BS}.
This is expected, since the number of bound states with zero angular momentum scales as $\sqrt{\alpha_{\rm eff} M_\chi \mW}$, neglecting $\mathcal{O}(1)$ factors with $\alpha_{\rm eff} = \aW (2n+1)^2/4$ for an (2$n$+1)-plet of SU(2)~\cite{Mitridate:2017izz}, where $n$ denotes the maximum charge of the representation.

We also show in the figure the predicted thermal mass band from Ref.~\cite{Bottaro:2021snn}, updated to include the theoretical uncertainties for heavy (2$n$+1)-plets as estimated in Ref.~\cite{Bottaro:2023wjv}.
This mass range is particularly compelling, as it corresponds to the region where these states are expected to yield the correct relic abundance under the assumption of a conventional thermal cosmology.
For the triplet, the thermal mass is predicted with percent-level accuracy, resulting in a precise prediction for both the annihilation cross section and the spectrum.
In the case of the quintuplet, the uncertainty in the mass prediction is estimated to be at the few-percent level and is dominated by the approximate treatment of electroweak symmetry breaking effects in the computation of the bound state formation (BSF) cross sections.
This uncertainty could, in principle, be reduced through a more accurate treatment of the electroweak phase transition.
However, as we will show in Sec.~\ref{ssec:results}, the expected reach of CTAO derived here is able to cover the full quintuplet range, even with such uncertainties on the thermal mass.
For the higher representations, however, the uncertainties become order unity (due to relativistic corrections; see Ref.~\cite{Bottaro:2023wjv} for discussion), and especially for the tredecuplet there is then a wide range of line cross section values that could be consistent with the correct DM density.\footnote{The uncertainty on the mass for representations above the septuplet is dominated by relativistic corrections to the potentials, of the form $1/r^2$ and $1/r^3$.
Such potentials render the Hamiltonian unbounded from below, making it impossible to fix the boundary conditions for the Schr\"odinger equation at the origin.
The boundary condition can still be fixed at a finite distance by matching the scattering phase generated at shorter scales that can be computed within the Born approximation~\cite{Bellazzini:2013foa,Agrawal:2020lea}.
The uncertainty in the mass shown here arises from the uncertainty in the determination of this phase.
In Ref.~\cite{Bottaro:2023wjv} this procedure was carried over for a U(1) gauge theory. However, for very large representations, the relevant freeze-out dynamics occurs in the unbroken SU(2) phase and closely resembles that of a U(1) theory with an effective coupling given by $\alpha\simeq \aW n^2/4$.
As such, this simple scaling can be used to determine the expected theoretical uncertainty on the mass of the electroweak WIMPs.}
As we show, this will play a role in making the tredecuplet more challenging to constrain.

The line cross section provides a characterization of the overall strength of WIMP annihilation. However, another important ingredient for constraining these scenarios is the distribution of photons that emerge from the annihilation.
These are shown in Fig.~\ref{fig:Spectra}, where the photon spectrum for the septuplet ($n=\mathbf{7}$) and tredecuplet ($n=\mathbf{13}$) are given for several different masses near the central thermal value.
Here the spectrum is defined as,
\be
\frac{dN}{dE} = \frac{1}{\langle \sigma v \rangle_{\rm line}} \frac{d\langle \sigma v \rangle}{dE}.
\label{eq:spectra}
\ee
A notable aspect of the spectra shown is the strong variation as a function of mass.
As the line cross section has been factored out in Eq.~\eqref{eq:spectra}, the sharp variation of the spectrum with mass is not identical to the strong fluctuations in the line cross section seen in Fig.~\ref{fig:linexsec}.
Yet it can still be attributed to the Sommerfeld enhancement: the various contributions to the spectrum, such as from the line or continuum, are sensitive to different combinations of Sommerfeld channels, and interference between these channels generates a strong mass dependence in the relative size of the different contributions.
Further discussion can be found in Refs.~\cite{Montanari:2022buj,Baumgart:2023pwn}, where this effect was first observed in the context of quintuplet annihilation.
The figure also highlights that if CTAO only has sensitivity up to 200\,TeV, then for the higher representations the line-like feature that is expected to be the smoking gun signature of DM is beyond the range of the detector, although these scenarios can still be observed through the lower energy continuum emission.

\subsection{Hard photons from direct annihilation}
\label{ssec:DA}

Our first step is to determine the differential cross section for hard photons emerging from MDM annihilation.
The reference to hard photons implies that we are specifically focusing on those final states that carry an ${\cal O}(1)$ fraction of the available energy; in particular, if we define $z = E/M_{\chi}$, with $E$ the photon energy, our focus is photons with $(1-z) \ll 1$.
Lower energy photons also represent a key contribution to the spectrum; indeed, for the tredecuplet $M_{\chi} \simeq 325\,$TeV, the hard photons are above the nominal CTAO sensitivity window.
Nevertheless, those contributions are relatively straightforward to incorporate, and we return to them in Sec.~\ref{ssec:Cnt}.

As mentioned, our strategy to determine the differential cross section is to directly extend the quintuplet result of Ref.~\cite{Baumgart:2023pwn} to arbitrary real representations.
The results in that work involved the resummation of two sets of logarithms: electroweak logs due to the hierarchy between the DM mass and the electroweak scale, as well as endpoint logs that arise from our focus on photons with energies very near $M_{\chi}$.
Results of that resummation were provided at both leading logarithm (LL) and NLL precision, building on the equivalent calculations for the wino performed in Refs.~\cite{Baumgart:2017nsr,Baumgart:2018yed}.
We similarly provide both the LL and NLL results for arbitrary real representations.
Nevertheless, as the novelty of our calculation is largely an exercise in the group theory of the DM, we show only the LL result in this section.
The associated NLL cross section is deferred to App.~\ref{app:NLL}.

To establish our notation, we consider DM that is a (2$n$+1)-plet of Majorana fermions that transforms under the SM gauge group in the $\mathbf{2n\!+\!1}$ representation of SU(2).
In the limit where SU(2) is unbroken, all these states have a degenerate mass $M_{\chi}$.
In the broken phase, however, we can reorganize the multiplet into a single neutral Majorana fermion $\chi^0$, which constitutes the DM, and $n$ charged Dirac fermions, $\chi^+, \chi^{++}, \ldots, \chi^{n+}$, where by $\chi^{n+}$ we denote a state with electromagnetic charge $+n$ (i.e. $\chi^{2+}=\chi^{++}$).\footnote{In the broken theory, the charged fermions are only defined up to a phase associated with global transformations under U(1)$_{\scriptscriptstyle {\rm EM}}$, a degeneracy that one can exploit to rephase the couplings of the charged states to the $W^{\pm}$ bosons.
We mention this as different conventions for these couplings appear throughout the literature, which of course is perfectly acceptable so long as one uses a consistent choice throughout a given calculation.}
Radiative corrections provide a small lift in the charged state masses above $M_{\chi}$, ensuring that $\chi^0$ is the lightest state in the spectrum.

To begin the calculation, as shown in Ref.~\cite{Baumgart:2023pwn}, the differential annihilation cross section can be written as follows,
\be
\frac{d\sigma}{dz} =F_{\chi}^{a'b'ab}\,\frac{d\hat{\sigma}^{a'b'ab}}{dz},
\label{eq:F-sig-fact}
\ee
where the SU(2) adjoint indices $a'b'ab$ are summed over.
This expression decomposes the cross section of interest, $d\sigma/dz$ (recall $z = E/M_{\chi}$), into a term $F_{\chi}$ that depends only on the representation of the DM, as well as $d\hat{\sigma}/dz$, which is the cross section stripped of any knowledge of the representation.
In this sense $d\hat{\sigma}/dz$ can be computed once and for all, as was done in Ref.~\cite{Baumgart:2023pwn} by reorganizing the earlier wino results of Refs.~\cite{Baumgart:2017nsr,Baumgart:2018yed} into a form strictly independent of the representation.

The key term for us to consider, therefore, is $F_{\chi}$.
In order to define and evaluate this object, we first introduce the following matrix element,
\be
{\cal M}^{ab} = \Big\langle 0 \Big\vert \left(\chi^{T}_v i \sigma_2 \left\{ T_{\chi}^a,\,T_{\chi}^b \right\} \chi_v \right) \Big\vert (\chi^0 \chi^0)_S \Big\rangle,
\label{eq:calM-def}
\ee
in terms of which $F_{\chi}^{a'b'ab} = [{\cal M}^{a'b'}]^{\dag} {\cal M}^{ab}$~\cite{Baumgart:2023pwn}.
The matrix element is expressed in terms of the DM initial state: two neutral particles annihilating in a spin singlet, $(\chi^0 \chi^0)_S$.
The operator is written using $\chi_v$, a non-relativistic DM field operator, and the generators of SU(2) in the DM representation, $T_{\chi}$.
Heuristically, $\chi_v$ is a (2$n$+1) column vector, whereas $T_{\chi}$ is a (2$n$+1)$\times$(2$n$+1) matrix.

When evaluating $F_{\chi}$ it is convenient to work in the broken basis where we can rewrite expressions in the basis of electric charge.
Within that basis we can consider the various matrix elements of the form of ${\cal M}^{ab}$ that could provide a non-vanishing contribution.
In particular, the anticommutator will select out various elements from the DM field operators, and unless the combination of the two has charge zero, then there is no overlap with our charge neutral initial state.
It is convenient to express the non-zero matrix elements as follows,
\bea
\Big\langle 0 \Big\vert \left(\chi^{0T}_v i \sigma_2 \chi^0_v \right) \Big\vert (\chi^0 \chi^0)_S \Big\rangle 
=\,&4 \sqrt{2} M_{\chi} s_{00}, \\
\Big\langle 0 \Big\vert \left(\chi^{m+T}_v i \sigma_2 \chi^{m-}_v \right) \Big\vert (\chi^0 \chi^0)_S \Big\rangle 
=\,&4 M_{\chi} s_{0m}.
\label{eq:SomFac-Def}
\eea
Here $s_{0m}$ are the Sommerfeld factors associated with the initially $\chi^0 \chi^0$ state evolving into $\chi^{m+} \chi^{m-}$.
We compute these in Sec.~\ref{ssec:SE}, although we note here that in the absence of Sommerfeld enhancement we would have $s_{00} = 1$ and $s_{0m}=0$ for $m \geq 1$; further, we emphasize that in general these factors are unique between the representations, i.e. $s_{00}$ for the wino and quintuplet differ.
Equation~\eqref{eq:SomFac-Def} represents the only non-zero matrix elements; an operator of the form $\chi^{+T}_v i \sigma_2 \chi^{0}_v$ would have no overlap with the initially neutral state.

We can similarly rewrite the generators in the broken basis as,
\be
T^0 = T^3,\hspace{0.5cm}
T^{\pm} = \frac{1}{\sqrt{2}} (T^1 \pm i T^2).
\ee
Working in the charge basis, only anticommutators with net-zero charge project out the DM field operators in Eq.~\eqref{eq:calM-def} that carry zero-charge.
Following Eq.~\eqref{eq:SomFac-Def} these must be the only non-zero matrix elements and therefore we simply need to evaluate ${\cal M}^{00}$ and ${\cal M}^{+-}$.
For the $\mathbf{2n\!+\!1}$ representation, the neutral generator couples to the charges of the states and therefore takes the form $T_{\chi}^0 = \operatorname{diag}(n,n-1,\ldots,-n)$.
We conclude that
\be
{\cal M}^{00} = \Big\langle 0 \Big\vert \left(\chi^{T}_v i \sigma_2 \left\{ T_{\chi}^0,\,T_{\chi}^0 \right\} \chi_v \right) \Big\vert (\chi^0 \chi^0)_S \Big\rangle
= 16 M_{\chi} \sum_{m=1}^n m^2 s_{0m}.
\label{eq:M00}
\ee
To obtain the remaining matrix element we note that $T^a T^a = C_{\chi} \mathbb{1}$, with the quadratic Casimir given by $C_{\chi} = n(n+1)$.
It follows that $\{ T_{\chi}^+, T_{\chi}^-\} = C_{\chi} \mathbb{1} - T^0 T^0$ and consequently,
\bea
{\cal M}^{+-} =\, &\Big\langle 0 \Big\vert \left(\chi^{T}_v i \sigma_2 \left\{ T_{\chi}^+,\,T_{\chi}^- \right\} \chi_v \right) \Big\vert (\chi^0 \chi^0)_S \Big\rangle \\
=\, &4 \sqrt{2} M_{\chi} n (n+1) s_{00}
+ 8 M_{\chi} \sum_{m=1}^n [n(n+1)-m^2] s_{0m}.
\label{eq:Mpm}
\eea
Although it was not necessary for the calculation above, we note that the explicit form of the generators can be determined as,
\bea
T^+_{ij} & = T^-_{ji} = \frac{1}{\sqrt{2}} \sqrt{n(n+1) - \left(i - n \right) \left(i - n - 1\right)}\delta_{i,j-1}, \\
T^0_{ij} & = \delta_{i,j} (n-i+1), \quad 1 \le i \le 2n+1, \quad 1 \le j \le 2n+1.
\eea

Having computed all non-zero matrix elements from which we can construct $F_{\chi}$, the next step is to determine the differential cross section using Eq.~\eqref{eq:F-sig-fact}.
Other than the more general $F_{\chi}$, the calculation for this follows identically to in Ref.~\cite{Baumgart:2023pwn} and we therefore simply state the result,
\bea
\left.\frac{d\sigma}{dz}\right|_{\rm LL} = \,&(F_0+F_1) \,\sigma_0\, e^{-(4\alpha^{}_{\scaleto{W}{3.pt}}/\pi) L_{\chi}^2}\, \delta(1-z)\\
-\, & \frac{2\aW}{\pi}\sigma_0\, e^{-(4\alpha^{}_{\scaleto{W}{3.pt}}/\pi)L_\chi^2} \bigg\{
2 F_0{\cal L}_1^J(z) e^{(4\alpha^{}_{\scaleto{W}{3.pt}}/\pi) L_J^2(z)} \\
&\hspace{2.5cm}+ F_1 \Big( 2 {\cal L}_1^J(z) - 3 {\cal L}_1^S(z)\Big) e^{(4\alpha^{}_{\scaleto{W}{3.pt}}/\pi) \left(\Theta_JL_J^2(z)-\tfrac{3}{4}\Theta_S L_S^2(z)\right)} \bigg\}.
\label{eq:LL}
\eea
This is the differential cross section to produce two photons.
There are two physically distinct contributions to this expression.
The first, proportional to $\delta(1-z)$ with coefficient $\sigma_{\rm line} \equiv \sigma_{\gamma \gamma} + \sigma_{\gamma Z}/2$, is the cross section to produce exactly two photons with energy equal to $M_{\chi}$, to which the $\gamma Z$ final state contributes.
The second, which appears on the final two lines, is associated with the endpoint contribution, that is $\gamma+X$ final states where in order to ensure the photon is hard the invariant mass of $X$ is constrained to be near the lightcone---generically $X$ is an electroweak jet, hence the appearance of jet and soft quantities, labeled by $J$ and $S$.
We emphasize again that because this is the cross section to produce {\it two} photons, and the endpoint is associated with $\gamma + X$ rather than $\gamma \gamma$ (as for the line), we should expect that the endpoint is normalized to $\sigma_{\gamma+X}/2$ (similar to the prefactor of 1/2 for $\sigma_{\gamma Z}$ in $\sigma_\text{line}$).
This can be confirmed by matching our results to fixed order calculations to the $W^+ W^- \gamma$ final state~\cite{Bergstrom:2005ss,Bringmann:2007nk}.

The various notation introduced in Eq.~\eqref{eq:LL} is defined as follows.
(We refer to Refs.~\cite{Baumgart:2017nsr,Baumgart:2023pwn} for a detailed discussion of the origin of the various terms.)
The reference cross section is given by
\be
\sigma_0 = \frac{\pi \aW^2 \sW^2}{2M_{\chi}^2 v},
\label{eq:sigtree}
\ee
with $\aW$ the SU(2) fine structure constant, $\sW = \sin \thetaW$, and $v$ the relative velocity of the incident DM states.
The tree level cross section receives a correction from a massive Sudakov logarithm given by $L_{\chi} = \ln (\mW/2M_{\chi})$; this generates a suppression of the line-like cross section at larger masses compared to the tree level prediction, a trend that can be observed clearly in Fig.~\ref{fig:linexsec}.
The endpoint contribution depends on additional logarithms and thresholds associated with the jet and soft scales, which are defined through,
\bea
{\cal L}_1^J = \frac{L_J}{1-z} \Theta_J,\hspace{0.5cm}
L_J = \ln \!\left( \frac{\mW}{2M_{\chi}\sqrt{1-z}} \right)\!,\hspace{0.5cm}
\Theta_J = \Theta \!\left( 1 - \frac{\mW^2}{4M_{\chi}^2} - z \right)\!, \\
{\cal L}_1^S = \frac{L_S}{1-z} \Theta_S,\hspace{0.5cm}
L_S = \ln \!\left( \frac{\mW}{2M_{\chi}(1-z)} \right)\!,\hspace{0.5cm}
\Theta_S = \Theta \!\left( 1 - \frac{\mW}{2M_{\chi}} - z \right)\!.
\label{eq:LL-JS}
\eea
The jet and soft scales emerge as a result of our kinematic requirement of having a hard photon; indeed, if the final state is $\gamma+X$ the invariant mass back reacting against the photon is $m_X = 2 M_{\chi} \sqrt{1-z}$, which is none other than the jet scale.
Finally, the expressions $F_{0,1}$ that appear in both terms are the only contributions dependent upon the DM representation.
These are given by,
\be
F_0 = \frac{\left| {\cal M}^{00} \right|^2 + 2 \left| {\cal M}^{+-} \right|^2}{192 M_{\chi}^2},\hspace{0.5cm}
F_1 = \frac{\left| {\cal M}^{00} \right|^2 - \left| {\cal M}^{+-} \right|^2}{96 M_{\chi}^2},
\ee
which when combined with Eqs.~\eqref{eq:M00} and \eqref{eq:Mpm} allow us to evaluate the LL cross section for arbitrary real SU(2) representations.

The NLL expression is conceptually identical although technically more involved and we provide it in App.~\ref{app:NLL}.
Before moving on, however, let us make two comments about the above result.
Firstly, for the photons with exactly $E = M_{\chi}$, arising from the first line of Eq.~\eqref{eq:LL}, the group theory factors organize themselves in a particularly simple form,
\be
F_0 + F_1 = \frac{\left| {\cal M}^{00} \right|^2}{64 M_{\chi}^2} = 4 \left| \sum_{m=1}^n m^2 s_{0m} \right|^2\!.
\ee
In the absence of Sommerfeld enhancement there are no such contributions as $s_{0m} = 0$ for $m \geq 1$.
Beyond this, we see that the $n$th charged state contributes to the cross section by a factor of $n^4$ more strongly than the singly charged state (assuming comparable Sommerfeld factors) and further there is also interference between the terms.
Secondly, we emphasize that the cross section computed in this section -- as given by Eq.~\eqref{eq:LL} and the equivalent NLL expression in App.~\ref{app:NLL} -- is explicitly \textit{the annihilation rate to produce two photons}.
This is simply a choice, although we emphasize it here as this choice has been used implicitly at times in the earlier works of Refs.~\cite{Baumgart:2017nsr,Baumgart:2018yed,Baumgart:2023pwn}.
The motivation for this choice is that experimental collaborations commonly set a model agnostic limit on the process $\chi^0 \chi^0 \to \gamma \gamma$, the cross section for which is denoted $\sigma_{\rm line}$; in that sense our calculation determines the differential line cross section for the real MDM representations.

\subsection{Sommerfeld enhancement}
\label{ssec:SE}

Next we turn to the computation of the Sommerfeld factors $s_{0m}$, as introduced in Eq.~\eqref{eq:SomFac-Def}.
The Sommerfeld effect arises when a long-range interaction modifies the wave-function describing the system of the annihilating particles, which can no longer be approximated as plane waves as is usually assumed when computing cross sections in field theory~\cite{Iengo:2009ni}. 
As such, the final annihilation cross-section will depend on the value of the wave-function at the interaction point, which is obtained by solving the Schr\"odinger equation with a potential describing the long-range interaction.

For the problem at hand, the Schr\"odinger equation that we need to solve takes the form~\cite{Cirelli:2007xd,Hisano:2006nn,Hisano:2004ds}
\be
-\frac{1}{M_\chi}\nabla^2\phi_i(\mathbf{r})+\left(V_{ij}(r)+\Delta_i\delta_{ij}\right)\phi_j(\mathbf{r})=\frac{1}{4} \mdm v^2\phi_i(\mathbf{r}).
\label{eq:schroedinger}
\ee
Here $\phi_i$ is the two-body wave function in the center of mass of the state $i$, $V_{ij}(r)$ the potential induced by the electroweak interactions for the given annihilation channel, and $\Delta_i$ is the total mass splitting of the state $i$ with respect to $\left|\chi^0\chi^0\right\rangle$.\footnote{We assume the particles are always initially in the lowest-mass DM state.
For all cases except the tredecuplet, the thermal mass is below the threshold where the chargino states are collisionally excited.
Even for the tredecuplet, however, the chargino states promptly decay to the (lowest mass) neutralino state and there is no ambient bath of decay products to maintain a non-negligible chargino abundance (in contrast to the situation during freezeout), implying charginos provide a negligible contribution to the initial states.}
This initial state has charge zero, whilst its orbital angular momentum ($L$) and total spin ($S$) quantum numbers are constrained by spin-statistics for identical particles.
This second criterion implies that $L+S$ is even for the initial state; that is, for fermionic DM the spin configuration is singlet ($S=0$, antisymmetric) for even-$L$ modes and triplet ($S=1$, symmetric) for odd-$L$ modes.\footnote{For scalar MDM, $L$ must be even to ensure a symmetric wavefunction.}

We compute the potentials appearing in Eq.~\eqref{eq:schroedinger} by transforming from the basis of eigenstates of total isospin $j$ for the two-particle state. 
In this basis, the potential is diagonal and takes the form
\be
V_{j_1j_2} = \delta_{j_1,j_2}\left[n(n+1)-\frac{j_1(j_1+1)}{2}\right]\!,
\ee
where $0\leq j_1 \leq 2n $ is even (odd) for even (odd) $L+S$.
To convert the potentials to the usual $\chi^{(n-i+1+Q)+} \chi^{(n-i+1)-}$ two-state basis with total electric charge $Q$, we compute the transformation matrices whose elements are proportional to the Clebsh-Gordan coefficients
\be
C_{ji}^Q=\langle j;\,Q|n(n-i+1);\, n(-n+i-1+Q)\rangle\sqrt{2-\delta_{n-i+1,Q-n+i-1}},
\ee
where $1\leq i\leq n+1-Q$ for even $L+S$ and $1\leq i\leq n$ for odd $L+S$.
The second factor stems from the fact that, for real representations, $\left(\chi^{q+}\right)^c = \chi^{q-}$. 
The final potentials then read
\be
V_{i_1i_2}^Q=C_{j_1i_1}^QC_{j_2i_2}^QV_{j_1j_2}.
\ee

The most important potentials for our calculation are those for the initial state (net charge $Q=0$, $L+S$ even), relevant for both the Sommerfeld factors and capture rate computations, and those for the $Q=1$, $L+S$ odd states, relevant for bound states accessed by dipole transition from the initial state.
For the $2n+1$ representation, the initial-state potential is given at leading order (LO) by:
\bea
V_{ij} & = N_i N_j \left[-(n-i+1)^2 \delta_{i,j} V_{\gamma \smallZ}(r) + \left( (T_{ij}^+)^2 + (T_{ij}^-)^2 \right) V_{\smallW}(r) \right]\!, \\
N_i & =\begin{cases}1 & i \le n, \\ \sqrt{2} & i = n+1,\end{cases} \\
V_{\smallW}(r) & = \frac{\aW}{r} e^{-m^{}_{\scaleto{W}{3.pt}} r}, \quad V_{\gamma \smallZ}(r) = \frac{\aW}{r} \left( \cW^2 e^{-m^{}_{\scaleto{Z}{3.pt}} r} + \sW^2 \right)\!.\label{eq:VpotQ0LO}
\eea
The $V_{ij}$ potential describes the coupling between the two-particle states $\chi^{(n-i+1)+} \chi^{(n-i+1)-}$ and $\chi^{(n-j+1)+} \chi^{(n-j+1)-}$, where $i$ and $j$ run from 1 to $n+1$.
Note that here we use the simplification described as Method 2 in Ref.~\cite{Beneke:2014gja}, where we combine the $\chi^{m+} \chi^{m-}$ and $\chi^{m-} \chi^{m+}$ states rather than treating them separately, and so consequently the potential matrix has dimension $n+1$ rather than $2n+1$.

We include next-to-leading order (NLO) infrared corrections to the potentials using the analytic fitting functions derived by Refs.~\cite{Beneke:2019qaa, Urban:2021cdu}, which corresponds to the following replacements in the LO potential.
Defining $L = \ln(\mW r)$, the potential is described by
\begin{align}
e^{-m^{}_{\scaleto{W}{3.pt}} r} & \rightarrow e^{-m^{}_{\scaleto{W}{3.pt}} r} + \frac{2595}{\pi} \aW 
\begin{cases} - \text{exp}\left[-\frac{79 \left(L - \frac{787}{12}\right)\left(L - \frac{736}{373}\right) \left(L - \frac{116}{65}\right)\left(L^2 - \frac{286 L}{59} + \frac{533}{77}\right) }{34 \left(L - \frac{512}{19}\right) \left(L - \frac{339}{176}\right) \left(L - \frac{501}{281}\right) \left(L^2 -  \frac{268 L}{61} + \frac{38}{7}\right)} \right]\!, & \mW r < \frac{555}{94} \\
\text{exp}\left[-\frac{13267 \left(L - \frac{76}{43}\right)\left(L - \frac{28}{17}\right) \left(L + \frac{37}{30}\right)\left(L^2 - \frac{389 L}{88} + \frac{676}{129}\right) }{5 \left(L - \frac{191}{108}\right) \left(L - \frac{256}{153}\right) \left(L + \frac{8412}{13}\right) \left(L^2 -  \frac{457 L}{103} + \frac{773}{146}\right)} \right]\!, & \mW r > \frac{555}{94}\end{cases} 
\nn \\
\cW^2 e^{-m^{}_{\scaleto{Z}{3.pt}} r} & \rightarrow \cW^2 e^{-m^{}_{\scaleto{Z}{3.pt}} r} + \aW \left[ -\frac{80}{9} \frac{\sW^4 \left(\ln(\mZ r) + \gamma_E\right)}{2\pi (1 + (32/11) (\mW r)^{-22/9})}   + \frac{\left( \frac{19}{6}  \ln(\mZ r) - \frac{1960}{433} \right)}{2\pi(1 + (7/59) (\mW r)^{61/29})} \right. \nn \\
& \left. - \frac{\frac{s^{2}_{\scaleto{W}{3.pt}}}{\alpha^{}_{\scaleto{W}{3.pt}}} \left(-\frac{1}{30} + \frac{4}{135} L \right)}{1 + (58/79) (\mW r)^{-17/15} + (1/30) (\mW r)^{119/120} + (8/177) (\mW r)^{17/8}}\right]\!.
\label{eq:vnlo}     
\end{align}
As previously mentioned in the context of freeze-out predictions, for large electroweak (2$n$+1)-plets, the non-relativistic potentials receive additional corrections from relativistic ultraviolet (UV) threshold effects. These are induced by loops involving the heavy DM field~\cite{Bottaro:2023wjv}.
The leading corrections to the Yukawa potential in Eq.~\eqref{eq:VpotQ0LO} scale as
\be
\frac{\Delta V_{\text{NLO}}^{\text{UV}}}{\Delta V_{\text{LO}}} \sim \frac{\alpha_{\text{eff}}}{M_\chi r} \sim \frac{1}{4} (2n+1)^2 \aW v,
\ee
where in the second expression we estimated the relevant radius as the de Broglie wavelength of the DM, $r \sim 1 / (M_\chi v)$. 
Although these UV-induced corrections grow rapidly with representation, at the velocities relevant for indirect detection, $v \sim 10^{-3}$, they remain subleading compared to the corrections included in Eq.~\eqref{eq:vnlo} and can therefore be safely neglected.

To compute the Sommerfeld factors, we used the method described in Ref.~\cite{Garcia-Cely:2015dda}.
To extract the Sommerfeld factors $s_{0m}$, we can solve the Schr\"odinger equation in Eq.~\eqref{eq:schroedinger} in terms of the reduced wave function $\phi_i^{(m)}(\mathbf{r})=r\psi_i^{(m)}(r)/\sqrt{4\pi}$
\be
-\frac{1}{M_\chi}\frac{d^2\psi_i^{(m)}(r)}{dr^2}+\left(V_{ij}(r)+\Delta_i\delta_{ij}\right)\psi_j^{(m)}(r)=\frac{1}{4} \mdm v^2\psi_i^{(m)}(r)
\ee
with the following boundary conditions
\be
\psi_i^{(m)}(0)=\delta_{im},\quad \frac{d\psi_i^{(m)}}{dr}(r\rightarrow\infty)=i\frac{M_\chi v}{2} \sqrt{1-\frac{4\Delta_i}{M_\chi v^2}}\psi_i^{(m)}(r).
\ee
The Sommerfeld factors are then obtained from $\psi_0^{(m)}(r\rightarrow \infty)=s_{0m}e^{iM_\chi v r/2}$.
If we introduce the matrices $\mathbf{g}_{im}(r)=\psi_i^{(m)}(r)$ and $\mathbf{h}(r)=\mathbf{g}'(r)\mathbf{g}^{-1}(r)$, we can rewrite the Schr\"odinger equation as
\be
\frac{1}{M_\chi}\left(\mathbf{h}'(r)+\mathbf{h}^2(r)\right)+\mathbf{V}(r)-\mathbf{\Delta}+\frac{1}{4} \mdm v^2 \mathbb{1}=0
\ee
with now a single boundary condition
\be
\mathbf{h}_{ij}(r\rightarrow\infty)=i\frac{M_\chi v}{2}\sqrt{1-\frac{4\Delta_i}{M_\chi v^2}}\delta_{i,j}.
\ee
It is possible to show that $s_{0m}s_{0l}=\frac{2}{M_\chi v}\left[\mathbf{h}_{ml}(0)-\mathbf{h}^*_{lm}(0)\right]$.
Notice that this requires the rank of the matrix $\left[\mathbf{h}_{ml}(0)-\mathbf{h}^*_{lm}(0)\right]$ to be 1.
This, in turn, implies that the initial kinetic energy is such that no intermediate charged state goes on-shell in the annihilation process.
As discussed further below, by default we simply fix the velocity to $v=10^{-3}$, and then this condition is satisfied for almost all representations.
As a cross-check, we have computed the Sommerfeld enhancement using the version of the variable phase method outlined in App.~A of Ref.~\cite{Asadi:2016ybp} (which does not require this condition), finding good agreement.

For the highest-masses associated with the tredecuplet, the chargino states become kinematically allowed for velocities below the Galactic escape velocity.
For this region of parameter space, we switch to the variable phase method for our main calculation, and also average over the Galactic velocity distribution described by the following isotropic distribution~\cite{Boddy:2018ike}
\be
f(v)=\sqrt{\frac{27}{4\pi}}\frac{v^2}{v_{\rm disp}^3} \exp \left[-\frac{3v^2}{4v_{\rm disp}^2} \right]\!,
\label{eq:v_distr}
\ee
where $v_{\rm disp}\in [130,330]$\,km/s. We have compared the results for $v_{\rm disp}=130$\,km/s and $v_{\rm disp}=330$\,km/s with the usual choice of fixed $v=10^{-3}$.
As we show in Sec.~\ref{ssec:results}, the results are only mildly sensitive to the velocity distribution, with the most conservative bounds expected for $v_{\rm disp}=330$\,km/s.

\subsection{Contribution from bound states}
\label{ssec:BS}

A second consequence of the potential experienced between the initial DM states is the possible formation of bound states.
This contribution was studied for the quintuplet in Ref.~\cite{Baumgart:2023pwn} and was found to generally only provide a small contribution to the hard photon spectrum.
That work also argued that the contribution to the hard line-like part of the spectrum from bound state decays (in the context of indirect detection) should also be suppressed relative to direct annihilation for higher odd-numbered representations.
However, the same level of suppression is not expected for contributions to the lower-energy continuum spectrum, which can be important especially for higher representations (cf. Fig.~\ref{fig:Spectra}).
Furthermore, to the degree that there are special points in parameter space where the direct annihilation contribution is significantly suppressed (see Fig.~\ref{fig:linexsec}), the intuition that bound-state contributions to the endpoint are small may not hold.
For those reasons it is important to compute the bound state contribution as we do so below.

We focus on dipole-mediated capture from a ($S=1$) $p$-wave initial state into the $s$-wave ground state, which we expect to be the dominant contribution, particularly for the continuum signal (as discussed in Ref.~\cite{Baumgart:2023pwn}).
The ground state then annihilates directly to SM particles, producing a continuum signal as discussed in Sec.~\ref{ssec:Cnt}. 
The capture cross section receives contributions both from photon/$Z$ emission (leading to a $Q=0$ final state) and from $W$ emission (leading to a $Q=\pm 1$ final state).
Let us denote the components of the initial ($Q=0$) state by $\phi_{mC}$, where $m$ is the absolute value of the charge of the individual particles, or $\phi_N$ for the $\chi^0\chi^0$ component.
We correspondingly denote the components of the final state by $\psi_{q_1+,q_2-}$, or $\psi_{qC}$ if $q_1=q_2$. There is no final-state component with $q_1=q_2=0$ due to the selection rules for states with identical fermions, as discussed above.

Accordingly, for the $2n+1$ representation, the contribution to the capture cross section from emission of a photon or a $Z$ boson is given by,
\begin{align}
\begin{split}
\sigma v=&\frac{2\aW k}{\pi M_\chi^2}\int d\Omega_k\,\left|\int d^3\mathbf{r}\,\hat{\boldsymbol\epsilon}(\hat{\mathbf{k}})\cdot\left[\sum_{m=1}^{n}m\psi^*_{mC} \nabla \phi_{mC}\right.+\frac{\aW M_\chi}{2}\hat{\mathbf{r}}e^{-m^{}_{\scaleto{W}{3.pt}} r}\frac{n(n+1)}{\sqrt{2}}\psi_C^*\phi_N\right.\\
&\left.\left.+\frac{\aW M_\chi}{2}\hat{\mathbf{r}}e^{-m^{}_{\scaleto{W}{3.pt}} r} \sum_{m=1}^{n}\left(\frac{n(n+1)}{2}-\frac{m(m+1)}{2}\right)\left(\psi^*_{(m+1)C}\phi_{mC}-\psi^*_{mC}\phi_{(m+1)C}\right)\right]\right|^2\!,    
\end{split} \nonumber \\
k & = \sW^2 (M_\chi v^2/4 - E_n) + \begin{cases} \cW^2 \sqrt{(M_\chi v^2/4 - E_n)^2 - \mZ^2}, & \mZ <  M_\chi v^2/4 - E_n \\
0, & \text{otherwise} \end{cases}
\end{align}
Here $\hat{\boldsymbol\epsilon}(\hat{\mathbf{k}})$ denotes the polarization direction of the outgoing (purely transverse) gauge boson and $\int d\Omega_k$ indicates the integral over the direction of the boson momentum, $\mathbf{k}$.
$E_n$ indicates the (negative) binding energy of the final state.

For capture through emission of a $W$ boson, the equivalent expression is,
\begin{align}
\begin{split}
\sigma v=&\frac{2\aW k}{\pi M_\chi^2}\int d\Omega_k\,\left|\int d^3\mathbf{r}\,\hat{\boldsymbol\epsilon}(\hat{\mathbf{k}})\cdot\left[\sqrt{\frac{n(n+1)}{4}}\psi^*_{+0} \nabla \phi_N\right.\right.+\frac{\aW M_\chi}{2}\hat{\mathbf{r}}\zeta(r)\sqrt{\frac{n(n+1)}{2}}\psi^*_{+0}\phi_C\\
&+\sum_{m=1}^{n}\frac{\nabla\phi_{mC}}{2\sqrt{2}}\left(\sqrt{n(n+1)-m(m+1)}\psi^*_{(m+1)+,m-}-\sqrt{n(n+1)-m(m-1)}\psi^*_{m+,(m-1)-}\right)\\
&+\left.\frac{\aW M_\chi}{2}\hat{\mathbf{r}}\zeta(r)\sum_{m=1}^{n}\sqrt{\frac{n(n+1)}{2}-\frac{m(m+1)}{2}}\psi^*_{(m+1)+,m-}\left((m+1)\phi_{(m+1)C}+m\phi_{mC}\right)\right]\!. 
\end{split}, \nonumber \\
k & = \begin{cases}  \sqrt{(M_\chi v^2/4 - E_n)^2 - \mW^2}, & \mW <  M_\chi v^2/4 - E_n \\ 
0 & \text{otherwise} \end{cases}, \nonumber \\
\zeta(r) & \equiv \frac{2}{r^2} \left[ \frac{\cW^2}{\mZ^2 - \mW^2} \left(e^{-m^{}_{\scaleto{W}{3.5pt}} r} (1 + \mW r) - e^{-m^{}_{\scaleto{Z}{3.5pt}} r}(1 + \mZ r) \right) \right.\nonumber \\
& \hspace{0.2cm}\left.+ \frac{\sW^2}{\mW^2} \left( 1- e^{-m^{}_{\scaleto{W}{3.5pt}} r} (1 + \mW r)\right) \right]\!.
\end{align}
These expressions are valid for any final state, not only the ground state, provided the appropriate final-state wavefunction $\psi$ is inserted.
Choosing specific angular momenta for the initial and final states allows for further simplification of the angular integral.

The remaining required ingredients are the final-state wavefunctions.
We derive these wavefunctions by adapting the method discussed in App.~B of Ref.~\cite{Asadi:2016ybp}, i.e.~by modeling the wavefunction as a linear combination of Coulombic basis wavefunctions and solving for their coefficients.
We introduce two important refinements on the method presented there.
Firstly, for the wino \cite{Asadi:2016ybp} and quintuplet \cite{Baumgart:2023pwn} cases studied previously in the literature, in the limit of unbroken SU(2) symmetry, there is only a single isospin eigenstate that experiences an attractive potential for odd $L+S$.
For the septuplet and higher representations, there are multiple such eigenstates.
We expand the basis of Coulombic wavefunctions to include all the attractive isospin eigenvalues.
This ensures that all the numerically computed bound state wavefunctions become exact in the high-mass limit where the SU(2) symmetry becomes approximately valid.

The second refinement is that for each of these isospin eigenvalues $j$, we replace the electroweak coupling $\aW$ with an ``effective'' coupling  $\epsilon_j \aW$ when defining our basis of Coulombic wavefunctions, and treat $\epsilon_j$ as a variational parameter.
This accounts first of all for the fact that we do not have only diagonal Coulombic potentials and, secondly, that the finite range of the Yukawa potentials effectively reduces the strength of the corresponding Coulomb potential.
More concretely, we compute the matrix elements of the Hamiltonian using the following Coulombic radial wavefunctions
\be
R_{n_p}^j(r)=\left(\frac{\epsilon_j \lambda_j \aW}{n_p}\right)^\frac{3}{2}\sqrt{\frac{(n_p-1)!}{2n_p~n_p!}}e^{-M_\chi r\frac{\epsilon_j\lambda_j\alpha^{}_{\scaleto{W}{3.pt}}}{2n_p}}L_{n_p-1}^{(1)}\!\left(M_\chi r\frac{\epsilon_j\lambda_j\aW}{n_p}\right)\!,
\ee
where $n_p$ is the principal quantum number, $\lambda_j\aW$ is the effective coupling in the $j^{\rm th}$ isospin channel with $\lambda_j=\min\left[n(n+1)-j(j+1)/2,\,1\right]$, and $L_n^{(m)}(x)$ the generalized Laguerre polynomials.
For odd $L+S$, we select all odd values of $j$ up to $2n-1$. The lowest lying bound states in the spectrum realize almost exact isospin eigenstates.
For each $j$, we select the deepest bound state in the corresponding approximate isospin eigenstate configuration, and determine $\epsilon_j$ by maximizing its binding energy.
As an example of this procedure, for the septuplet and assuming $M_\chi=48.8$\,TeV, we have $\epsilon_1= 0.88$ and $\epsilon_3= 0.89$.

The relevant LO potential for $Q=0$ states with $L+S$ odd matches Eq.~\eqref{eq:VpotQ0LO}, but with the removal of the rows and columns corresponding to the $\chi^0\chi^0$ state. 
For $Q=1$ states the LO potential couples states of the form $\chi^{m+}\chi^{(m-1)-}$.
We take $V_{ij}$ to describe the coupling between the two-particle states $\chi^{(n-i+1)+} \chi^{(n-i)-}$ and  $\chi^{(n-j+1)+} \chi^{(n-j)-}$, where $i$ and $j$ run from 1 to $n$.
Then in general the entries of the potential matrix are:
\begin{align} 
V_{ij} & = -(n - i + 1)(n-i) \delta_{ij} V_{\gamma Z}(r) + \frac{1}{2} V_{\smallW}(r) \nonumber \\
& \times \left[ \delta_{i,j-1}  \sqrt{\left(n(n+1) - (i-n)(i-n-1) \right)\left(n(n+1) - (i-n)(i-n + 1) \right)} \right. \nonumber \\
& \left. + \delta_{j,i-1} \sqrt{\left(n(n+1) - (i-n-1)(i-n-2) \right)\left(n(n+1) - (i-n - 1)(i-n) \right)}\right. \nonumber \\
& \left. + (-1)^{L+S} \delta_{i,n}\delta_{j,n} n(n+1) \right]\!.
\end{align}
The potential for $Q=-1$ states is identical (with the appropriate redefinition of the states coupled by the potential, i.e. from $\chi^{(n-i+1)+} \chi^{(n-i)-}$ to $\chi^{(n-i)+} \chi^{(n-i+1)-}$).
In all cases the NLO corrections are implemented as in Eq.~\eqref{eq:vnlo}.

Note that higher-net-charge states will generally exist in the spectrum, but are not directly accessible by radiative capture from the $Q=0$ initial state.
They can in principle be populated by decays from more shallowly bound states, but from a different channel than the $p\rightarrow s$ capture considered in this paper.
In fact, the $p\rightarrow s$ capture implies $S=1$ since the initial scattering state must be a combination of even isospin states.
A bound state with $Q=2$ and $S=1$ must have odd $L$ and, if there is enough phase space, it will decay to a lower-lying $L=0,~ S=1$ state which, with the exceptions we discuss later in this section, is an almost exact isospin eigenstate with $I=1$.
Such states have at most $Q=1$ and their dominant annihilation channel is into pairs of fermions.
$Q=2$ states are expected to affect the spectrum if produced from $d\rightarrow p$ channels with $S=0$.
These states are dominantly $I=2$ isospin eigenstates which, however, decay into vectors rather than fermions.
The spectrum resulting from the annihilation of $Q=2$ bound states requires matching to a different set of hard operators than those needed for the annihilation channels considered so far.
In this work, we neglect these contributions since, as we show in Sec~\ref{ssec:results}, our conservative signal is already sufficient to exclude the undecuplet and most of the tredecuplet mass range.

We conclude this section by reconsidering the assumption that all the $L=0$, $S=1$ bound states with $Q=0,1$ dominantly annihilate into SM fermions and longitudinal vectors.
In fact, this assumption is not valid for all the bound states of multiplets greater than the quintuplet.
In the following, we will discuss in detail the example of the septuplet and then summarize the general properties for the higher multiplets.
In particular, in Tab.~\ref{tab:seven_BS} we show, for the case of the thermal septuplet with $M_\chi=48.8$\,TeV, the binding energy $E_B^i$ of the $i^{\rm th}$ $L=0$, $S=1$ bound state, its BSF cross section $\langle\sigma_i v\rangle=\sum_{X=\gamma, W, Z}\langle\sigma(\chi_0\chi_0\rightarrow \mathrm{BS}_i+X)v\rangle$, the projection of the bound states onto the isospin eigenstates, and the fraction of the total BSF cross section.
As we can see, the total BSF cross section is saturated by the deepest states, which are almost exact isospin eigenstates. Due to the selection rules, such bound states must necessarily be odd-isospin states, $I=1,3$ in the case of the septuplet.
In the following, we denote these states as $n_s s_{1,3}$, respectively, $n_s$ being the principal quantum number.

\renewcommand{\arraystretch}{1.2}
\begin{table}[t!]
\centering
\begin{tabular}{c|c|c|c|c}
\(E_B^i \, [{\rm TeV}]\) & \(\langle \sigma_i v \rangle \,[{\rm cm}^3/{\rm s}]\) & \(|\langle {\rm BS}_i | I=1 \rangle|^2\) & \(|\langle {\rm BS}_i | I=3 \rangle|^2\) & \(\frac{\langle \sigma_i v \rangle}{\sum_i\langle \sigma_i v \rangle}\) \\ \hline
$-1.32$   & $4.7 \times 10^{-23}$  & $0.999996$        & $4.2 \times 10^{-6}$   & $0.57$ \\
$-0.40$  & $1.6 \times 10^{-23}$ & $0.00002$    & $0.999979$                & $0.196$ \\
$-0.33$  & $4.7 \times 10^{-24}$ & $0.999973$        & $0.000027$            & $0.057$ \\
$-0.14$   & $8.2 \times 10^{-25}$  & $0.999757$        & $0.00024$             & $0.00998$ \\
$-0.093$ & $1.2 \times 10^{-23}$ & $0.00024$     & $0.999748$                & $0.147$ \\
$-0.068$ & $1.1 \times 10^{-27}$ & $0.99938$        & $0.00062$             & $0.000013$ \\
$-0.037$ & $4.9 \times 10^{-26}$ & $0.9925$        & $0.0075$              & $0.0006$ \\
$-0.036$ & $7.3 \times 10^{-26}$ & $0.00667$      & $0.99328$                 & $0.00088$ \\
\end{tabular}
\caption{Properties of the BS spectrum for the thermal septuplet with $M_{\chi}=48.8$\,TeV. The columns in order show the binding energy of the BS, its formation cross-section, the projection of the BS onto isospin eigenstates (for the septuplet, the only attractive channels that support BS and that are consistent with odd-($L+S$) states are $I=1$ and $I=3$). Finally, we show the fraction of the total BSF cross-section corresponding to each BS.}
\label{tab:seven_BS}
\end{table}

The $n_ss_1$ states directly annihilate into SM particles with a rate $\Gamma_{\rm ann}\sim \aW^5 M_\chi$, while the channels in which higher isospin $n_ss_I$ states can annihilate are suppressed at least by $(\aW/16\pi^2)^{\frac{I-1}{2}}$.
Therefore, the question is whether they can dominantly decay into lower $n_ss_1$ or not.
A possible decay chain is given by $n_s s_{3}\rightarrow n_pp_2\rightarrow n_s' s_{1}$, where $n_pp_2$ denotes a $L=S=1$ state with $I=2$ and $n_p$ its principal quantum number.
While $n_pp_2$ annihilates into SM particles with $\Gamma_{\rm ann}\sim \aW^7 M_\chi$, both decays are mediated by electric dipole transitions with a rate that scales as $\Gamma_{\rm dec}\sim \aW^5M_\chi$; thus they are more likely to occur (relative to annihilation) if there is enough phase space.
Taking the Coulombic limit, this condition can be written as
\be
E_B(n_ss_3)<E_B(n_pp_2)\;\Rightarrow\; n_s>\frac{2}{3}n_p>1,
\ee
which means that only for the first $I=3$ state, $1s_3$, there is no available phase space for electric dipole decay channels.
The $1s_3$ can still decay to $1s_2$ (with $S=0$) through a magnetic transition whose rate scales like $\aW^7M_\chi$~\cite{Bottaro:2021srh}.
Therefore, a fraction $\sim\aW/(\aW+1/16\pi^2)$ of the initially produced $1s_3$ states are expected to decay to $1s_2$ states which, in turn, annihilate into pairs of transverse vectors with a fast $\mathcal{O}(\aW^5M_\chi)$ rate, thus contributing to all the line, endpoint, and continuum spectra, as in the direct annihilation case as discussed in Ref.~\cite{Baumgart:2023pwn}.
For general multiplets beyond the quintuplet, these dynamics are common to all the $1s_I$ bound states with $I>1$ present in the spectrum.
Therefore, the final spectrum arising from $S=1$ bound state annihilations is a sum of two contributions:
\begin{enumerate}
\item Continuum spectrum from final fermions and longitudinal vectors produced by the annihilation of all the $L=0$, $S=1$ states but $1s_{I>1}$ states; and 
\item Line, endpoint and continuum spectrum from final transverse $VV$ final states, with $V=\gamma, Z, W$ from $1s_{I>1}$ annihilations.
\end{enumerate}
In this work, we have again opted for the more conservative choice of neglecting the latter contribution.

\subsection{Continuum photons}
\label{ssec:Cnt}

The final contribution to consider is the production of photons with energy far below the DM mass.
There are three primary contributions to this continuum spectrum: final states involving a $Z$, final states involving a $W$, and the decay of bound states.

Consider the two electroweak boson contributions.\footnote{The treatment of continuum emission from $W$ and $Z$ final states presented here differs to that in Ref.~\cite{Baumgart:2023pwn}.
The approach in that work was more ad hoc than the procedure outlined herein, nevertheless at the thermal wino and quintuplet masses the difference between prescriptions is marginal.}
We need to compute the rate at which $W$ and $Z$ bosons are produced per annihilation and then weight those cross sections by the appropriate photon spectra these final state produce.
Unlike for the case of the hard photons studied in Sec.~\ref{ssec:DA}, observing a continuum photon is insufficient to provide an effective restriction on the kinematics of the initial $Z$ or $W$ boson.
Accordingly, the appropriate cross sections to compute are those for the semi-inclusive processes $W+X$ and $Z+X$.
As the initial and final states of the annihilation both carry electroweak charge, in general we should expect these cross sections to involve electroweak logs.

The cross sections are computed as follows.
For simplicity of presentation we restrict our attention to the wino; results for general representations are obtained in a conceptually identical manner, with results provided in App.~\ref{app:WZ-general}.
We start with the inclusive $Z+X$ cross section as this can be extracted immediately from the equivalent calculations for $\gamma+X$ provided in Refs.~\cite{Baumgart:2017nsr,Baumgart:2018yed}.
Exchanging $\sW$ for $\cW$, we have the following LL result
\bea
\langle \sigma v \rangle_{\smallZ+\smallX} = \frac{\pi \aW^2 \cW^2}{3M_{\chi}^2} 
&\left\{ 2 |s_{00}|^2 \left( 1 - e^{-(3/2\pi) \alpha^{}_{\scaleto{W}{3.pt}} L_{\chi}^2} \right) 
+ 3 |s_{01}|^2 \left( 1 + e^{-(3/2\pi) \alpha^{}_{\scaleto{W}{3.pt}} L_{\chi}^2} \right) \right. \\
&\left.\hspace{0.2cm}+ 2\sqrt{2} \textrm{Re}[s_{00} s^*_{01}] \left( 1 - e^{-(3/2\pi) \alpha^{}_{\scaleto{W}{3.pt}} L_{\chi}^2} \right) \right\}\!.
\label{eq:Wino-ZX}
\eea
In principle this cross section can be immediately extended to NLL.
Yet unlike the hard photon calculation, this result must be weighed by the continuum spectrum for a final state $Z$ boson, which has not been computed to NLL accuracy, rendering the additional complexity unnecessary.

Slightly more effort is required to obtain the analogous result for $W+X$.
This has not been computed previously in the literature, although it can be extracted indirectly.
The fully inclusive cross section can be decomposed as follows,
\be
\langle \sigma v \rangle_\textrm{inc.}
= \langle \sigma v \rangle_{\gamma+\smallX} + \langle \sigma v \rangle_{\smallZ+\smallX} + \langle \sigma v \rangle_{\smallW+\smallX},
\ee
where we imagine distinguishing the bosons in a two boson final state to avoid double counting between the three channels.
For the fully inclusive result on the left of the above expression, the final state is now an electroweak singlet, although the initial state, $\chi^0 \chi^0$, is not.
Nevertheless, as shown in Ref.~\cite{Baumgart:2023pwn}, as the initial state is non-relativistic, once we are fully inclusive over the final states the result cannot involve large logarithmic terms.
We can therefore simply compute the result at tree level, obtaining
\bea
\langle \sigma v \rangle_\textrm{inc.} 
= \frac{\pi \aW^2}{M_{\chi}^2} \left\{ 2 |s_{00}|^2 + 3 |s_{01}|^2 
+ 2\sqrt{2} \textrm{Re}[s_{00} s^*_{01}]\right\}.
\eea

Combining the above results, we arrive at the second cross section
\bea
\langle \sigma v \rangle_{\smallW+\smallX} = \frac{\pi \aW^2}{3M_{\chi}^2} 
&\left\{ 2 |s_{00}|^2 \left( 2 + e^{-(3/2\pi) \alpha^{}_{\scaleto{W}{3.pt}} L_{\chi}^2} \right) 
+ 3 |s_{01}|^2 \left( 2 - e^{-(3/2\pi) \alpha^{}_{\scaleto{W}{3.pt}} L_{\chi}^2} \right) \right. \\
&\left.\hspace{0.2cm}+ 2\sqrt{2} \textrm{Re}[s_{00} s^*_{01}] \left( 2 + e^{-(3/2\pi) \alpha^{}_{\scaleto{W}{3.pt}} L_{\chi}^2} \right) \right\}\!.
\label{eq:Wino-WX}
\eea
Once more, Eqs.~\eqref{eq:Wino-ZX} and \eqref{eq:Wino-WX} hold only for the wino.
The equivalent results for an arbitrary real representations take a similar form, simply with more involved Sommerfeld contributions, and are given in App.~\ref{app:WZ-general}.

These cross sections now need to be combined with the spectrum per annihilation of continuum photons that arises from a final state $ZZ$ or $WW$ (the presence of additional final states such as $Z\gamma$ are accounted for in the $Z+X$ cross section computed above).
There are many tools available for computing these spectra, including \texttt{PPPC4DMID}~\cite{Cirelli:2010xx}, \texttt{HDMSpectra}~\cite{Bauer:2020jay}, and \texttt{CosmiXs}~\cite{Arina:2023eic}.
In our case we use \texttt{HDMSpectra}, and to ensure there is no double counting between the continuum spectrum and the electroweak corrections we include in our hard photon calculation, we simply evaluate the spectra at a relatively low center-of-mass scale of 1\,TeV.
Whilst this approach is imperfect, we checked that our results are comparable to those obtained with \texttt{PPPC4DMID}, which can have the electroweak corrections turned off (although that code does not extend to our full mass range).

Combining the two ingredients above, we can evaluate the appropriate cross section for continuum photons from a final state $Z$ or $W$ as,
\be
\frac{d \langle\sigma v \rangle_{\smallZ\smallW}}{dz}
=
\langle \sigma v \rangle_{\smallZ+\smallX}\frac{dN_{\smallZ}}{dz}
+ \langle \sigma v \rangle_{\smallW+\smallX}
\frac{dN_{\smallW}}{dz}.
\ee

Finally, there is the contribution from the decay of metastable bound states.
In particular, bound states with $Q=0$ and $Q=1$ that have $L=0$ and $S=1$ dominantly annihilate into fermions and longitudinal vectors through $Z$- and $W$-mediated $s$-channel diagrams, respectively.
With this in mind, we denote the cumulative capture cross section into $L=0$ and $S=1$ states with $Q=0$ -- proceeding via the emission of a $Z$ or photon -- as $\langle \sigma v\rangle_{\scriptscriptstyle {\rm BS}}^{\smallZ}$; the analogous cross section for $Q=1$ states where a $W$ is emitted is written $\langle \sigma v\rangle_{\scriptscriptstyle {\rm BS}}^{\smallW}$.
Neutral bound states decay to pairs of oppositely charged fermions and vectors, whereas charged bound states decay into final states with total charge one.
There are half as many degrees of freedom in the available charge one final states, although this factor is exactly compensated by the presence of bound states with $Q=+1$ and $-1$.
Therefore the charged and neutral bound states produce an equal contribution to the continuum spectrum,
\be
\frac{d\langle \sigma v \rangle_{\scriptscriptstyle {\rm BS}}}{dz}=\left(\langle \sigma v\rangle_{\scriptscriptstyle {\rm BS}}^{\smallZ}+\langle \sigma v\rangle_{\scriptscriptstyle {\rm BS}}^{\smallW}\right)\frac{dN_{\scriptscriptstyle {\rm BS}}}{dz}.
\ee
The common spectrum is determined by a sum over all the fermionic and longitudinal vector SM states weighted by their branching ratios, which in this case coincide with the fraction of degrees of freedom.
Summing over all leptons, $l$, and quarks, $q$, the spectrum is given by
\be
\frac{dN_{\scriptscriptstyle {\rm BS}}}{dz} = 
\frac{1}{25}\sum_l \frac{dN_l}{dz}+\frac{3}{25}\sum_q \frac{dN_q}{dz}
+ \frac{1}{50}\frac{dN_{W_L}}{dz}+\frac{1}{100}\left(\frac{dN_{Z_L}}{dz}+\frac{dN_{h}}{dz}\right)\!,
\ee
where again all $dN/dz$ are spectra per annihilation.

\section{Forecasts for Real WIMPs at CTAO}
\label{sec:forecasts}

Having determined the expected signal strength and spectrum for the photons that emerge from the annihilation of real WIMPs we next consider the CTAO sensitivity to the corresponding DM models.
The first step in connecting the predictions to an experiment is the canonical indirect detection expression for the flux resulting from DM annihilations,
\be
\frac{d\Phi}{dE} = \frac{1}{8\pi M_{\chi}^2} \frac{d \langle \sigma v \rangle}{d E}\, J_{\Omega}.
\label{eq:IDflux}
\ee
This expression codifies the expected number of photons an instrument will observe per unit detector area per unit time per unit energy; correspondingly, $d\Phi/dE$ carries units of [photons/cm$^2$/s/TeV] (see e.g. Ref.~\cite{Lisanti:2017qoz}).
For the LL prediction, we can obtain the differential cross section in Eq.~\eqref{eq:IDflux} from Eq.~\eqref{eq:LL} although the term proportional to the $\delta$-function must be doubled to account for the presence of two photons in the exclusive final state that could contribute to the flux (see e.g. the discussion in Ref.~\cite{Rinchiuso:2018ajn}).
The equivalent NLL result follows similarly from the expressions in App.~\ref{app:NLL}.
Examples of the difference spectra were shown in Fig.~\ref{fig:Spectra}.
Beyond this, fully specifying the flux requires providing a value for the so-called $J$-factor in the notation of Ref.~\cite{Bergstrom:1997fh}: a measure of the DM density squared, integrated along the line-of-sight and over the full region of interest (ROI), labeled $\Omega$.
In Sec.~\ref{ssec:Rho} we outline our procedure for modeling the DM content of the Galactic Center which can then be mapped to a definitive prediction for $J_{\Omega}$.

Of course, no telescope sees the DM flux directly.
Firstly, the flux in Eq.~\eqref{eq:IDflux} must be convolved with the instrument response of CTAO before we arrive at a prediction for the DM signal at the telescope.
Further, any DM signal does not arrive in isolation and must be disentangled from unavoidable background emissions.
With this in mind, in Sec.~\ref{ssec:CTAOresponse} we detail our procedure for implementing the CTAO instrument response, modeling the background, and our likelihood analysis for teasing out a DM signal.
Putting these together our results and corresponding forecasts for real WIMPs are provided in Sec.~\ref{ssec:results}.
By default our results are presented using a conventional binned analysis, although we also confirm the impact of an unbinned approach in Sec.~\ref{ssec:unbinned}.

\subsection{The dark matter profile of the Milky Way}
\label{ssec:Rho}

The rate of DM annihilations occurring in the galaxy depends upon the number density of DM particles squared.
As gravitational probes of DM allow us to constrain the mass rather than number density, it is conventional to encode this contribution to the DM flux in the $J$-factor that appeared in Eq.~\eqref{eq:IDflux},
\be
J_{\Omega} = \int_{\Omega}  d\Omega\,\int_{\rm los}  ds\,\rho^2(r[s,\Omega]).
\ee
Here $\rho(r)$ is the mass density of DM, which we assume to be spherically symmetric and therefore dependent purely on the distance $r$ from the Galactic Center.
To infer the total flux for a given observation, we integrate the density squared over all $s$, the line-of-sight (los), and the solid angle of the ROI, $\Omega$.

Evaluating $J_{\Omega}$ is therefore reduced to a choice of ROI and model for $\rho(r)$.
Our choice of ROI is motivated in Sec.~\ref{ssec:CTAOresponse}, although in brief we analyze the inner galaxy with the Galactic plane masked to partially remove the brightest astrophysical emission.
In detail, we take the region of the celestial sphere with galactic latitude, $b$, restricted to have magnitude greater than $0.3^{\circ}$, and the angle from the Galactic Center $\theta$ to be less than $2^{\circ}$.
For $\rho(r)$, we take the classic Einasto profile that has been broadly adopted for DM searches with IACTs~\cite{1965TrAlm...5...87E,Pieri:2009je}.
The profile is defined by,
\be
\rho(r) = \rho_0 \exp \left[ - \frac{2}{\alpha} \left( \left( \frac{r}{r_s} \right)^{\alpha}-1 \right) \right]\!,
\label{eq:Einasto}
\ee
with the parameters fixed to $\alpha=0.17$, $r_s=20\,$kpc, and $\rho_0$ adjusted to ensure the density at the solar radius ($r_{\odot}=8.5\,$kpc) is 0.39\,GeV/cm$^3$.

\begin{figure*}[!t]
\centering
\includegraphics[width=0.45\textwidth]{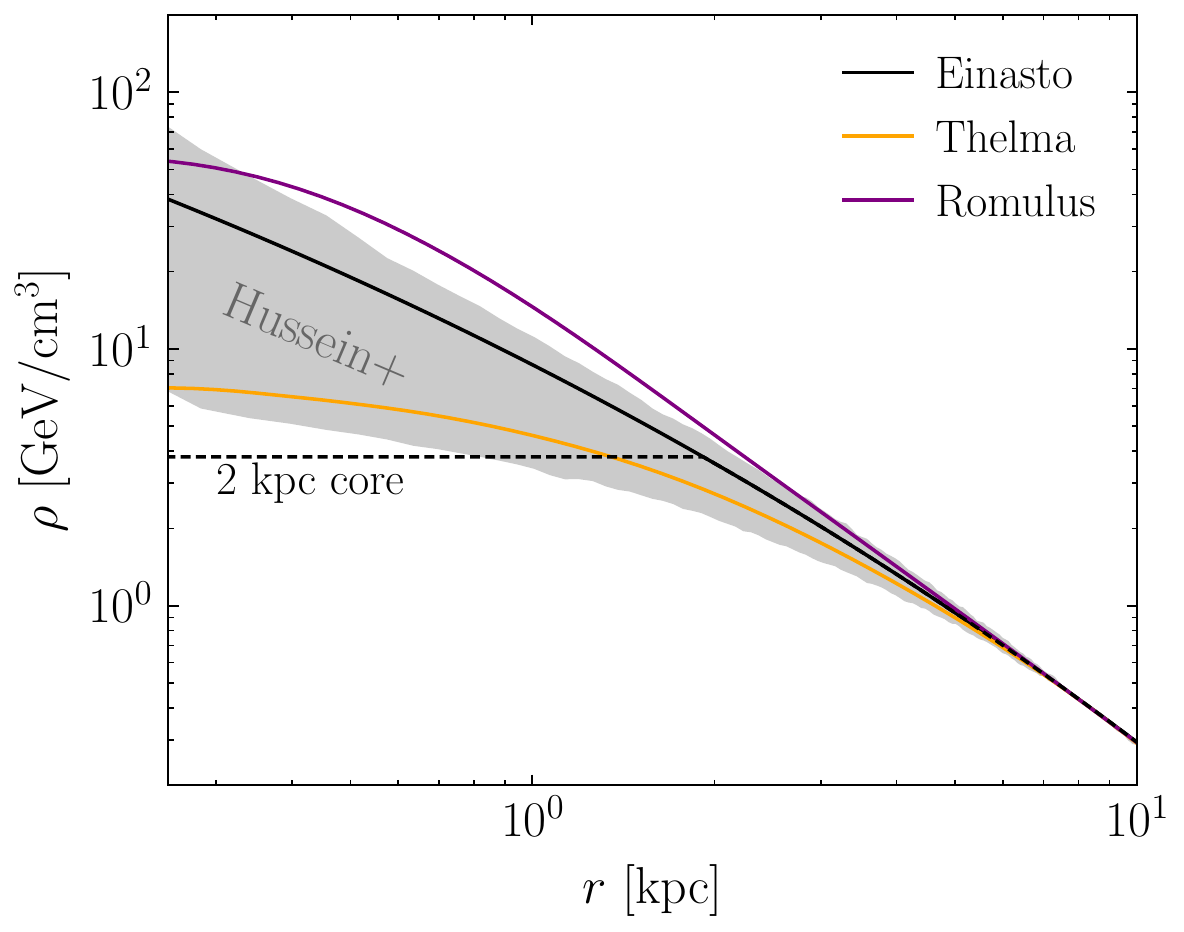}
\hspace{0.5cm}
\includegraphics[width=0.45\textwidth]{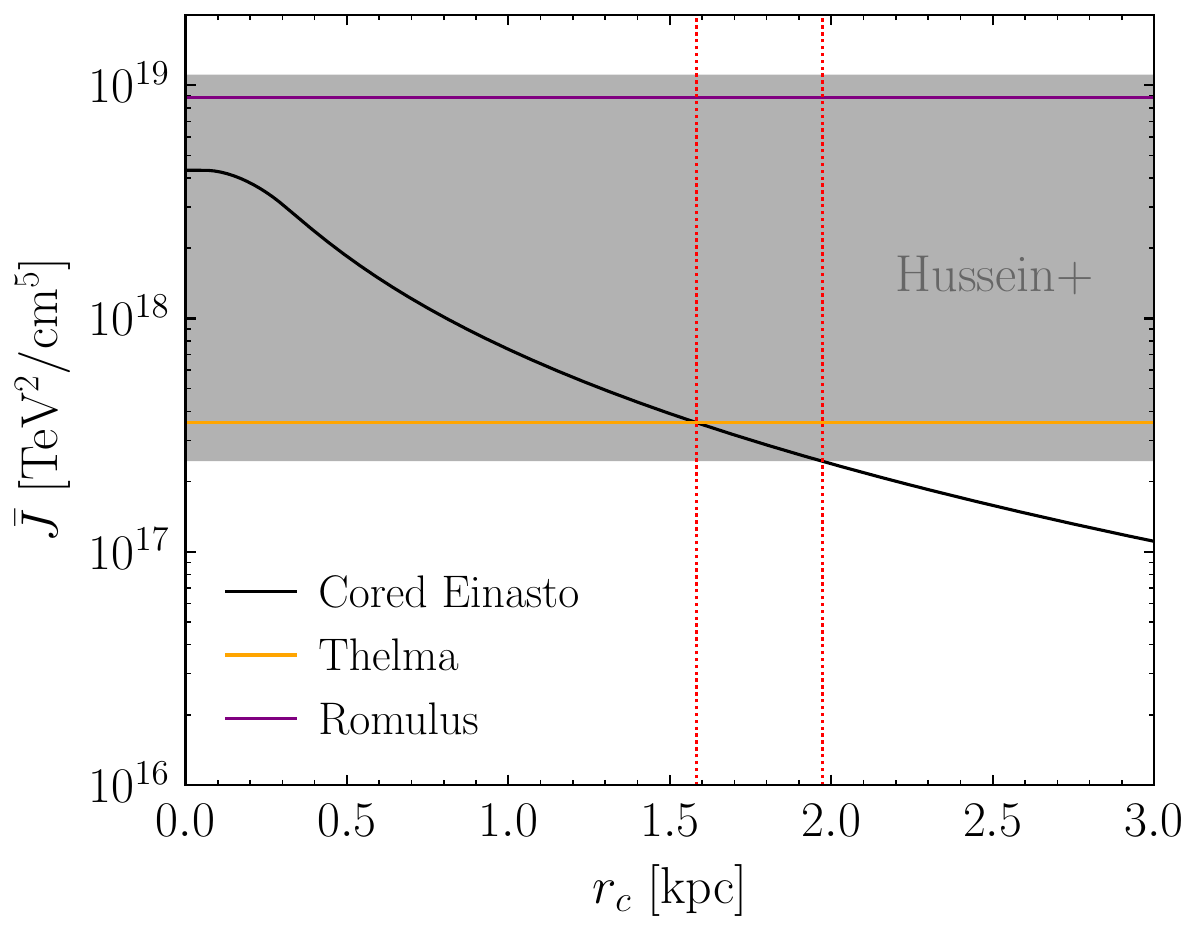}
\vspace{-0.2cm}
\caption{A comparison of various DM density profiles to a cored Einasto distribution.
(Left) We show the DM density computed for three profiles, our default Einasto (black, including in dashed the impact of a 2\,kpc core), and the Thelma (orange) and Romulus (purple) profiles that bracket the range of densities found in FIRE-2 simulations~\cite{Hopkins:2017ycn,McKeown:2021sob}.
The gray band corresponds to the uncertainty band on the Milky Way DM density profile proposed in Ref.~\cite{Hussein:2025xwm}.
All profiles are aligned to a local density of $\rho = 0.39$\,GeV/cm$^3$ at 8.5\,kpc.
(Right) The average $J$-factor in our ROI for these same profiles.
We now show how the Einasto DM flux decreases as the core radius increases.
Comparing this to the alternative profiles, we find that Thelma maps to a $\simeq$1.6\,kpc core, whereas the lower edge of the band from Ref.~\cite{Hussein:2025xwm} corresponds closely to our chosen 2\,kpc exclusion.}
\vspace{-0.2cm}
\label{fig:DM-Comparison}
\end{figure*}

With the above choices, $J_{\Omega}$ is a single number, for instance we can compute the average $J$-factor over the ROI to be $\bar{J} \simeq 4.3 \times 10^{18}\,$TeV$^2$/cm$^5$.
Here $\bar{J} = J_{\Omega} / \int_{\Omega} d\Omega$ and for our ROI $\int_{\Omega} d\Omega \simeq 3.1 \times 10^{-3}\,{\rm sr}$.
The simplicity of this procedure is deceptive: we do not know the DM distribution of the Milky Way -- particularly near the Galactic Center -- to anywhere near the precision with which we computed the spectrum.
There are various approaches one can take to this uncertainty.
When considering whether CTAO can make a definitive statement regarding real WIMPs, the primary concern is whether the Einasto profile could lead to an overestimate of $J_{\Omega}$.
Indeed, the Einasto profile is relatively cuspy and it is known from numerical simulations that in principle the profile can become cored, see e.g. Ref.~\cite{Chan:2015tna}.
A simple parameterization of this adopted in previous studies, e.g. Refs.~\cite{Rinchiuso:2018ajn,Rinchiuso:2020skh}, is to explicitly core the profile; in detail, we take $\rho(r)$ as in Eq.~\eqref{eq:Einasto} if $r \geq r_c$ and $\rho(r) = \rho(r_c)$ for $r < r_c$.\footnote{In addition to imposing a flat core for $r<r_c$, the parameters of the Einasto profile in Eq.~\eqref{eq:Einasto} could be modified in order to keep the mass of the Galactic halo unchanged.
In particular, for a given $r_c$ we can modify $r_s$ and $\rho_0$ in such a way as to keep the mass enclosed within 60 kpc~\cite{SDSS:2008nmx} constant, as well as to maintain the local DM density at 0.39\,GeV/cm$^3$.
We have checked that performing such a rescaling only marginally affects the final results and so throughout we choose the conservative option of not performing this rescaling.}
To anchor this procedure, we note that $r_c \gtrsim 2\,$kpc appears to be loosely disfavored~\cite{2015MNRAS.448..713P,Hooper:2016ggc}, and we take that value as our conservative lower limit on the DM density in our ROI.
Quantitatively this reduces the $J$-factor by a factor of $\sim$20.

The above approach is admittedly crude.
A more physically motivated procedure is to take the DM profile extracted from cutting edge numerical simulations of Milky Way-like galaxies and look to the lower edge of such values for a conservative limit.
For instance, the suite of FIRE-2 simulations provide the DM density profiles of fourteen Milky Way like galaxies~\cite{Hopkins:2017ycn,McKeown:2021sob}.
On the upper and lower end of those profiles are the densities labeled as Romulus and Thelma, and we show a comparison to those in Fig.~\ref{fig:DM-Comparison}.
(All profiles are normalized to the same density at the solar radius.)
In particular, we see that the DM density in our ROI for the Thelma profile is comparable to an Einasto profile with a $\simeq$1.6\,kpc core.

An alternative was recently introduced in Ref.~\cite{Hussein:2025xwm}.
The approach taken in that work was to build models calibrated to numerical galaxy simulations, which were then combined with the observed stellar kinematics of the Milky Way in the inner galaxy to generate an uncertainty band for $\rho(r)$.
The band derived in that work is also shown in Fig.~\ref{fig:DM-Comparison}.
The upper edge arises from scenarios where the adiabatic contraction of the DM halo as a result of the larger density of baryons in the inner galaxy is the dominant effect.
The lower edge is from the possibility of strong baryonic feedback, where effects such as supernovae, radiation pressure, and stellar winds drive baryons from the galactic center, which in turn can drive a coring of the DM profile.
As shown in Fig.~\ref{fig:DM-Comparison}, the lower limit derived in Ref.~\cite{Hussein:2025xwm} corresponds very closely to a 2\,kpc core Einasto.
In summary, the 2\,kpc core value appears to be a relatively robust lower limit on the DM density adding further weight to its use for our most conservative sensitivity estimates.
We emphasize, however, that an emerging DM signal would most likely be associated with a larger $J$-factor.

\subsection{CTAO instrument response and backgrounds}
\label{ssec:CTAOresponse}

With a prescription for $J_{\Omega}$ we have a complete model for the expected rate that various real WIMPs can produce photons hitting the Earth's atmosphere as given in $d\Phi/dE$ from Eq.~\eqref{eq:IDflux}.
These photons initiate a cascade in the atmosphere that includes charged particles whose Cherenkov radiation IACTs such as CTAO seek to detect.
The incident photons can only be reconstructed approximately and with finite efficiency from the Cherenkov light; both factors are encapsulated in the CTAO instrument performance that we must model as discussed in this section.
Further, DM annihilations will not be the only source of events at TeV energies.
Astrophysical sources of photons are expected, but even more importantly for IACTs is the far higher rate of charged cosmic-rays---dominantly protons and helium nuclei.
Although the charged cosmic-ray showers can be distinguished from photons to a degree, even with a rejection efficiency of 99\%, these events produce the dominant background that appears spatially isotropic.
This background must therefore be accounted for and represents a foreground through which we look for a signal of real WIMPs.

To begin with, we focus our attention on CTAO-South and its expected observation of the innermost part of the Milky Way.
In order to define our region of interest, we follow the approach adopted for projections by the CTAO collaboration~\cite{CTAO:2024wvb}.
In particular, we define our ROI as the region within $2^{\circ}$ of the Galactic Center and mask the plane of the Galaxy at latitudes less than $0.3^{\circ}$.
This region is covered by taking nine equally spaced observations obtained by setting the galactic latitude or longitude to $\{-1^{\circ},\,0^{\circ},\,+1^{\circ}\}$.
Each observation will be taken for $\simeq$58.3\,hrs, for a total observation time of 500\,hrs.
With this observation strategy, the effective area of CTAO is essentially uniform across the ROI.
The exact choice of ROI is unlikely to significantly impact our conclusions; for instance, Ref.~\cite{Rodd:2024qsi} considered observations to cover an ROI extending out to $5^{\circ}$ and with this obtained sensitivity qualitatively identical to Ref.~\cite{CTAO:2024wvb}.

\begin{figure}[!t]
\centering
\includegraphics[width=0.45\textwidth]{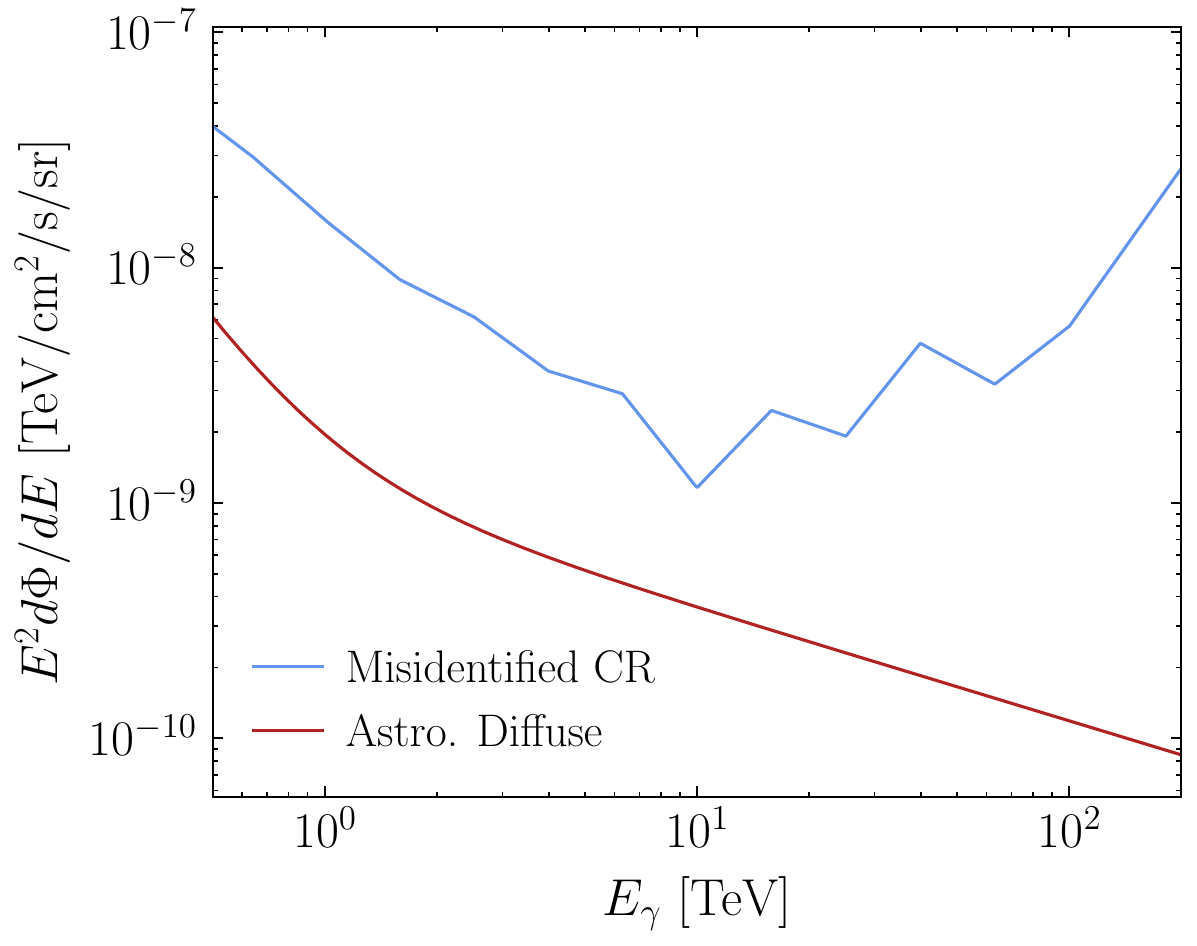}
\vspace{-0.2cm}
\caption{The expected background flux from misidentified charged cosmic-rays and astrophysical diffuse emission in our ROI.
See text for details.}
\vspace{-0.2cm}
\label{fig:background}
\end{figure}

As mentioned, there are two relevant background sources to consider: misidentified charged cosmic-rays and the contribution from astrophysical photons.
The CTAO collaboration provides a public model for the expected charged cosmic-ray contribution as this depends on the instrument's ability to reject such events.
We use the public projected performance for the Alpha Configuration of the array~\cite{CTA_performance}, taking the results for the expected CTAO-South array of 14 medium-sized and 37 small-sized telescopes, with an expected zenith angle for the observations of $20^{\circ}$.
The dominant astrophysical background is expected to be diffuse emission associated with our Milky Way Galaxy and is not isotropic.
This emission is observed at TeV energies by Fermi and may also be being observed by H.E.S.S. as discussed in Ref.~\cite{Rodd:2024qsi}.
To model this emission we adopt the simplified GDE scenario 2 introduced in Ref.~\cite{Rinchiuso:2020skh}.
We show the expected flux in our ROI from the two background sources in Fig.~\ref{fig:background}.

Although there is considerable uncertainty as to what the astrophysical background will look like across the CTAO energy range, this uncertainty does not impact the results of our default analysis; see App.~\ref{app:Syst}.
The reason for this is that for our most conservative DM profile, the Einasto distribution cored at 2\,kpc, the signal is essentially flat across our ROI.
Consequently, we have confirmed that spatial information does not help distinguish the most conservative DM signal from the dominant background, implying that the astrophysical diffuse emission remains a strictly subdominant effect.
When forecasting for discovery, the spatial information can be key and a more detailed consideration of the astrophysical contribution can be relevant; for a discussion on this point we refer to Refs.~\cite{CTAO:2024wvb,Rodd:2024qsi}.

In summary, the above procedure provides a prediction for the expected background flux in our ROI.
Combined with a signal model, we obtain a model for the total flux, $d\Phi/dE$, in units of [photons/cm$^2$/s/TeV] we expect to arrive at the Earth's atmosphere.
The expected number of events the instrument would then observe in an energy bin of size $\Delta E^i_r$ can then be computed as,
\be
N_i = T_{\rm obs} \int_{\Delta E^i_r} dE_r\,\int_0^{\infty} dE_t\, {\cal A}_{\rm eff}(E_t)\,{\cal R}(E_t,E_r)\,\frac{d\Phi}{dE_t}(E_t).
\label{eq:pred-Ni}
\ee
Let us break down the various contributions here.
Firstly, the flux $d\Phi/dE_t (E_t)$ is evaluated at the true incident photon energy $E_t$.
CTAO cannot reconstruct $E_t$ exactly, with that imprecision modeled by a Gaussian energy resolution,
\be
\mathcal{R}(E_t,E_r) = \frac{1}{\sqrt{2\pi\sigma^2(E_t)}}\exp\left(-\frac{(E_r-E_t)^2}{2\sigma^2(E_t)}\right)\!.
\ee
Here $\sigma(E_t)$ captures the energy-dependent width of the response that is provided for CTAO as part of the publicly available Alpha Configuration information. 
At the energies of interest for this study $\sigma(E_t)/E_t \simeq (5-7)\%$.
From here, the flux is further weighted by the instrumental effective area ${\cal A}_{\rm eff}(E_t)$, which captures the efficiency with which CTAO can detect photons of a given true energy and carries units of cm$^2$.
Again, we adopt the Alpha Configuration model for this parameter, recalling as above that for our ROI the effective area is approximately spatially uniform.
After integrating over all true energies and the reconstructed energy bin of interest, the result is finally weighted by the observation time $T_{\rm obs}=500\,$hrs to provide an expected number of measured counts.

The above prescription is sufficient to determine the expected number of signal and background events to occur in a defined set of energy bins.
We bin the data into 46 logarithmic spaced energy bins from 100\,GeV to 200\,TeV and have confirmed that our results are insensitive to the use of an even finer binning.  In Sec.~\ref{ssec:unbinned}, we show how this method gives consistent results with an unbinned analysis, which corresponds to the limit of infinitesimally fine bins.
With this binning, Eq.~\eqref{eq:pred-Ni} then provides a predicted number of counts in each bin, $\mathbf{N} = \{ N_i \}$.
In this context, CTAO will ultimately provide a set of integer observed counts in each bin $\mathbf{d} = \{d_i\}$, which we can compare to the prediction using the Poisson likelihood ${\cal L}(\mathbf{d}|\mathbf{N}) = \prod_i N_i^{d_i} e^{-N_i}/d_i!$.
To determine the expected reach of CTAO one approach is mock datasets, although it is more convenient to adopt the Asimov procedure~\cite{Cowan:2010js} where the expected limit can be determined from the expected dataset.
Under this approach we fix the data to the background only hypothesis, $\mathbf{d} = \mathbf{N}^{\rm bkg}$.
We then define a test statistic for upper limits,
\be
q = 2 [ \ln {\cal L}(\mathbf{N}^{\rm bkg}|\mathbf{N}^{\rm sig}+\mathbf{N}^{\rm bkg}) - \ln {\cal L}(\mathbf{N}^{\rm bkg}|\mathbf{N}^{\rm bkg})] \simeq -\sum_i \frac{(N_i^{\rm sig})^2}{N_i^{\rm bkg}},
\label{eq:tsq}
\ee
where the final approximation holds in the limit where $N_i^{\rm sig} \ll N_i^{\rm bkg}$, an excellent approximation for our limit setting procedure.
The 95\% limit occurs at $q \simeq -2.71$ and therefore using Eq.~\eqref{eq:tsq} we can determine the expected limit for the signal strength, as parameterized by the size of the $J$-factor.
This procedure is repeated for each real WIMP representation and we discuss the results in the following section.
The impact of a systematic uncertainty on the background rate is considered in Sec.~\ref{sec:systematics}.
Further, by default we fix the background amplitude in our computation of the sensitivity, but as we show in App.~\ref{app:Syst} floating the amplitude of the background does not alter our qualitative conclusions.

\subsection{Results}
\label{ssec:results}

\begin{figure*}[!t]
\centering
\includegraphics[width=0.5\textwidth]{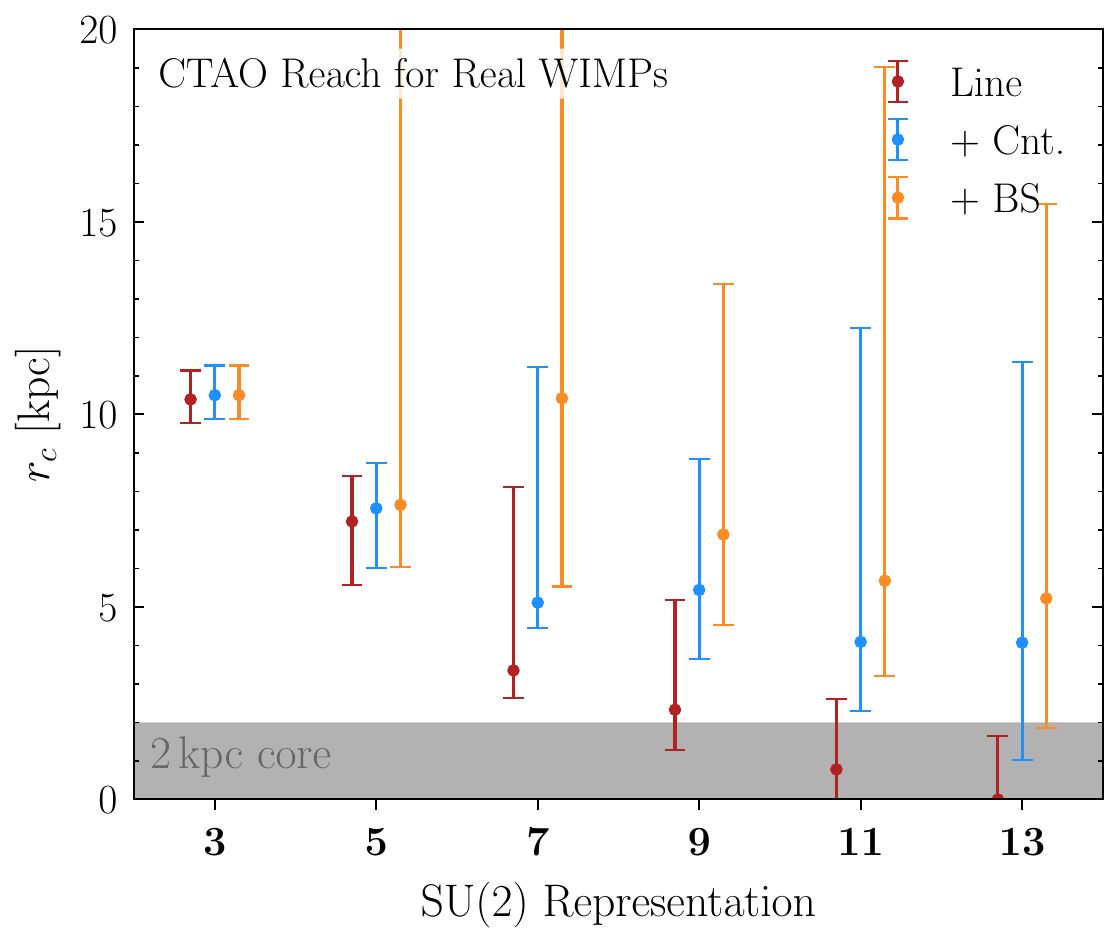}
\vspace{-0.2cm}
\caption{The projected reach of CTAO to real WIMPs assuming 500\,hrs of inner galaxy observations.
The sensitivity is quantified with how much the inner galaxy $J$-factor would need to be reduced for the scenario to avoid exclusion, which we quantify with an increasing core radius $r_c$.
Results are shown for the contribution of the line and endpoint spectrum (red, Line), together with the sensitivity gained when adding the continuum (blue, Cnt.) and bound states (orange, BS).
For a given representation the spectrum and cross section are fixed up to the uncertainty on the thermal mass.
The points correspond to the central mass prediction, whereas the error bars correspond to the strongest and weakest constraints obtained across the entire thermal mass window.
As discussed in Sec.~\ref{ssec:Rho}, we adopt a 2\,kpc core to set a lower limit on the amount of DM in the inner galaxy (see also Fig.~\ref{fig:DM-Comparison}): core sizes above this value are deemed excluded.
The results show that CTAO is poised to test all real WIMPs with the exception of possibly the tredecuplet, although as discussed in the text there are paths to probe this scenario also.}
\vspace{-0.2cm}
\label{fig:CoreReach}
\end{figure*}

\begin{figure}[t!]
\centering
\includegraphics[width=0.475\linewidth]{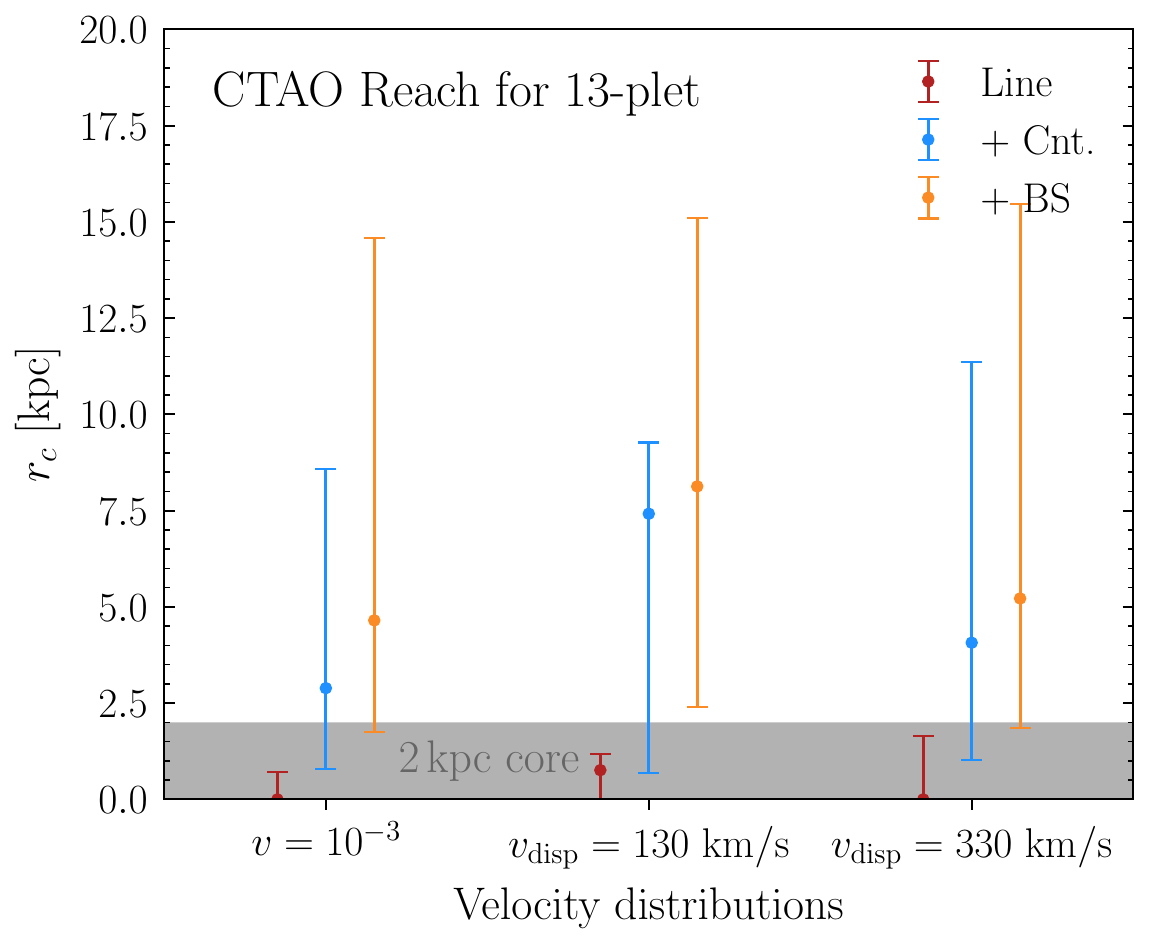}~\includegraphics[width=0.475\linewidth]{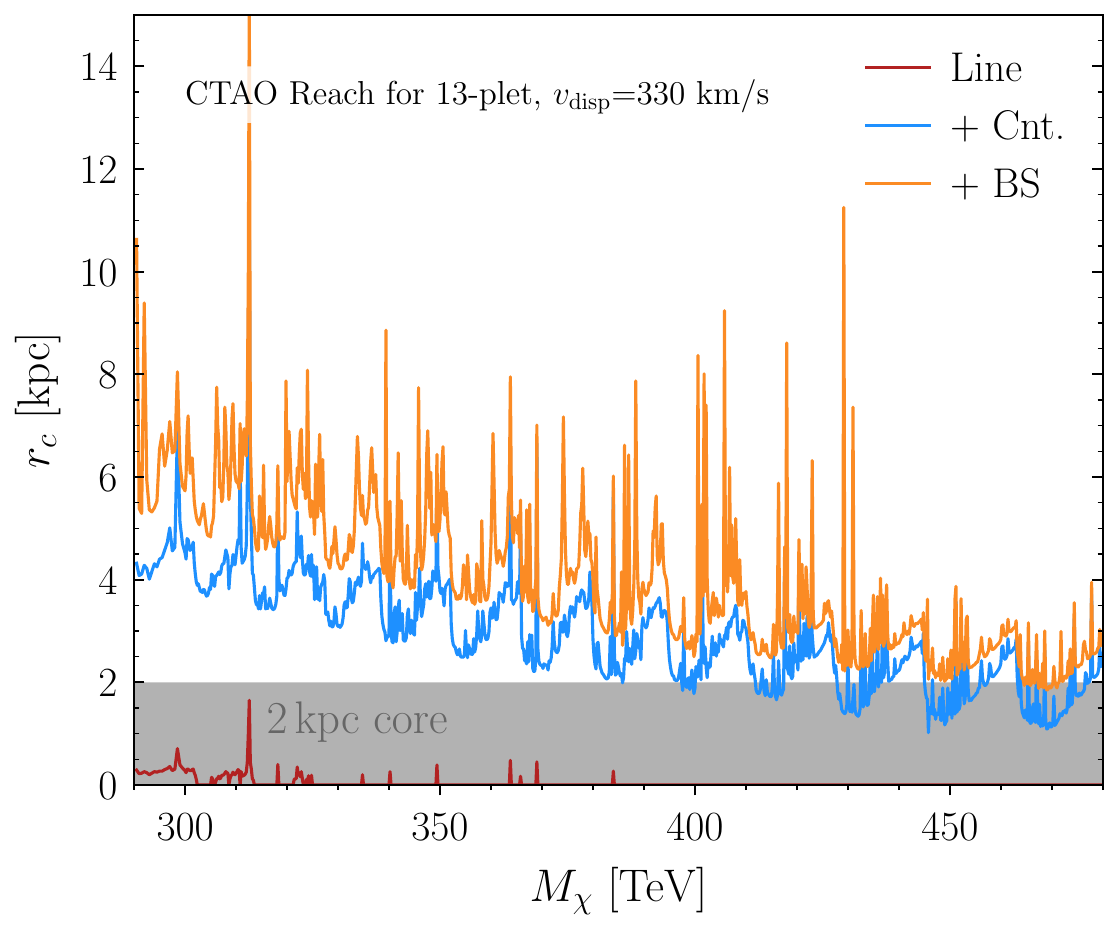}
\vspace{-0.2cm}
\caption{The projected reach of CTAO to the tredecuplet.
(Left) Comparison of the CTAO sensitivity for three different choices of velocity distributions.
We show, in particular, the case of fixed $v=10^{-3}$ as opposed to assuming the distribution in Eq.~\eqref{eq:v_distr} with $v_{\rm disp}=130$\,km/s and $v_{\rm disp}=330$\,km/s.
The sensitivity with the inclusion of BSF is quantitatively very similar and, as we explain in the main text, we adopt the case with $v_{\rm disp}=330$\,km/s as a conservative reach.
(Right) CTAO sensitivity to the tredecuplet as a function of the mass and assuming the most conservative velocity distribution.
With the inclusion of the BS, only few masses fall below the 2\,kpc threshold.
Such masses could likely be straightforwardly tested if CTAO were to extend their energy range to 500\,TeV.}
\vspace{-0.2cm}
\label{fig:13plet_vdistr}
\end{figure}

In Fig.~\ref{fig:CoreReach} we show the projected CTAO sensitivity to real electroweak WIMPs based on the analysis outlined so far.
We vary the core size for the Einasto profile and determine the core size at which the expected $95\%$ confidence limit on the cross section would equal the theoretically predicted value.
In particular, we compare the sensitivities obtained by including progressively different components of the spectrum, in order, the line plus endpoint contribution, the continuum from direct annihilations into pairs of vectors ($WW$, $ZZ$ and $Z\gamma$), and the continuum from bound state annihilations.
The points in the figure represent the central prediction for the thermal mass, whereas the upper and lower edges of the uncertainty bands correspond from the strongest and weakest constraints obtained across the mass range.
(These do not necessarily occur at the largest or smallest masses in the window.)
While the line analysis is sufficient to probe all the multiplets up to the septuplet, the nonuplet needs the addition of the continuum from direct annihilations and the undecuplet the further emission resulting from the bound state continuum.
The tredecuplet would marginally escape exclusion even with all contributions and deserves additional discussion.
While for the typical velocity distribution of the Milky Way assumed in Eq.~\eqref{eq:v_distr}, the kinetic energy of the initial $\chi^0\chi^0$ system is essentially never enough to produce a pair of on-shell charged particles, this is not the case for the tredecuplet for which a $\mathcal{O}(20\%)$ fraction of the DM is able to excite on-shell charged states.
For this reason, we study the impact of the velocity distribution on the reach of the tredecuplet, while fixing $v=10^{-3}$ for the remainder of the multiplets.
On the left of Fig.~\ref{fig:13plet_vdistr}, we show the projected reach for the tredecuplet assuming three different choices for the velocity distribution.
In particular, we compare the results obtained by assuming the distribution in Eq.~\eqref{eq:v_distr} with $v_{\rm disp}=130$\,km/s and $v_{\rm disp}=330$\,km/s, with the result for fixed $v=10^{-3}$.

The reach for the complete spectrum that includes BS is quantitatively very similar for all three choices.
In particular, the lower bound on the core radius varies by about 0.6 kpc.
In Fig.~\ref{fig:CoreReach}, we used the result for $v_{\rm disp}=330$\,km/s that is expected to give the more conservative bound.
In fact, the BSF cross section as a function of velocity is peaked at $v_p\simeq \mW/M_\chi$, below which the centrifugal potential term dominates over the electroweak potentials, leading to a $v^{2L}$ suppression, where $L$ is the angular momentum of the initial scattering wave function.
For the range of mass of the tredecuplet, $v_p\in[1.7,2.8] \times 10^{-3}$, so that the closer the peak of the velocity distribution to $v_p$ the larger the BSF cross section will be.
This ultimately explains the result shown on the left of Fig.~\ref{fig:13plet_vdistr} and justifies our choice for $v_{\rm disp}=330$\,km/s as conservative.

Finally, we notice that, conversely to the other multiplets, the tredecuplet is the only one whose mass range cannot be fully probed by CTAO even once the bound states are accounted for. 
However, as we can see from the right of Fig.~\ref{fig:13plet_vdistr}, only a very small interval within the tredecuplet thermal mass range falls below the 2\,kpc threshold.
For such large masses, our approach of setting the CTAO sensitivity to zero above 230\,TeV cuts out a significant part of the tredecuplet annihilation spectrum.
The possibility of expanding the energy range to 500\,TeV, as currently under discussion, would likely open the whole tredecuplet mass range to be testable with CTAO.

\subsection{Comparison to an unbinned analysis}
\label{ssec:unbinned}

Unbinned analyses have in principle the greatest statistical sensitivity.
They correspond to the limit of infinitesimally fine bins and do not suffer from binning artifacts.
In order to compare the impact of an unbinned analysis of our results, we perform one for three mass points in two representations, the triplet and tredecuplet, and compare these to the results of our binned analysis.
This serves as both an independent cross-check on the results of Fig.~\ref{fig:CoreReach} and a test of this approach's strength.

Our unbinned likelihood function uses a modified version of that adopted by VERITAS in Ref.~\cite{Acharyya:2023ptu}.\footnote{The modifications arise from the fact that the experimental analysis in Ref.~\cite{Acharyya:2023ptu} uses an ON/OFF procedure to differentiate background from possible signal data.
For the present work, we have a background model instead of real telescope data, and thus there is no equivalent of the OFF region data utilized by VERITAS.}
For the signal plus background hypothesis, it is
\begin{align}
\mathcal{L}(\mathbf{d} | \mu) &= \frac{(\mu\, N^{\rm sig}+N^{\rm bkg})^{N^{\rm obs}}e^{-(\mu\, N^{\rm sig} + N^{\rm bkg})}}{N^{\rm obs}!} \nn \\
&\quad \times \prod_{i=1}^{N^{\rm obs}} \frac{\mu\, N^{\rm sig} \, p_s(E_i) + N^{\rm bkg} \, p_b(E_i)}{\mu\, N^{\rm sig} + N^{\rm bkg}}.
\label{eq:unbin}
\end{align}
In the unbinned case, the dataset $\mathbf{d}$ consists of the total number of observed photons, $N^{\rm obs}$, and each of their reconstructed energies, $E_i$.
The expected number of signal and background events is $\mu N^{\rm sig}$ and $N^{\rm bkg}$, with $\mu$ a parameter introduced to control the relative signal strength.  We obtain them from Eq.~\ref{eq:pred-Ni} integrated over the whole energy range of the analysis. $N^{\rm sig}$ is determined from $d\Phi/dE_t$ given by Eq.~\ref{eq:IDflux} and the $J$-factor corresponding to a 2 kpc core.  Similarly, we get $N^{\rm bkg}$ from the background flux model described in Sec.~\ref{ssec:CTAOresponse}. 
Finally, $p_s$ and $p_b$ are the probability distributions in energy of the signal and background events.

Using the above likelihood, we construct a test-statistic to perform a one-sided 95\% confidence-level upper limit,
\begin{equation}
q_\text{unbin.}(\mu) = 2 \ln\!\left( \frac{\mathcal{L}(\mathbf{d}|\mu)}{\mathcal{L}(\mathbf{d}|\mu=0)} \right)\!.  
\end{equation}
The denominator of the $\ln$ argument is the background-only hypothesis likelihood.
In principle, one should compute the upper limit on $\mu$ relative to the value that maximizes the likelihood.
However, for our particular signal and background models, the likelihood-maximizing value of $\mu$ ($\simeq$$0.01$) is quite small.
We can thus test against the naive background-only hypothesis to a good approximation.
The upper limit on $\mu$ is given when the median value of $q_\text{unbin.} = -2.71$ when run over our full set of pseudoexperiments.
Since $\mu$ adjusts the number of expected signal events, we can thus convert its limit to one on $J$-factor or core size, $r_c$.  
As mentioned above, we adopt a convention that $\mu=1$ corresponds to $r_c = 2$ kpc.

Looking at Eq.~\eqref{eq:unbin} in detail, we see that the unbinned likelihood is built from a one-bin Poisson probability times the weighted probability of each event in both signal and background probability distribution functions (PDFs).
We obtain these by normalizing $dN/dE$ for WIMP annihilation and our background model, respectively, for energies between 92 GeV and 200 TeV.
%
%
Unlike the binned analysis, this computation requires a list of events.
Since we are testing the limits on exclusion, we run our pseudoexperiments by drawing from the background-only PDF.
The number of draws in each run is given by a Poisson fluctuation about the mean number of expected background events ($\simeq 3.2 \times 10^5$) for 10\,hrs of observation.
This provides the $N^{\rm obs}$ used in Eq.~\eqref{eq:unbin} for a given pseudoexperiment.  Our pseudo-dataset consists of 500 such runs.

\begin{table}[!t]
\centering
\begin{tabular}{c|c|c||c}
Representation (Mass) & $\mu$-limit & $r_c$-limit [kpc] & Binned $r_c$-limit [kpc] \\ \hline
$\mathbf{3}$ (2.86 TeV)       & 0.14 & 5.2 & 5.1 \\
$\mathbf{13}$ (324.6 TeV)       & 0.94 & 2.1 & 2.0 \\
$\mathbf{13}$ (469.0 TeV)       & 7.7 & 0.5 & 0.5 \\
\end{tabular}
\caption{Limit on $\mu$ (see Eq.~\eqref{eq:unbin}) and core size for the unbinned likelihood method, compared to the binned approach used in Sec.~\ref{ssec:results} for 10 hours of observation.
A value of $\mu=1$ corresponds to the $J$-factor for a 2 kpc core.}
\label{tab:unbin}
\end{table}

We give the results of the unbinned analysis in Tab.~\ref{tab:unbin}, along with a comparison to the core-size limits using the binned analysis of Section \ref{ssec:results}. 
We tested the central thermal relic mass values for both the 3- and tredecuplet, along with 469 TeV for the latter.
This is because this mass is the value within the error bar for the tredecuplet's thermal relic range that minimizes the core-size limit.
We find excellent agreement for all three points that we tested, which serves as an independent cross-check on the method of Sec.~\ref{ssec:results}.

\subsection{The impact of background systematics}
\label{sec:systematics}

\begin{figure*}[!t]
\centering
\includegraphics[width=0.45\textwidth]{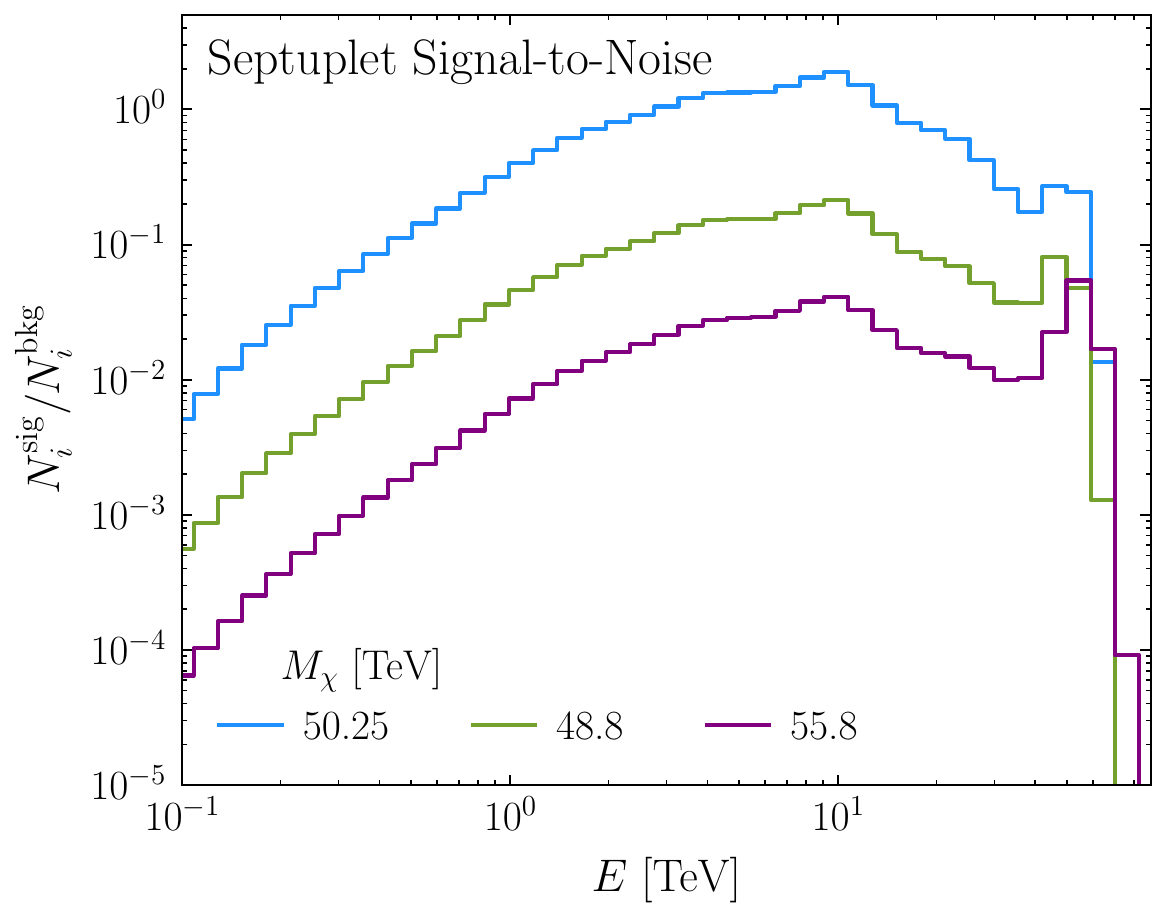}
\hspace{0.5cm}
\includegraphics[width=0.45\textwidth]{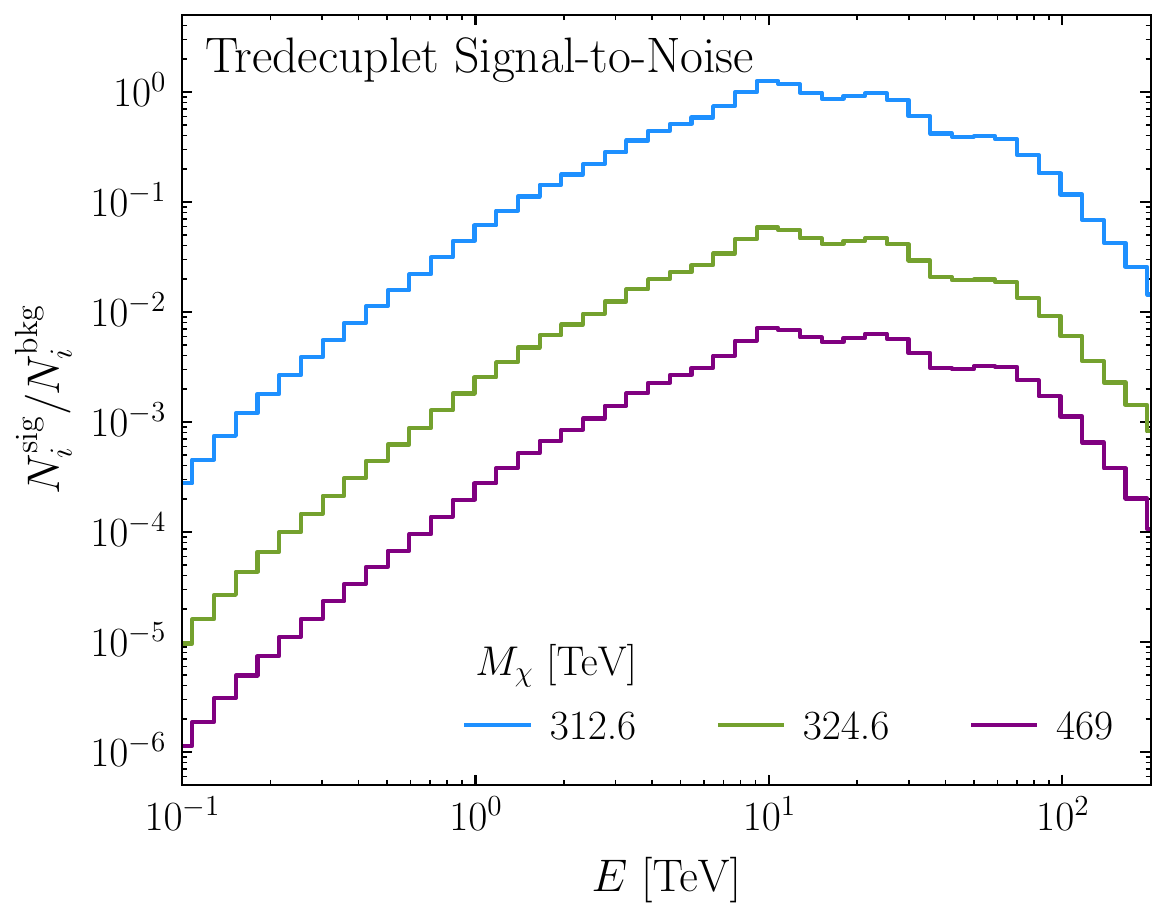}
\vspace{-0.2cm}
\caption{Ratio of signal over background for the septuplet and for the tredecuplet for different mass choices.
The green line is the central value of of the thermal mass prediction while the blue line and the purple one correspond to the mass which maximize and minimize the expected reach respectively.
In all cases, the core radius is fixed to $r_c=2$\,kpc.}
\vspace{-0.2cm}
\label{fig:SpectraSoverB}
\end{figure*}

In this section, we assess the potential impact of background systematics on the projected sensitivity presented in Fig.~\ref{fig:CoreReach}.
A comprehensive evaluation of systematic uncertainties associated with background modeling lies beyond the scope of this work and is ultimately the responsibility of the CTAO collaboration.
Our aim here is to provide a quantitative estimate of their potential effect using a simplified parametrization where we conservatively allow the background shape to vary across energy bins.
If instead the background spectrum is known and only its amplitude is uncertain, then as we demonstrate in App.~\ref{app:Syst} our qualitative results presented in the last section are unchanged.

To this end, we compare the binned test statistic defined in Eq.~\eqref{eq:tsq} with the following modified statistic:
\be
q(\epsilon_{\rm{sys}}) = -\sum_i \frac{(N_i^{\rm sig})^2}{N_i^{\rm bkg} + \epsilon_{\rm{sys}}^2(N_i^{\rm bkg})^2},
\label{eq:testsys}
\ee
where $\epsilon_{\rm{sys}}$ parameterizes the relative size of systematic uncertainties in the background.
To understand the modification, note that when $\epsilon_{\rm{sys}} < (N_i^{\rm bkg})^{-1/2}$, statistical fluctuations dominate the background uncertainty, $q(\epsilon_{\rm{sys}}) \simeq q(\epsilon_{\rm{sys}}=0)$, and the results in Fig.~\ref{fig:CoreReach} remain unaffected.
In the opposite regime, systematics dominate, with the sensitivity now determined by $\sum_i (N_i^{\rm sig})^2/(\epsilon_{\rm{sys}} N_i^{\rm bkg})^2$ so that there is no improvement with observation time.

Consequently, for the bins contributing significantly to the test statistic, the condition $\epsilon_{\rm{sys}} \lesssim N_i^{\rm sig} / N_i^{\rm bkg}$ must be satisfied to retain sensitivity.
Figure~\ref{fig:SpectraSoverB} illustrates this ratio, providing a sense of how small the systematics must be to achieve the desired performance.
As an example, we show the signal spectra for the septuplet and tredecuplet models—previously shown in Fig.~\ref{fig:Spectra}—assuming a core radius fixed to the conservative value of $r_c = 2\,$kpc (see Sec.~\ref{ssec:Rho}) and displaying three representative mass values within the allowed thermal band.

\begin{figure*}[!t]
\centering
\includegraphics[width=0.45\textwidth]{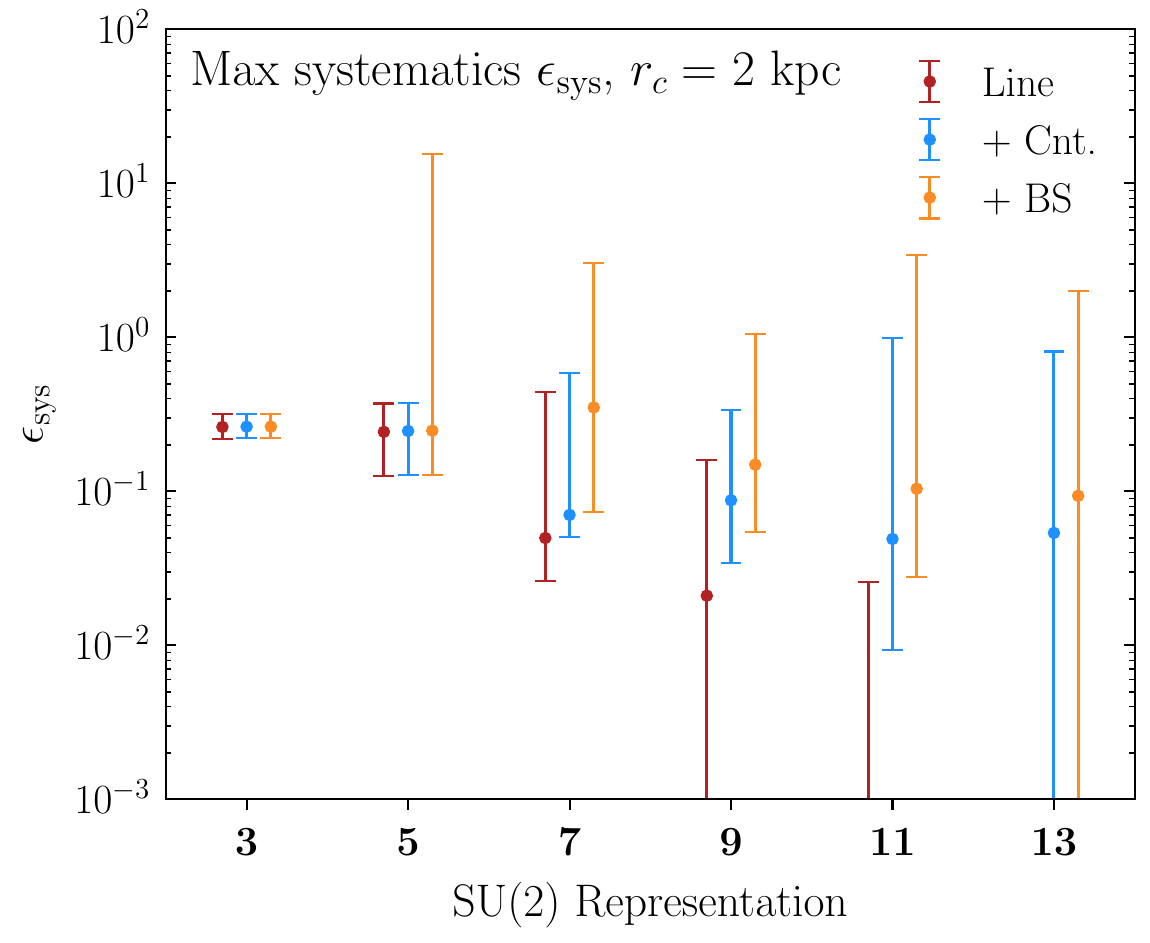}
\vspace{-0.2cm}
\caption{Summary of the impact of systematic uncertainties, as defined in Eq.~\eqref{eq:testsys}, on the expected reach.
We fix \( r_c = 2\,\mathrm{kpc} \) and show the maximal systematic uncertainty on the background allowed in order to retain sensitivity to a given electroweak multiplet.
As before, we distinguish the signal contributions between the line (red), line plus continuum (blue), and the further inclusion of bound state effects (orange).}
\vspace{-0.2cm}
\label{fig:syssummary}
\end{figure*}

In Fig.~\ref{fig:syssummary} we summarize the range of systematic uncertainties required to probe each MDM scenario with the same signal decomposition as in Fig.~\ref{fig:CoreReach}.
For all results we have again fixed $r_c = 2\,$kpc.
With systematic uncertainties around 10\%, only the triplet and the quintuplet are fully within the projected sensitivity reach of CTAO.
For these multiplets, the reach of the pure line search is comparable to that of the continuum-based search across most of the mass range.
An exception occurs in the high-mass region for the quintuplet, within the theoretically allowed thermal mass window, where the line cross section drops significantly---an effect that is further pronounced by the inclusion of NLL corrections (see the second panel of Fig.~\ref{fig:linexsec}).

\begin{figure*}[!t]
\centering
\includegraphics[width=0.45\textwidth]{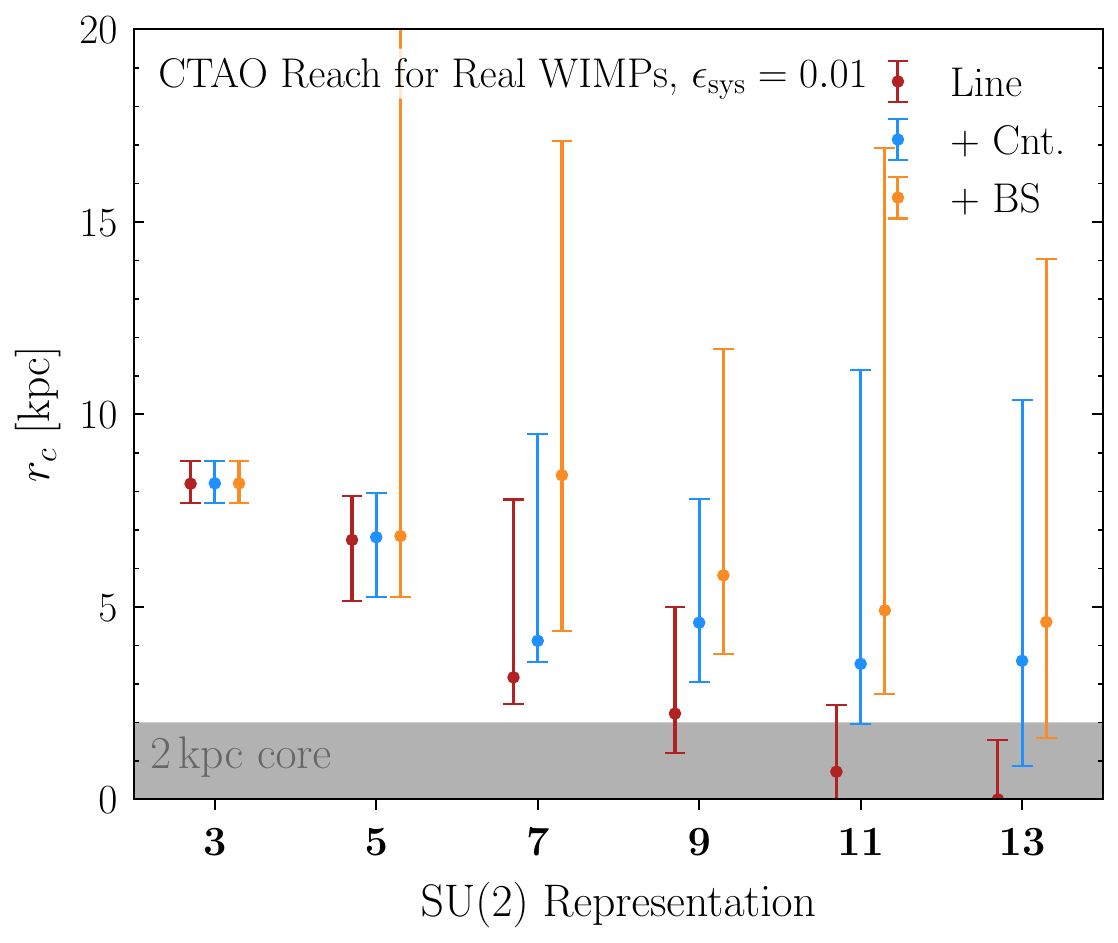}\hspace{0.5cm}
\includegraphics[width=0.45\textwidth]{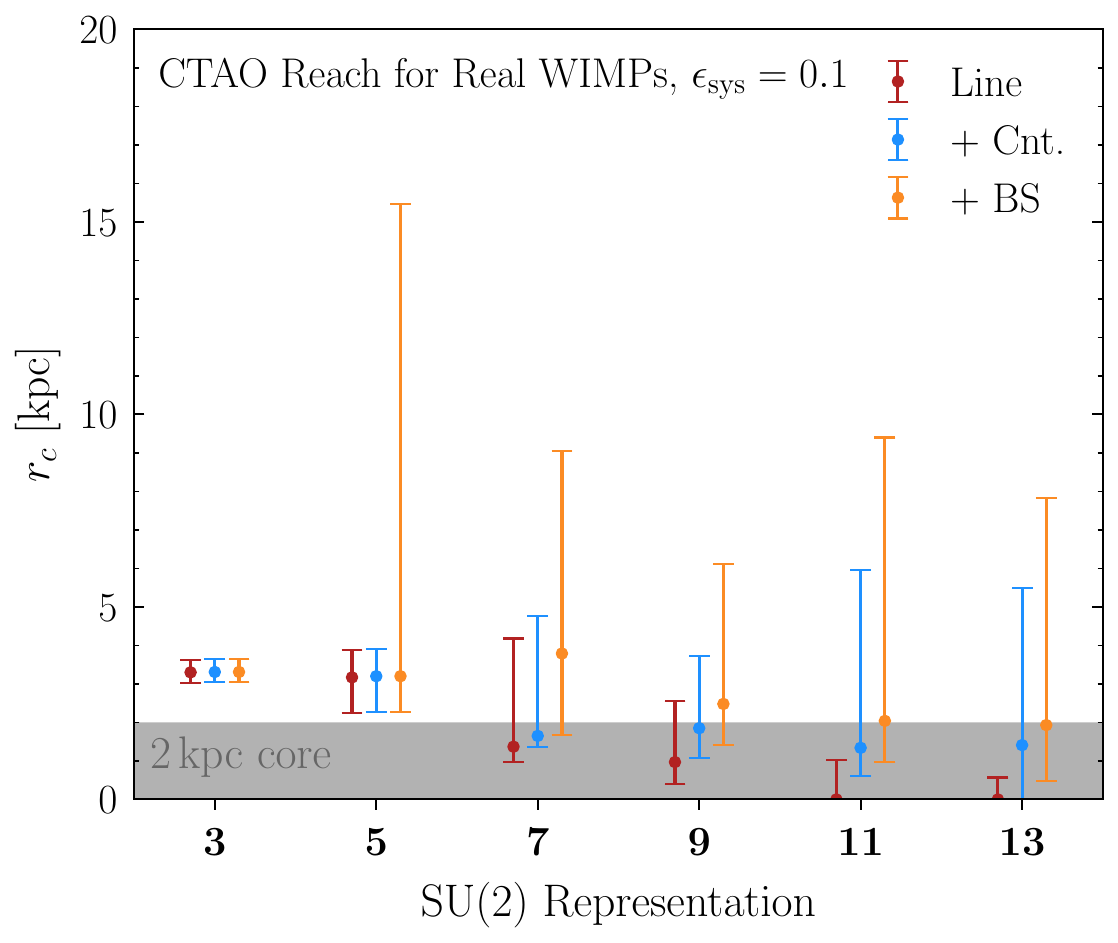}
\vspace{-0.2cm}
\caption{Projected sensitivity for the different multiplets as in Fig.~\ref{fig:CoreReach} but assuming a fixed systematic uncertainty of $\epsilon_{\rm sys} = 1\%$ (right) and $\epsilon_{\rm sys} = 10\%$ (left). }
\vspace{-0.2cm}
\label{fig:sys1vs10}
\end{figure*}

For heavier multiplets, the line signal loses its discriminating power, and sensitivity becomes dominated by the continuum component.
Probing these higher-dimensional candidates across their full mass range at CTAO would require systematic uncertainties on the background to be tamed at or below 1\%.
This is demonstrated on the left of Fig.~\ref{fig:sys1vs10}, where we show the expected reach of CTAO in terms of the minimal allowed core radius fixing $\epsilon_{\mathrm{sys}}=1\%$.
For comparison, we show on the right of Fig.~\ref{fig:sys1vs10} how the expected reach would change if instead the background was affected by systematic uncertainties of 10\%.
Although undoubtedly challenging, HESS has achieved ${\cal O}(1\%)$ systematic uncertainties~\cite{HESS:2022ygk}.
We note, however, that HESS achieved this sensitivity using an ON/OFF analysis, which is not ideal for large DM core scenarios that lie at the edge of our sensitivity.
Indeed, our ROI probes radii far smaller than our threshold 2\,kpc core.
Consequently, within this ROI, morphological information cannot effectively be used to distinguish the background from the signal. 
Therefore, the background rate must be constrained otherwise, either using a dedicated data-taking period in regions sufficiently far from the Galactic Center that the DM signal is suppressed, or else determined through Monte Carlo simulations.
As discussed in App.~\ref{app:Syst}, if the spectrum of the background can be well predicted and only the normalization is unknown, that would be sufficient.
If the cosmic-ray background is well-constrained, uncertainties in the diffuse astrophysical background may also become a consideration, especially at lower energies where it is less suppressed relative to the cosmic-ray contribution (see Fig.~\ref{fig:background}).
The astrophysical background has spatial structure and so cannot be calibrated by observations away from the Galactic Center, but by the same token it could potentially be separated from a DM signal by an analysis including spatial morphology.
For a study of the impact of these effects for the $\simeq 1\,{\rm TeV}$ higgsino see Ref.~\cite{Rodd:2024qsi}.

\section{Conclusions}
\label{sec:conclusions}

For decades, indirect detection has been a central pillar in the search for DM. The Cherenkov Telescope Array Observatory (CTAO) will mark the next major advance in this effort, and the results of this work show that the significant boost in sensitivity anticipated from CTAO can translate into concrete progress across critical milestones of the WIMP parameter space.

For MDM, this progress becomes even more compelling when combined with insights from direct detection searches~\cite{Bloch:2024suj}.
While indirect detection typically has the strongest sensitivity to the lightest electroweak multiplets, as shown in Fig.~\ref{fig:CoreReach}, direct detection experiments are generally more sensitive to larger representations.
This distinction arises because the scattering rate $\Gamma_{\scriptscriptstyle{\rm DD}}$ of WIMPs off nuclei scales linearly with the DM number density, whereas the annihilation rate relevant for indirect detection scales quadratically.
This can be quantified as follows.
Thermal freeze-out achieves the correct relic abundance for $M_\chi \sim (2n+1)^{5/2}$.\footnote{This scaling derives from the fact that freeze-out predicts a fixed value for the thermally averaged annihilation cross-section, $\Omega_{\rm DM}\propto1/\langle\sigma v\rangle$.
The cross-section $\langle\sigma v\rangle$ can be schematically written as $\langle\sigma v\rangle\sim \langle S_E \sigma_{\rm hard} v\rangle$, where $S_E$ is the Sommerfeld enhancement factor which can be estimated as $\alpha_{\rm eff}/v\sim \aW(2n+1)^2/v$, while the scaling of the hard cross-section $\sigma_{\rm hard}$ depends on the isospin channel.
For the real electroweak WIMP, the dominant channels are the singlet and the quintuplet, for which $\sigma_{\rm hard}\sim \aW^2(2n+1)^5/(g_\chi^2M_\chi^2)$, where $g_\chi\propto 2n+1$ is the number of degrees of freedom of the multiplet.
Combining these results, we find $\langle\sigma v\rangle\sim \aW^3(2n+1)^5/M_\chi^2$, from which we arrive at the claimed scaling for the thermal mass predicted by freeze-out, $M_\chi\sim (2n+1)^{5/2}.$}
The indirect detection rate scales as $\langle \sigma v \rangle/M_{\chi}^2 \sim [(2n+1)^5 / M_\chi^2][1 / M_\chi^2] \sim (2n+1)^{-5}$, where the first factor arises from the annihilation cross section and the second from the number density, while the direct detection rate scales as $\Gamma_{\scriptscriptstyle{\rm DD}} \sim \sigma_{\chi {\scriptscriptstyle {\rm N}}}/M_{\chi} \sim (2n+1)^{3/2}$, with $\sigma_{\chi {\scriptscriptstyle {\rm N}}} \sim (2n+1)^4$ the one-loop spin-independent WIMP-nucleon cross section.

Although these parametric scalings suggest that direct detection could be ideal for heavier multiplets, Fig.~\ref{fig:CoreReach} demonstrates that CTAO has the potential to probe the full range of real electroweak multiplets.
Yet for the largest representations, that conclusion is contingent on whether the background systematics can be controlled at the percent level.
Furthermore, should a signal begin to emerge for a model like the tredecuplet, a definitive discovery at CTAO would be challenging.
In either case, a definitive statement may require direct detection, although of course it faces its own challenges such as suppressing radioactive backgrounds like radon decay in liquid xenon~\cite{Kravitz:2022mby,Chen:2023llu}.
In summary, with complementarity between direct and indirect detection strategies, the long-standing question of whether DM resides in a real representation of SU(2) may ultimately find a definitive answer.

Looking more broadly, one can ask what other DM milestones CTAO is poised to cover.
The most obvious of these would be to test the MDM paradigm as a whole.
A conceptually simple extension of the present work would be to consider the complex MDM representations, i.e. the generalizations of the higgsino as in the present work we have considered the generalizations of the wino.
Although challenging, the fact that CTAO is forecast to reach the nominal higgsino prediction as shown in Refs.~\cite{Rinchiuso:2020skh,Rodd:2024qsi} (and potentially in the near term using CTAO-North~\cite{Abe:2025lci}) suggests the telescope is likely well placed to test all these models, although this remains an interesting question to resolve in detail.

Beyond CTAO and LACT there are no definitive plans for a next generation gamma-ray telescope.
As such, the onus is on the community to determine the full set of models these instruments can reach and the optimal way to search for DM with these telescope.  We devoted this study to CTAO due to its greater maturity and the easier accessibility of relevant information for analysis. 
An important facet of the optimal strategy will be further improvements in our understanding of the DM distribution in the inner galaxy.
In our fiducial analysis we adopted a simplistic approach to this problem by simply coring the canonical Einasto profile and adopting a 2\,kpc core as our threshold for exclusion.
As noted, this crude approach leads to a comparable $J$-factor as one finds at the lower end of the FIRE-2 simulations, or even from more recent refined analyses as in Ref.~\cite{Hussein:2025xwm}.
Nevertheless, numerical galaxy simulations continue to improve and may evolve substantially on the timescale that it takes CTAO-South to collect the 500\,hrs dataset we envisioned for our analyses.
It will be important to interface these developments with the CTAO search strategy and more broadly to work to maximize the chance that CTAO could finally discover DM.

\begin{acknowledgments}

Our work benefited from discussions with Alessandro Dondarini, Roberto Franceschini, Daniele Gaggero, Abdelaziz Hussein, Tomohiro Inada, Lina Necib, Paolo Panci, Michele Redi, Ben Safdi, and Linda Xu.
MB is supported by the DOE (HEP) Award DE-SC0019470.
SB is supported by the Israel Academy of Sciences and Humanities \& Council for Higher Education Excellence Fellowship Program for International Postdoctoral Researchers.
The work of DR is supported in part by the European Union - Next Generation EU through the PRIN2022 Grant n. 202289JEW4.
The work of NLR was supported by the Office of High Energy Physics of the U.S. Department of Energy under contract DE-AC02-05CH11231.
TRS' work is supported by the Simons Foundation (Grant Number 929255, T.R.S) and by the U.S. Department of Energy, Office of Science, Office of High Energy Physics of U.S. Department of Energy under grant Contract Number DE-SC0012567.
During the course of this work, T.R.S.~was supported in part by a Guggenheim Fellowship; the Edward, Frances, and Shirley B.~Daniels Fellowship of the Harvard Radcliffe Institute; and the Bershadsky Distinguished Fellowship of the Harvard Physics Department. 
The work of MB is supported by the U.S. Department of Energy, Office of Science, Office of High Energy Physics of U.S. Department of Energy under grant Contract Number DE-SC019470.
Part of this work was performed at Aspen Center for Physics, which is supported by National Science Foundation grant PHY-2210452.

\end{acknowledgments}

\appendix
\addtocontents{toc}{\protect\setcounter{tocdepth}{1}}
\section*{Appendix}

\section{Impact of Background Model Uncertainty}
\label{app:Syst}

The statistical procedure we adopt in the main text (as detailed in Sec.~\ref{ssec:CTAOresponse}) assumes a perfect knowledge of the background spectrum.
Accordingly, the results derived from the default analysis could prove too optimistic. In Sec.~\ref{sec:systematics} we showed how the result are modified assuming a constant background systematic in a binned analysis. Here we show that knowledge of the spectral shape, but not amplitude, is sufficient for our conclusions to remain largely unchanged.

In detail, we now imagine that the data in each bin is now specified as a combination of signal and background with variable amplitude,
\be
N_i=\alpha  N^{\rm sig}_i+\beta_{\rm CR}N^{\rm bkg}_{{\rm CR},i}+\beta_{\rm GDE}N^{\rm bkg}_{{\rm GDE},i}.
\ee
Here $N^{\rm sig}_i$ is the expected number of signal events assuming $r_c=2$ kpc, so that $\alpha<1$ defines the exclusion region.
The parameters $\beta$ are introduced to to allow the amplitudes of each background component to float, such that our fiducial background model is obtained by fixing $\alpha=0$ and $\beta_{\rm CR}=\beta_{\rm GDE}=1$.
Then, if $d_i$ is the observed number of events per bin, we define as usual the Poisson likelihood $\mathcal{L}(\textbf{d}|\textbf{N}(\alpha,\beta_{\rm CR},\beta_{\rm GDE}))=\prod_i N_i^{d_i} e^{-N_i}/d_i!$, exactly as previously.
We again adopt the Asimov procedure, that is we assume that the $d_i$ values coincide with the expected events from the default background model, and define $P(\alpha,\beta_{\rm CR},\beta_{\rm GDE})=\mathcal{L}(\textbf{N}(0,1,1)|\textbf{N}(\alpha,\beta_{\rm CR},\beta_{\rm GDE}))$.
The test statistic in Eq.~\eqref{eq:tsq} is then generalized to
\be
q(\alpha)= 2 \left[ \ln P(0,1,1) - \ln P(\alpha,\hat{\beta}_{\rm CR},\hat{\beta}_{\rm GDE}) \right]\!,
\ee
with $\hat{\beta}_{\rm CR}$ and $\hat{\beta}_{\rm GDE}$ determined by maximizing $P(\alpha,\beta_{\rm CR},\beta_{\rm GDE})$ with respect to $\beta$ for a fixed $\alpha$; in other words, we are using the profile likelihood technique.

\begin{figure*}[!t]
\centering
\includegraphics[width=0.475\textwidth]{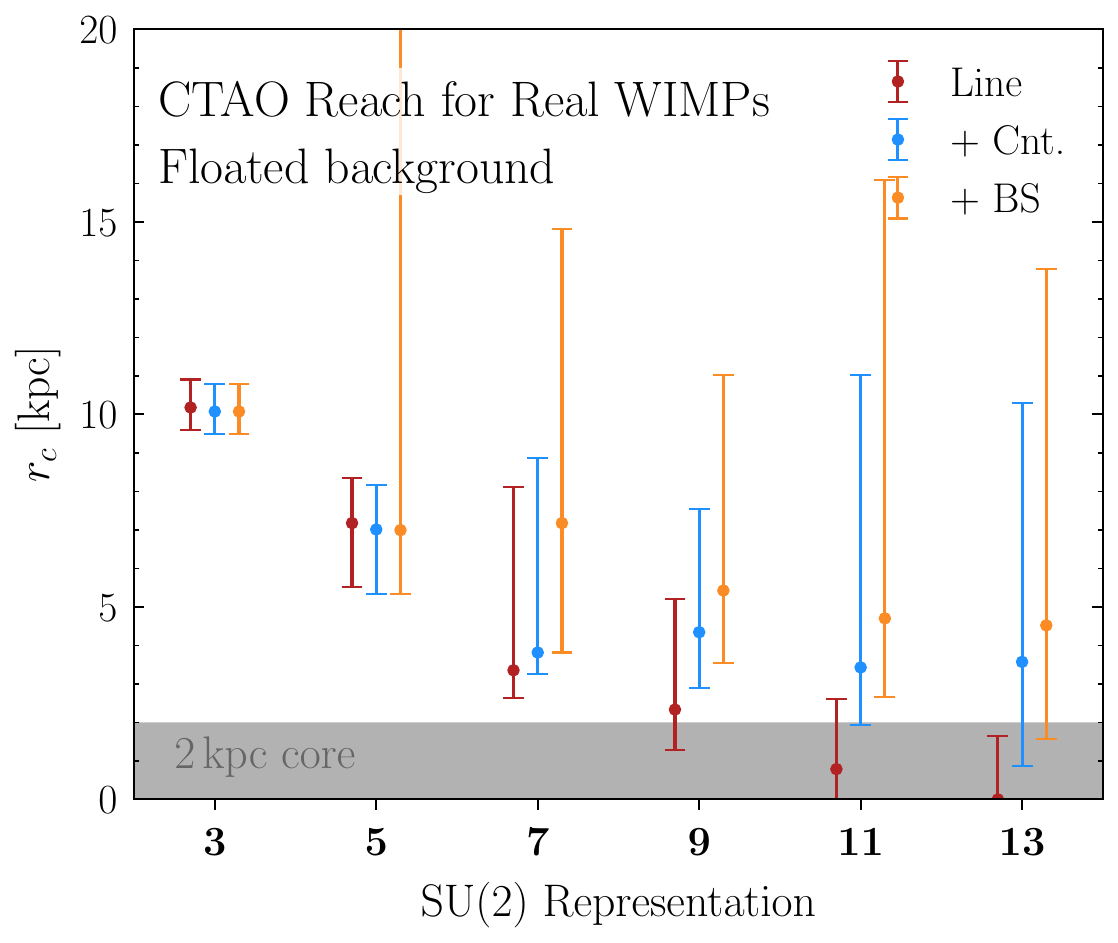}\hspace{0.5cm}
\includegraphics[width=0.475\linewidth]{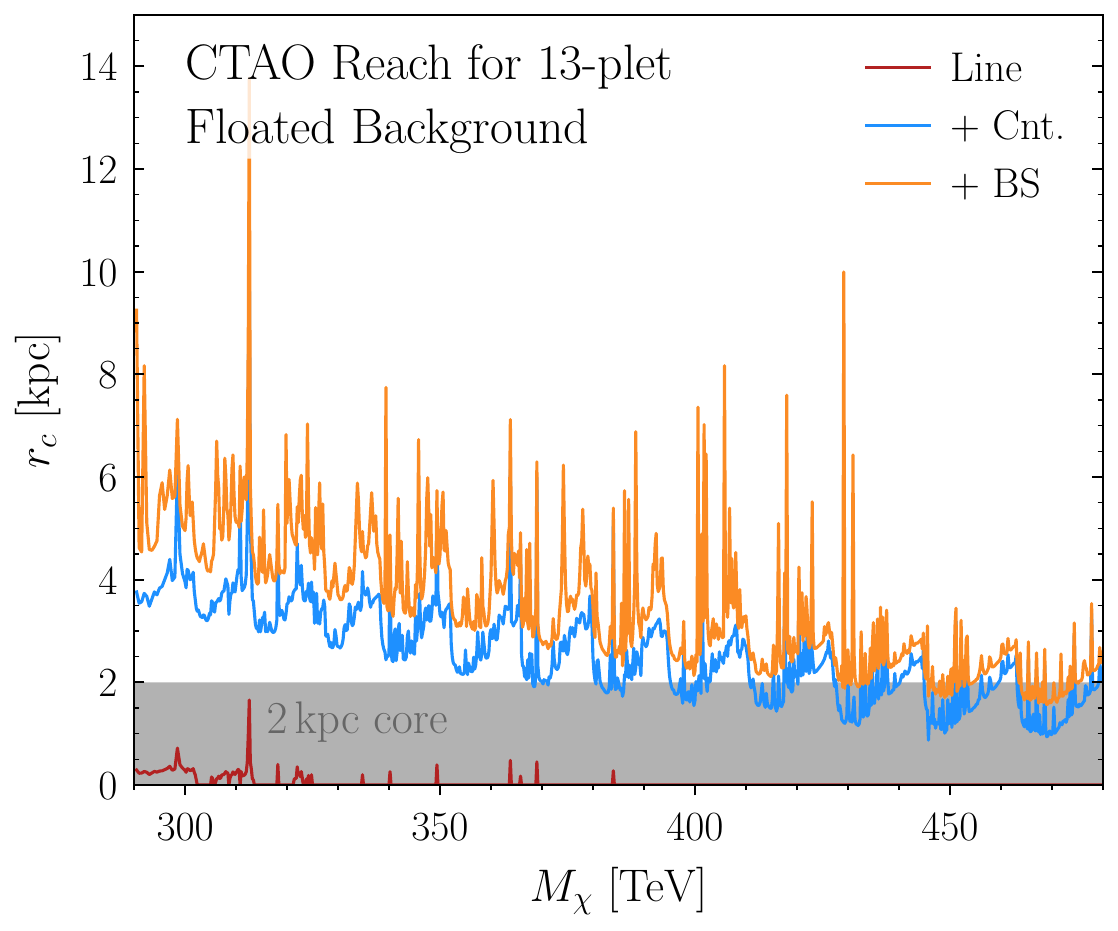}
\vspace{-0.2cm}
\caption{The impact of floating the background normalizations on the projected sensitivity to real WIMPs with CTAO.
(Left) We show the analog of Fig.~\ref{fig:CoreReach}.
As there, all representations except for the tredecuplet can be robustly tested.
(Right) To show the impact on the $\mathbf{13}$ specifically, we show the analog of Fig.~\ref{fig:13plet_vdistr} right.
A slightly larger range of masses cannot be excluded, but again our conclusions remain largely unchanged.}
\vspace{-0.2cm}
\label{fig:CoreReachML}
\end{figure*}

In Fig.~\ref{fig:CoreReachML} we show how this modified background treatment impacts our projected WIMP sensitivity.
The results for all representations are almost identical to those shown in Fig.~\ref{fig:CoreReach}, while quantitatively the reach is only slightly worse.
All representations outside the tredecuplet remain within reach of a definitive test, and in that case we also show the analog of the results in Fig.~\ref{fig:13plet_vdistr}, demonstrating that only a few more masses fall below the 2 kpc limit compared to Fig. \ref{fig:13plet_vdistr} right.
Broadly, the line signal is barely affected while a more consistent reduction in the signal constraining power is observed when the continuum is added.

\begin{figure*}[!t]
\centering
\includegraphics[width=0.475\textwidth]{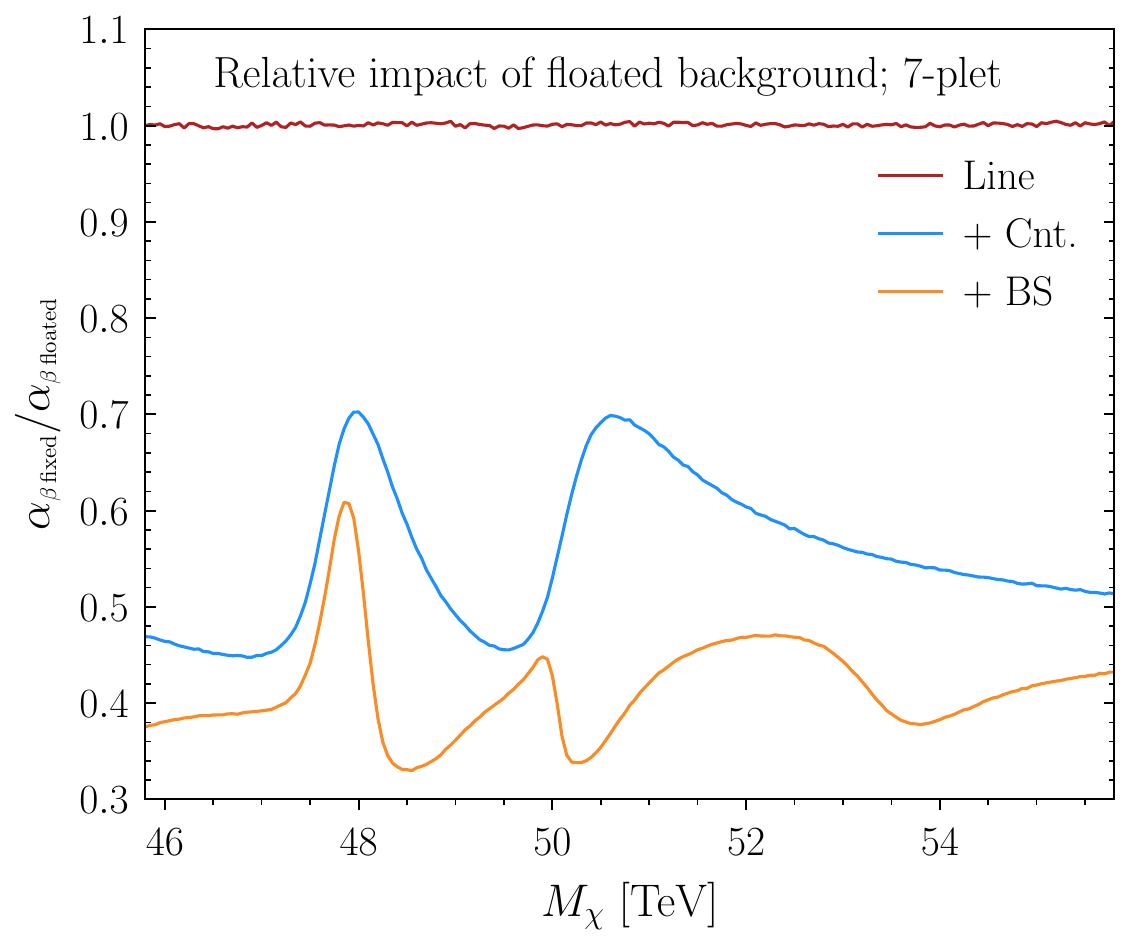}
\hspace{0.5cm}
\includegraphics[width=0.475
\linewidth]{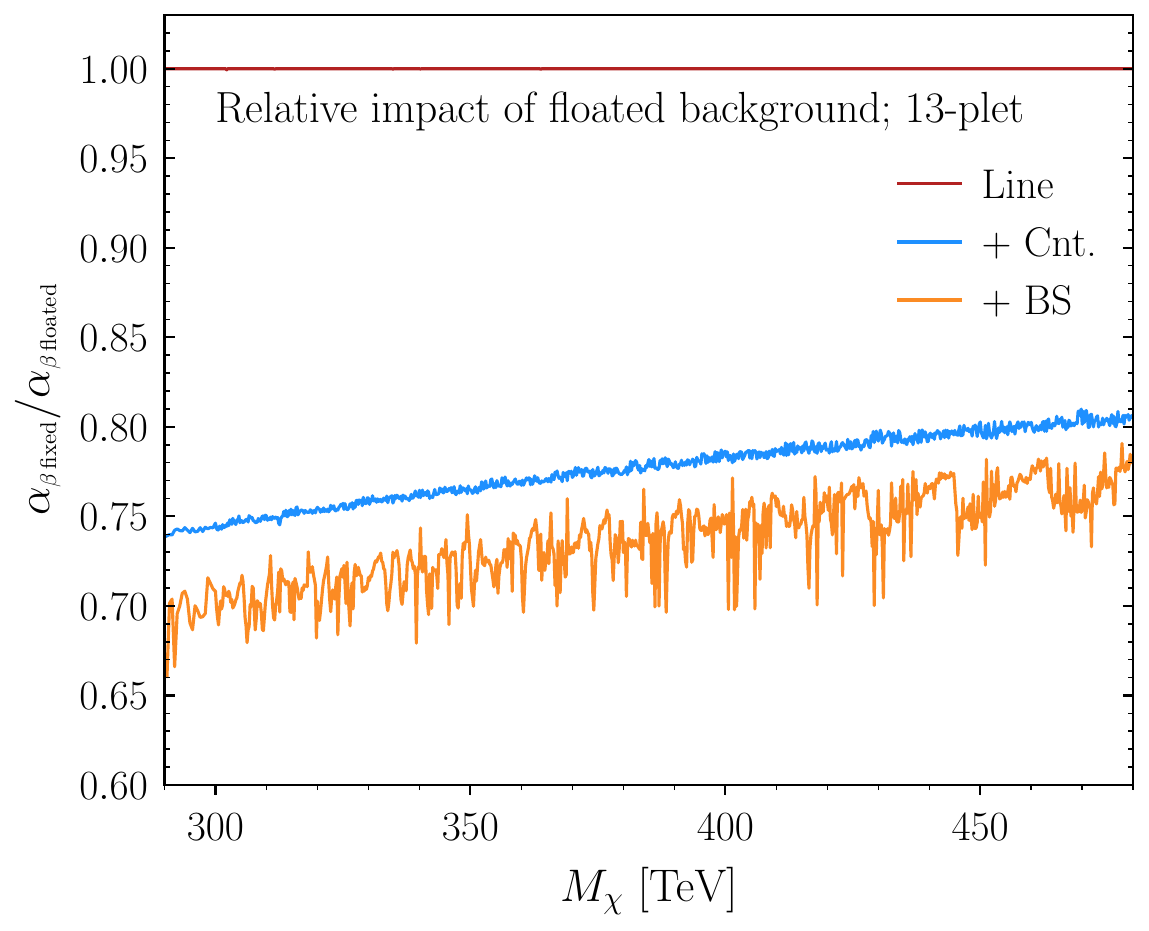}
\vspace{-0.2cm}
\caption{Ratio of the expected 95\% limit on the DM signal strength when the background is fixed compared to when its amplitude is floated. 
Results are shown for the septuplet (left) and tredecuplet (right).}
\vspace{-0.2cm}
\label{fig:RatioReach}
\end{figure*}

\begin{figure*}[!t]
\centering
\includegraphics[width=0.475\textwidth]{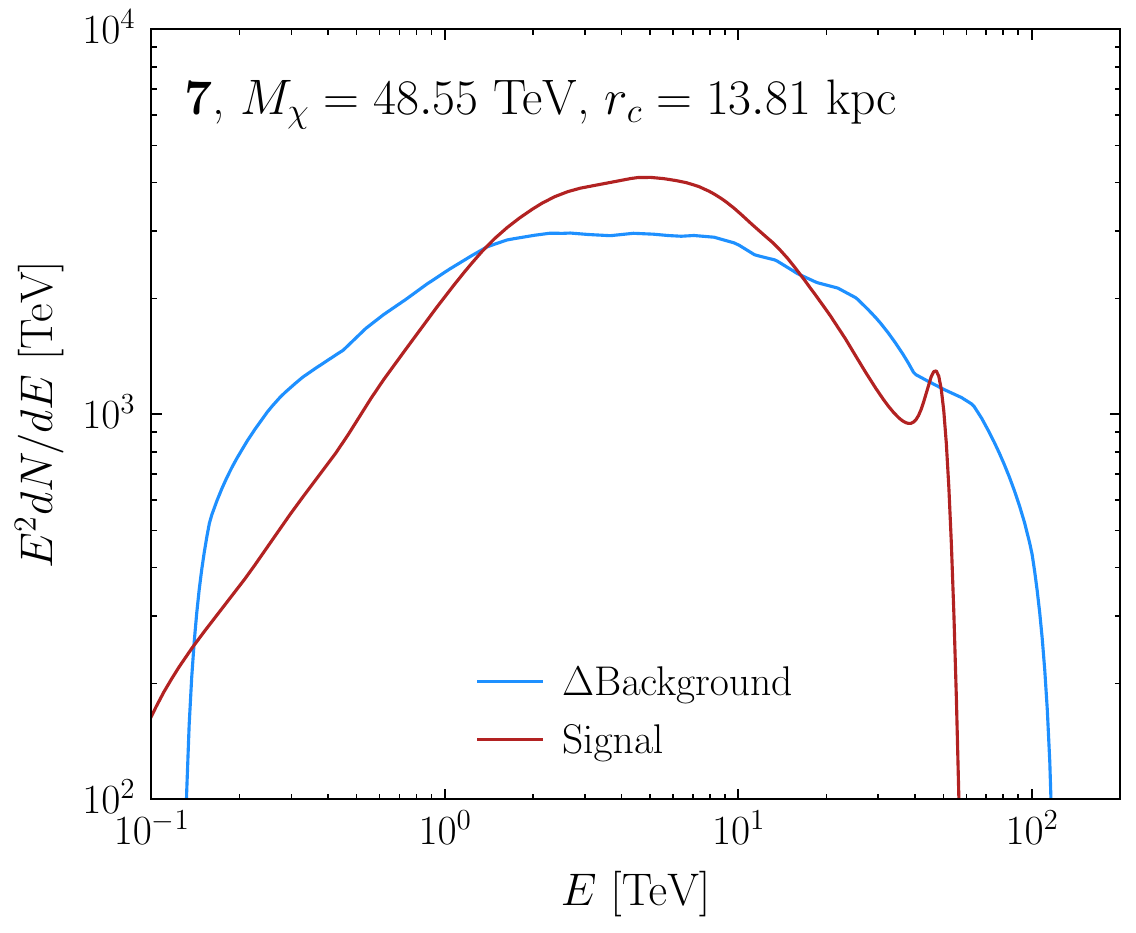}\hspace{0.5cm}
\includegraphics[width=0.475
\linewidth]{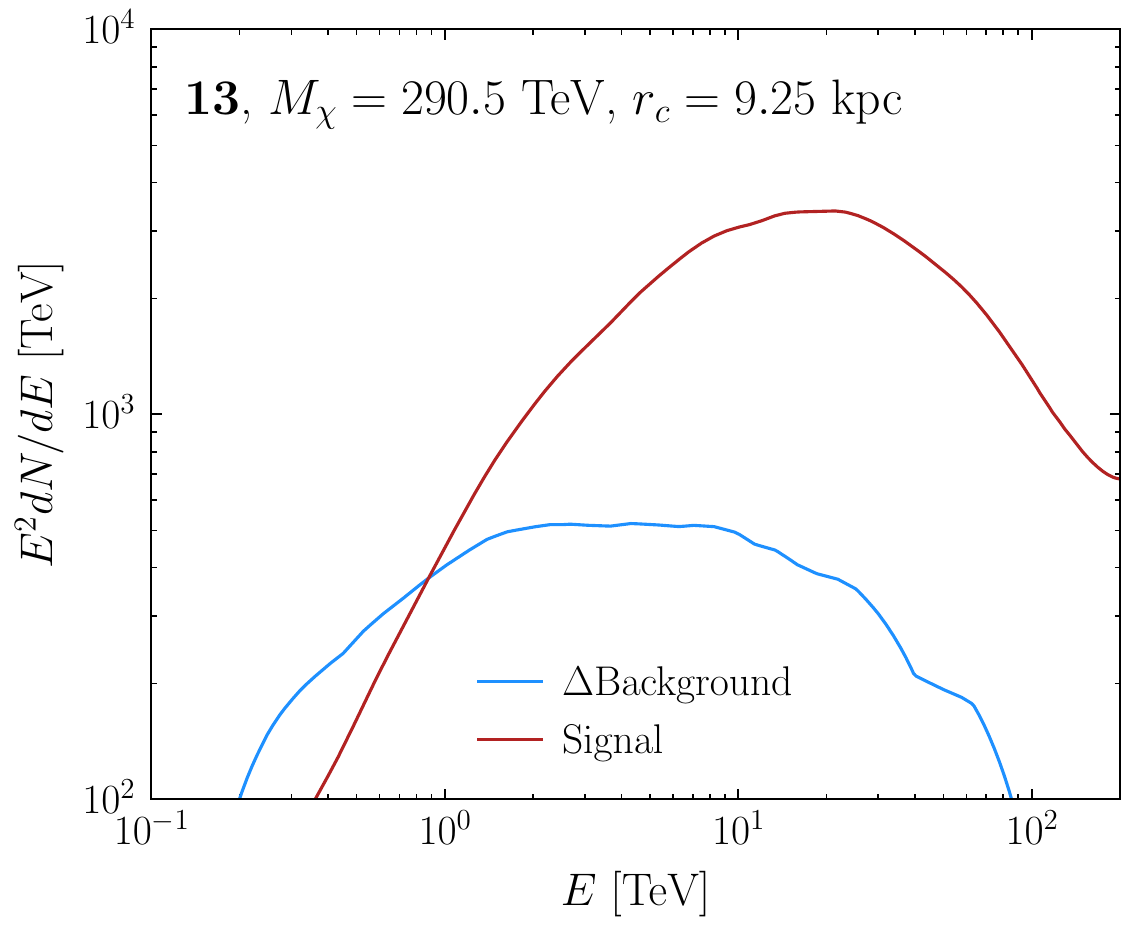}
\vspace{-0.2cm}
\caption{Spectrum of the signal (red) and that of the difference between background models, labeled $\Delta$Background (blue).
In detail, $\Delta$Background is the difference between the fiducial background $(\beta_1=\beta_2=1)$ and the one obtained by floating the background amplitudes.
Results are shown for the septuplet (left) and tredecuplet (right).
The larger degeneracy for the septuplet helps explain why this representation is more impacted than others by the floated background.}
\vspace{-0.2cm}
\label{fig:SigvsB}
\end{figure*}

One representation that is noticeably impacted by the change is the septuplet.
We demonstrate this in Fig.~\ref{fig:RatioReach} where we show the ratio of results with and without the floated background for the $\mathbf{7}$ and $\mathbf{13}$ representations.
The reason for this behavior lies in the fact that the GDE background is in general negligible compared to the CR one except for a window around 10\,TeV, where the suppression is only a factor of a few, as shown in Fig.~\ref{fig:background}.
This is also the range where the continuum spectrum of the septuplet typically peaks, so that it is easier to reproduce the signal excess simply by varying the amplitude of the GDE background than for other multiplets.
This is evident in Fig.~\ref{fig:SigvsB}, where we show the annihilation spectrum for a septuplet with mass $M_\chi=48.55$ TeV, corresponding to the mass with the largest reduction in constraining power, and the difference between the total background spectrum with the background fixed and floated.
For comparison, in Fig.~\ref{fig:SigvsB} right we show the corresponding plot in the case of the tredecuplet, where the degeneracy is far less pronounced.

\section{The Real MDM Spectrum at NLL}
\label{app:NLL}

In Sec.~\ref{ssec:DA} we provided the LL line cross section for an arbitrary real MDM representation.
As mentioned there, the NLL cross section is qualitatively identical although technically more involved and was therefore relegated to the present appendix.
We state the full result below and note that this follows immediately from Refs.~\cite{Baumgart:2018yed,Baumgart:2023pwn} combined with the group theory analysis of the main text.
We refer to the earlier references for a full discussion of how this cross section is derived and how one can include scale-variation uncertainties.

With the above comments in mind, the NLL cross section is given by
\begin{align}
\left.\frac{d\sigma}{dz}\right|_{\rm NLL}
&= \frac{2}{9}\, \sigma_0\, U_H \left[ \left. \left( {\cal F}_0 + {\cal F}_1 \right) \right|_{\Lambda \to 1} \right] \delta(1-z) 
\label{eq:NLL}
\\
& + \frac{2}{9} \frac{\sigma_0}{1-z} U_H \left( (V_J - 1) \Theta_J +1 \right) \left\{ {\cal F}_0\, \frac{e^{\gamma_E \omega_J}}{\Gamma(-\omega_J)} \vphantom{\frac{e^{\gamma_E (\omega_J+2\omega_S)}}{\Gamma(-\omega_J-2\omega_S)}} 
+ \left( (V_S - 1) \Theta_S +1 \right) {\cal F}_1\, \frac{e^{\gamma_E (\omega_J+2\omega_S)}}{\Gamma(-\omega_J-2\omega_S)} \right\}\!. \nn
\end{align}
As claimed, the result is qualitatively identical to the LL expression in Eq.~\eqref{eq:LL}: the first line controls the rate to produce photons with exactly $E = M_{\chi}$ whereas the second determines the endpoint contribution.
Nevertheless, the result involves a number of new functions.
Firstly, $\sigma_0$ is as given in Eq.~\eqref{eq:sigtree}, although as at NLL we account for running couplings, $\aW$ is evaluated at the hard scale $2M_{\chi}$, whereas $\sW$ is evaluated at $\mW$.
Additionally, $\Gamma(x)$ is the Euler gamma function and and $\gamma_E$ the Euler-Mascheroni constant.
The terms resumming the various logarithms are most conveniently expressed in terms $\beta_0 = 19/6$, $\beta_1 = -35/6$, $\Gamma_0 = 4$, and $\Gamma_1 = (8/3)(35/3-\pi^2)$; the two leading perturbative orders of the $\beta$ function and cusp anomalous dimension.
In terms of these, the NLL analog of $e^{-(4\aW/\pi) L_{\chi}^2}$ is
\be
U_H = r_H^2 \exp \left\{ - \frac{2\Gamma_0}{\beta_0^2} \left[ \frac{4\pi}{\aW} \left( \ln r_H + \frac{1}{r_H} - 1 \right) + \left( \frac{\Gamma_1}{\Gamma_0} - \frac{\beta_1}{\beta_0} \right) (r_H - 1 - \ln r_H) - \frac{\beta_1}{2\beta_0} \ln^2 r_H \right] \right\}\!.
\ee
Here $\aW$ should be evaluated at $2M_{\chi}$, whereas the ratio of the coupling between scales is encoded in $r_H = \aW(\mW)/\aW(2M_{\chi})$.
Turning to the endpoint, $\Theta_{J,S}$ are defined as in Eq.~\eqref{eq:LL-JS}, whereas we have the following new expressions,
\begin{align}
V_J&=\exp\left\{\frac{2\Gamma_0}{\beta_0^2}\left[\frac{4\pi}{\aW}\left(\ln r_J+\frac{1}{r_J}-1\right)+\left(\frac{\Gamma_1}{\Gamma_0}-\frac{\beta_1}{\beta_0}\right)(r_J-1-\ln r_J)-\frac{\beta_1}{2\beta_0}\ln^2r_J\right]-\ln r_J\right\}\!,\nn\\
V_S&=\exp\left\{-\frac{3\Gamma_0}{2\beta_0^2}\left[\frac{4\pi}{\aW}\left(\ln r_S+\frac{1}{r_S}-1\right)+\left(\frac{\Gamma_1}{\Gamma_0}-\frac{\beta_1}{\beta_0}\right)(r_S-1-\ln r_S)-\frac{\beta_1}{2\beta_0}\ln^2r_S\right]\right\}\!,\nn\\
\omega_J&=-\frac{2\Gamma_0}{\beta_0}\left[\ln r_J+\frac{\aW}{4\pi}\left(\frac{\Gamma_1}{\Gamma_0}-\frac{\beta_1}{\beta_0}\right)(r_J-1)\right]\Theta_J,\\
\omega_S&=\frac{3\Gamma_0}{2\beta_0}\left[\ln r_S+\frac{\aW}{4\pi}\left(\frac{\Gamma_1}{\Gamma_0}-\frac{\beta_1}{\beta_0}\right)(r_S-1)\right]\Theta_S.\nn
\end{align}
The couplings appearing in the jet (soft) functions are evaluated at the scale $\mu^0_J = 2M_{\chi}\sqrt{1-z}$ ($\mu^0_S = 2M_{\chi}(1-z)$), and in direct analogy to the hard case we have $r_J = \aW(\mW)/\aW(\mu^0_J)$ ($r_S = \aW(\mW)/\aW(\mu^0_S)$).

As at LL, the only terms that track the DM representation are ${\cal F}_{0,1}$.
Following the discussion in the main text, these expressions can only depend on the matrix elements ${\cal M}^{00}$ and ${\cal M}^{+-}$ and explicit evaluation finds they take the form,
\bea
{\cal F}_0 =\, &\frac{1}{128 M_{\chi}^2} \left[ 
|{\cal M}^{33} + 2 {\cal M}^{+-}|^2 \Lambda^d 
+ 2 |{\cal M}^{33} - {\cal M}^{+-}|^2 r_{HS}^{12/\beta_0} \Lambda^c \right]\!, \\
{\cal F}_1 =\, &\frac{r_H^{6/\beta_0}}{64 M_{\chi}^2} \left[ 
2 \left( |{\cal M}^{33}|^2 - 2 |{\cal M}^{+-}|^2 + \textrm{Re}[{\cal M}^{33}({\cal M}^{+-})^*] \right) c_H \Lambda^b\right.\\
&\left.\hspace{1.2cm}
+ |{\cal M}^{33} - {\cal M}^{+-}|^2 r_{HS}^{6/\beta_0} \Lambda^a+ 6 \textrm{Im}[{\cal M}^{33}({\cal M}^{+-})^*] s_H \Lambda^b
\right]\!.
\eea
We have introduced several more expressions capturing the ratio of couplings, $r_{HS} = r_H/r_S$, $s_H = \sin[(6\pi/\beta_0) \ln r_H]$ and $c_H = \cos[(6\pi/\beta_0) \ln r_H]$.
In the first line of Eq.~\eqref{eq:NLL}, the $\Lambda \to 1$ notation implies that we evaluate ${\cal F}_{0,1}$ with $\Lambda^{a-d} = 1$.
For the endpoint contribution, however, these have a more complex form as given below; these expressions are identical to those for the wino where they were evaluated in Ref.~\cite{Baumgart:2018yed}.
In detail,
\bea
\Lambda^a&=1+\frac{\aW(\mu^0_J)}{4\pi}\left[\Gamma_0\Delta^{(2)}_{JSJ}+\beta_0\Delta^{(1)}_{JS}\right]\Theta_J-\frac{3\aW(\mu_S^0)}{\pi}\left[\Delta^{(2)}_{JSS}-\Delta^{(1)}_{JS}\right]\Theta_S,\\
\Lambda^b&=1+\frac{\aW(\mu^0_J)}{4\pi}\left[\Gamma_0\Delta^{(2)}_{JSJ}+\beta_0\Delta^{(1)}_{JS}\right]\Theta_J-\frac{3\aW(\mu_S^0)}{\pi}\Delta^{(2)}_{JSS}\Theta_S,\\
\Lambda^c&=1+\frac{\aW(\mu^0_J)}{4\pi}\left[\Gamma_0\Delta^{(2)}_{J}+\beta_0\Delta^{(1)}_{J}\right]\Theta_J+\frac{6\aW(\mu_S^0)}{\pi}\Delta^{(1)}_{J}\Theta_S,\\
\Lambda^c&=1+\frac{\aW(\mu^0_J)}{4\pi}\left[\Gamma_0\Delta^{(2)}_{J}+\beta_0\Delta^{(1)}_{J}\right]\Theta_J,
\eea
which we have written in terms of
\bea
\Delta_J^{(1)}&=\gamma_E+\psi^{(0)}(-\omega_J),\\
\Delta_J^{(2)}&=\left[\gamma_E+\psi^{(0)}(-\omega_J)\right]^2-\psi^{(1)}(-\omega_J),\\
\Delta_{JS}^{(1)}&=\gamma_E+\psi^{(0)}(-\omega_J-2\omega_S),\\
\Delta_{JSJ}^{(2)}&=\Delta_{JSS}^{(2)}=\left[\gamma_E+\psi^{(0)}(-\omega_J-2\omega_S)\right]^2-\psi^{(1)}(-\omega_J-2\omega_S).
\eea
Note $\psi^{(m)}$ is the polygamma function of order $m$.

\section{Semi-Inclusive Cross Sections for General Representations}
\label{app:WZ-general}

In Eqs.~\eqref{eq:Wino-ZX} and \eqref{eq:Wino-WX} of the main text we provided the semi-inclusive cross sections required to compute the wino continuum photon spectrum.
Here we state the equivalent results for an arbitrary real representation,
\bea
\hspace{-0.3cm}\langle \sigma v \rangle_{\smallZ+\smallX} &\!=\!\frac{\pi \aW^2 \cW^2}{3M_{\chi}^2} \!\!\sum_{j,j'=0}^n \!\!N_j N_{j'} \{ 2(1 \!+\! 2u) j^2 j^{\prime 2} \!+\! (1\!-\!u) [n(n\!+\!1)\!-\!j^2] [n(n\!+\!1)\!-\!j^{\prime 2}] \} s_{0j} s_{0j'}^*, \\
\hspace{-0.3cm}\langle \sigma v \rangle_{\smallW+\smallX} &\!=\!\frac{\pi \aW^2}{3M_{\chi}^2}\!\sum_{j,j'=0}^n \!\!N_j N_{j'} \{ (2\!+\!u) [ n(n\!+\!1)\!-\!j^2][ n(n\!+\!1)\!-\!j^{\prime 2}] \!+\! 4 (1\!-\!u) j^2 j^{\prime 2} \} s_{0j} s_{0j'}^*.
\eea
In both cases we have adopted the shorthand $u = e^{-(3/2\pi) \alpha^{}_{\scaleto{W}{3.pt}} L_{\chi}^2}$ and $N_j = 1/\sqrt{1+\delta_{j0}}$.

\section{Scalar Representations}
\label{app:scalar}

\begin{figure*}[!htp]
\centering
\includegraphics[width=0.5\textwidth]{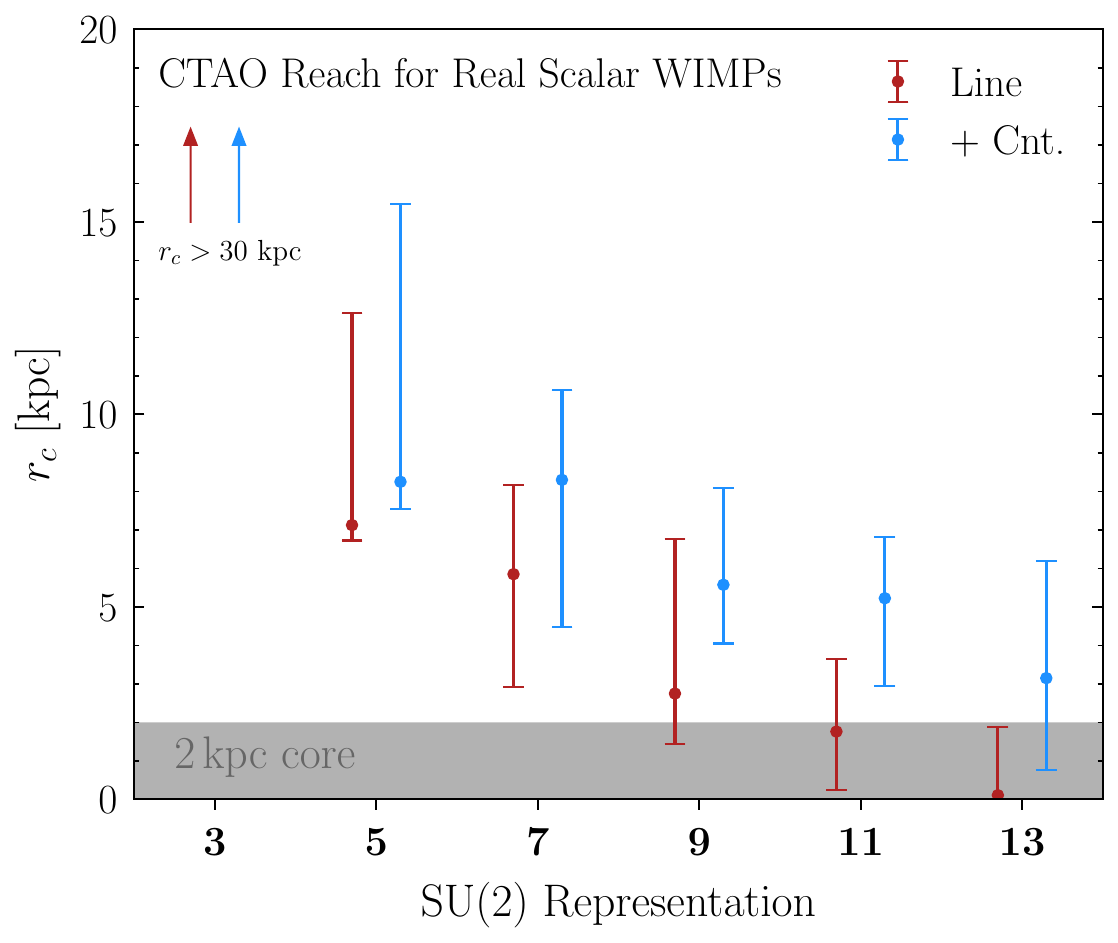}
\vspace{-0.2cm}
\caption{The projected reach of CTAO to real scalar WIMPs assuming 500\,hrs of inner galaxy observations; cf.~Fig.~\ref{fig:CoreReach} for the fermionic analog.
Sensitivity is determined as for the fermions but here we show results solely for the line plus endpoint spectrum (red, Line), together with the sensitivity gained when adding the continuum (blue, Cnt.) alone.
Despite this, the results show that, even without bound states, CTAO can test all real scalar WIMPs with the exception of the tredecuplet.
Adding the bound state contribution and extending the energy sensitivity of the instrument will likely allow CTAO to reach all real multiplets.}
\label{fig:CoreReachScal}
\vspace{-0.2cm}
\end{figure*}

Here we summarize the sensitivity of CTAO to scalar electroweak WIMP representations, in contrast to the fermionic MDM scenario studied in the main text.
The thermal masses of the scalar representations are reported in Tab.~\ref{tab:scal_mass} and are taken from Refs.~\cite{Bottaro:2021snn,Bottaro:2023wjv}.

\begingroup
\renewcommand{\arraystretch}{1.2}
\begin{table}[b!]
\centering
\begin{tabular}{c|c}
$2n+1$ & $M_\chi$ [TeV] \\
\hline
$\mathbf{3}$ & $2.54^{+0.03}_{-0.03}$\\
$\mathbf{5}$ & $15.4^{+0.6}_{-0.3}$\\
$\mathbf{7}$ & $54.2^{+7.0}_{-3.0}$\\
$\mathbf{9}$ & $118^{+26}_{-13}$\\
$\mathbf{11}$ & $199^{+77}_{-21}$\\
$\mathbf{13}$ & $338^{+150}_{-36}$
\end{tabular}
\caption{Predicted thermal masses of the scalar real electroweak representations.}
\label{tab:scal_mass}
\end{table}
\endgroup

The computation of the annihilation rate and spectrum for all aspects except the bound states follows almost directly as in the fermionic case.
In particular, the spin-statistics selection rules allow the same direct $s$-wave annihilation channels for scalars and fermions, with the annihilation cross-section being exactly a factor two larger for scalars~\cite{Cirelli:2007xd}.
At the same time, the non-relativistic potentials for $L=0$ scalars are equal to those for $L=S=0$ fermions~\cite{Cirelli:2007xd}, thus yielding identical Sommerfeld enhancement factors.
Similarly, the details of the EFT formalism used to compute the NLL spectrum is insensitive to the spin and the isospin of DM~\cite{Bauer:2014ula}.
Therefore, we can simply use
\be
\left.\frac{d\sigma_{\rm scalar}}{dz}\right|_{\rm no \,BS}=2\left.\frac{d\sigma_{\rm fermion}}{dz}\right|_{\rm no\,BS}.
\ee

The inclusion of bound states is less direct and is instead heavily affected by the different spin-statistics selection rules.
In fact, the $p\rightarrow s$ channel relevant in the fermionic case is forbidden for scalars.
Therefore, bound states form mostly through $s\rightarrow p$ and $d\rightarrow p$ capture to $L=1$ bound states with odd isospin.
These bound states eventually decay to $L=0$ states with $I=0$ and $I=2$, which dominantly annihilate into SM vectors.
This is similar to the dynamics of the exceptional bound state discussed in Sec.~\ref{ssec:BS} for fermionic multiplets higher than the quintuplet.

We have not computed the scalar bound state contribution, although in Fig.~\ref{fig:CoreReachScal} we demonstrate that with the line and continuum alone CTAO could test all representations except the tredecuplet.
It is likely that if the bound state contribution were included, the full set of scenarios could be reached.

\bibliography{MDMCTA}
\bibliographystyle{JHEP}

\end{document}